\documentclass[hyper]{JHEP3}  

\usepackage{amsmath,amssymb,amsfonts,amsxtra, mathrsfs, makeidx,graphics,graphicx,amsthm,epsfig}
\usepackage[all]{xy}
\pdfoutput=1

\newcommand{\be}{\begin{equation}}
\newcommand{\ee}{\end{equation}}
\newcommand{\beq}{\begin{equation}}
\newcommand{\eeq}{\end{equation}}
\newcommand{\ba}{\begin{array}}
\newcommand{\ea}{\end{array}}
\newcommand{\bi}{\begin{itemize}}
\newcommand{\ei}{\end{itemize}}
\newcommand{\bea}{\begin{eqnarray}}
\newcommand{\eea}{\end{eqnarray}}
\newcommand{\ben}{\begin{enumerate}}
\newcommand{\een}{\end{enumerate}}
\newcommand{\bean}{\begin{eqnarray*}}
\newcommand{\eean}{\end{eqnarray*}}
\newcommand{\eref}[1]{(\ref{#1})}
\newcommand{\sref}[1]{\S\ref{#1}}
\newcommand{\tref}[1]{Table~\ref{#1}}
\newcommand{\fref}[1]{Figure~\ref{#1}}
\newcommand{\nn}{\nonumber}

\newcommand{\tr}{\mathop{\rm Tr}}

\newcommand{\BC}{\mathbb{C}}

\newcommand{\BP}{\mathbb{P}}
\newcommand{\BZ}{\mathbb{Z}}
\newcommand{\sC}{\mathscr{C}}
\newcommand{\sD}{\mathscr{D}}
\newcommand{\sH}{\mathscr{H}}
\newcommand{\sS}{\mathscr{S}}
\newcommand{\comment}[1]{}

\newcommand{\CT}{{\cal T}}

\newcommand{\CMm}{{\cal M}^{\mathrm{mes}}}
\newcommand{\gm}{ g^{\mathrm{mes}}}

\newcommand{\CN}{{\cal N}}

\newcommand{\CC}{{\cal C}}
\newcommand{\BF}{\mathbb{F}}

\newcommand{\perm}{\mathrm{perm}}

\newcommand{\ud}{\mathrm{d}}

\newcommand{\CP}{\mathbb P}

\newcommand{\PL}{\mathrm{PL}}

\newcommand{\f}{{\cal F}^{\flat}}

\newcommand{\firr}[1]{{}^{{\rm Irr}}\!{\cal F}^{\flat}_{#1}}

\title{Higgsing M2-brane Theories}

\author{John Davey, Amihay Hanany, Noppadol Mekareeya and Giuseppe Torri\\
Theoretical Physics Group, The Blackett Laboratory \\
Imperial College London, Prince Consort Road\\ 
London,  SW7 2AZ,  UK \\
Email: {\tt j.davey07, a.hanany, n.mekareeya07, giuseppe.torri08@imperial.ac.uk}}

\abstract{Connections between different M2-brane theories are established via the Higgs mechanism, which can be most efficiently studied on brane tilings. This leads to several M2-brane models, with brane tilings or Chern-Simons levels which have not been considered so far.  The moduli spaces of these models are identified and examined in detail.  The toric diagrams are constructed using Kasteleyn matrices and the forward algorithm.}

\preprint{Imperial/TP/09/AH/03}

\begin{document}

\section{Introduction and Summary}
\subsection*{Introduction}
Considerable progress in understanding theories for multiple M2-branes in various backgrounds has been taking place since the works by Bagger--Lambert \cite{BL} and Gustavsson \cite{gus}.  A key role was played by 3-algebras which, at first sight, do not have a usual field theory structure.  Later, it was understood that the theory could be recast as an ordinary field theory \cite{VanRaamsdonk:2008ft}.  A $U(N) \times U(N)$ Chern--Simons (CS) theory at level $(k, -k)$ with bi-fundamental matter fields was subsequently proposed by Aharony, Bergman, Jafferis and Maldacena (ABJM) \cite{Aharony:2008ug} as a model describing $N$ M2-branes in the $\BC^4/\BZ_k$ orbifold background.  The worldvolume theory of M2-branes on various backgrounds is now believed to be a $(2 + 1)$-dimensional quiver CS theory \cite{Martelli:2008si, Ueda:2008hx, Hanany:2008cd, Hanany:2008fj, phase, Hanany:2009vx}.

It should be noted that the M2-brane models known so far are \emph{not} given by the general class of quiver gauge theories; rather, they are {\bf brane tiling} models\footnote{There have also been studies on brane crystal models \cite{Lee:2006hw, Lee:2007kv, Kim:2007ic, Imamura:2008qs}, which are three-dimensional bipartite graphs. However, in this paper, we focus only on brane tilings.}.  We emphasise that every brane tiling gives rise to a quiver but not every quiver can be recast as a brane tiling.  Preceding to the developments of M2-brane theories, brane tilings have proved to be a very powerful tool in establishing the relation between $(3+1)$-dimensional gauge theories and their moduli spaces which are Calabi--Yau 3-folds \cite{Hanany:2005ve,Franco:2005rj} (see also \cite{Hanany:2005ss,Franco:2005sm, Feng:2005gw, Broomhead:2008an} for further developments and \cite{Kennaway:2007tq, Yamazaki:2008bt} for reviews).  With some modifications, brane tilings have also been successfully applied to $(2 + 1)$-dimensional CS theories \cite{Hanany:2008cd, Hanany:2008fj, phase}. 

One of the interesting aspects of $(2 + 1)$-dimensional CS theories is {\bf toric duality}\footnote{Toric dualities have been also studied in detail in the setup of D3-branes at singularities \cite{Feng:2000mi,Feng:2001xr,Feng:2002zw, Feng:2002fv, Feng:2001bn, Franco:2003ea, Franco:2003ja, Franco:2002mu, Feng:2002kk, Forcella:2008ng}.} \cite{Hanany:2008fj, phase, Franco:2008um, taxonomy, Franco:2009sp, Amariti:2009rb}.  It corresponds to a situation in which one singular Calabi--Yau variety has more than one quiver gauge theory, which we refer to as a \emph{(toric) phase} or a \emph{model}, that has this manifold as its mesonic moduli space of vacua.  In \cite{phase}, we studied a number of toric phases and their brane tilings were presented. Subsequently, we will follow closely the presentation as well as nomenclature in \cite{phase}. 

In this paper, connections between different models are established via the Higgs mechanism, which can be most efficiently studied on brane tilings. This leads to several M2-brane models, with brane tilings or CS levels which have not been considered so far.  In particular, a vacuum expectation value (VEV) is given to a gauge field in a known M2-brane model. Flowing to an energy scale much lower than the scale set by the VEV, we obtain a new field theory resulting from higgsing gauge groups and integrating out massive fields. This amounts to removing one of the edges that separate the corresponding faces in the brane tiling, as well as collapsing the two vertices adjacent to a bivalent vertex into a single vertex of higher valence \cite{Hanany:2005ve, Franco:2005rj}. The CS levels associated with the higgsed gauge groups are added.  

As a result of the Higgs mechanism, one or more points in the original toric diagram \emph{may} be removed. Such cases are said to be results of {\bf partial resolutions} of their original theories.  The methods of partial resolutions have been studied in detail for $(3 + 1)$-dimensional theories \cite{Hanany:2005ve, Franco:2005rj, Feng:2000mi, Feng:2001xr, Feng:2002zw, Beasley:1999uz, Park:1999ep}, and recently have been discussed in the context of M2-brane theories \cite{Franco:2008um, Franco:2009sp}.  In those papers, \emph{one or more points} in the toric diagram of the original theory are removed, subject to the condition that the resulting toric diagram must be a convex polygon (or a convex polyhedron), and the mesonic moduli space is then identified from the resulting toric diagram.  In this paper, instead of starting from removing points from the toric diagram, a quiver field is removed from the brane tiling as a result of the Higgs mechanism, then the forward algorithm \cite{phase, taxonomy} is applied to construct the toric diagram as well as to identify  the mesonic moduli space from the resulting tiling.  The method presented in this paper is clearly more efficient, especially when the original toric diagram is complicated, since the removal of points becomes a result of computations rather than a ``trial-and-error" method.

In Appendix \ref{massivedeform}, we discuss another type of relation between M2-brane theories via massive deformations \cite{Franco:2008um, Kim:2007ic}, where theories are connected by a renormalisation group flow triggered by adding adjoint masses.  In Appendix \ref{app:IIA}, we demonstrate that various theories on M2-branes can be lifted from Type IIA theory on Calabi-Yau 3-folds with RR fluxes.

Below, we summarise key results of this paper in the flow chart and diagrams.

\subsection*{Summary}
\begin{figure}[htbp]
\begin{center}
\vskip -1cm
\includegraphics[totalheight=11cm]{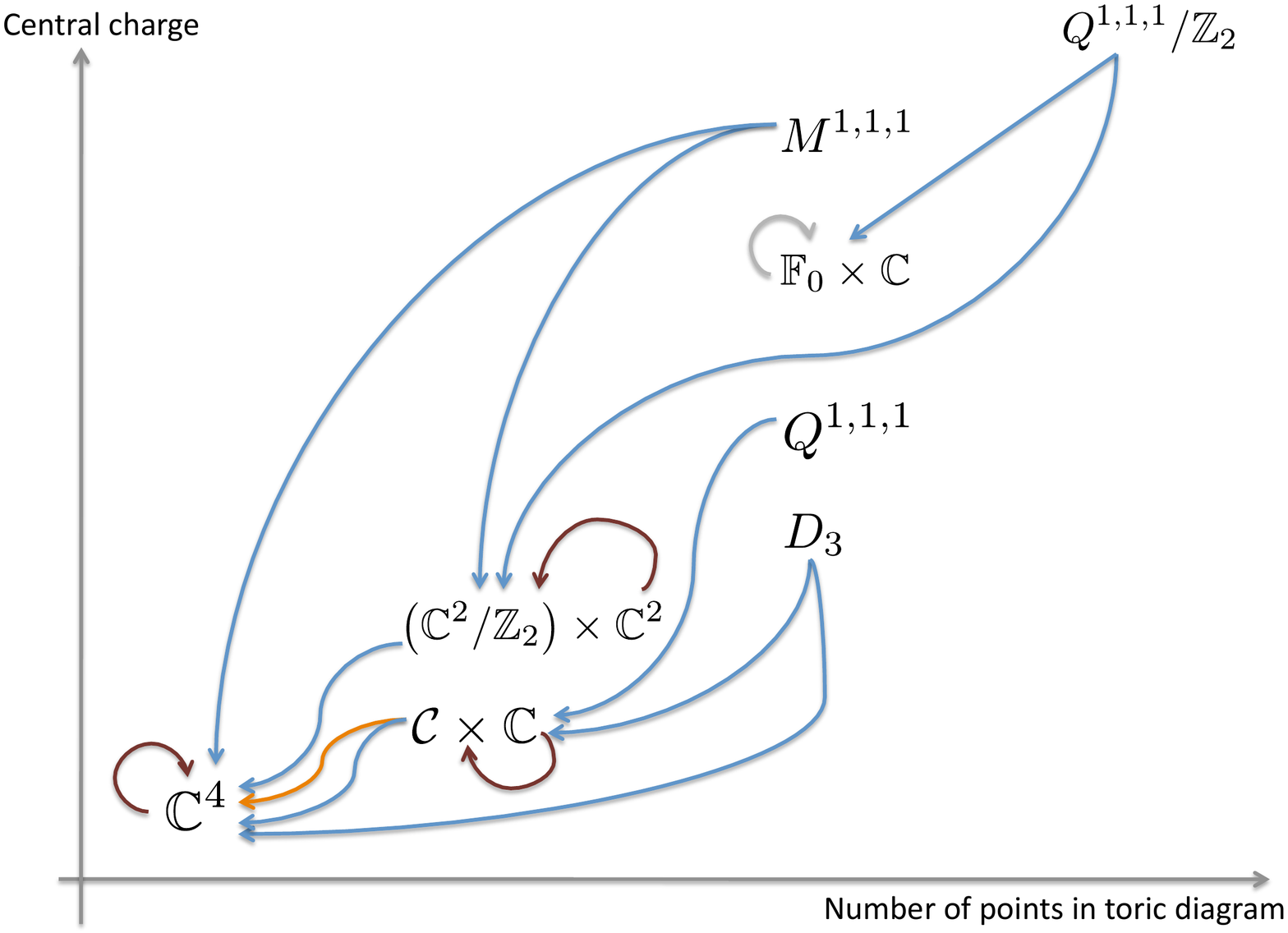}
\caption{A flow chart summarising the connections between various M2-brane theories. (Note that this diagram is not to scale.) A blue arrow from A to B indicates the Higgs mechanism from theory A to theory B.  An orange arrow from A to B indicates an RG flow from theory A to theory B in a massive deformation.  A maroon arrow, which goes from a model to itself, indicates that there is a branch of the moduli space which is $\mathrm{CY}_3$; in the presence of a gauge kinetic term, there is an additional complex degree of freedom and the mesonic moduli space is $\mathrm{CY}_3 \times \BC$. A grey arrow signifies that, under the Higgs mechanism, the central charge does not vary; this is a possible indication that one of the models, or both, does not compute correctly the properties of the SCFT in $(2+1)$-dimension.}
\label{f:flow}
\end{center}
\end{figure}

We draw a flow chart showing the connections between various M2-brane theories in \fref{f:flow}.  In the diagram, the central charges, which are inverse proportional to the volumes of the internal manifolds \cite{Gubser:1998vd, Butti:2005vn}, are plotted against the numbers of points in the toric diagram.  Note that the volumes of various internal manifolds can be found in \cite{Hanany:2008fj}.
As expected, the central charge as well as the number of points in the toric diagram of the resulting theory are less than or equal to those of the original theory. 

\begin{figure}[htbp]
\begin{center}
\vskip-2cm
\includegraphics[totalheight=7.5cm]{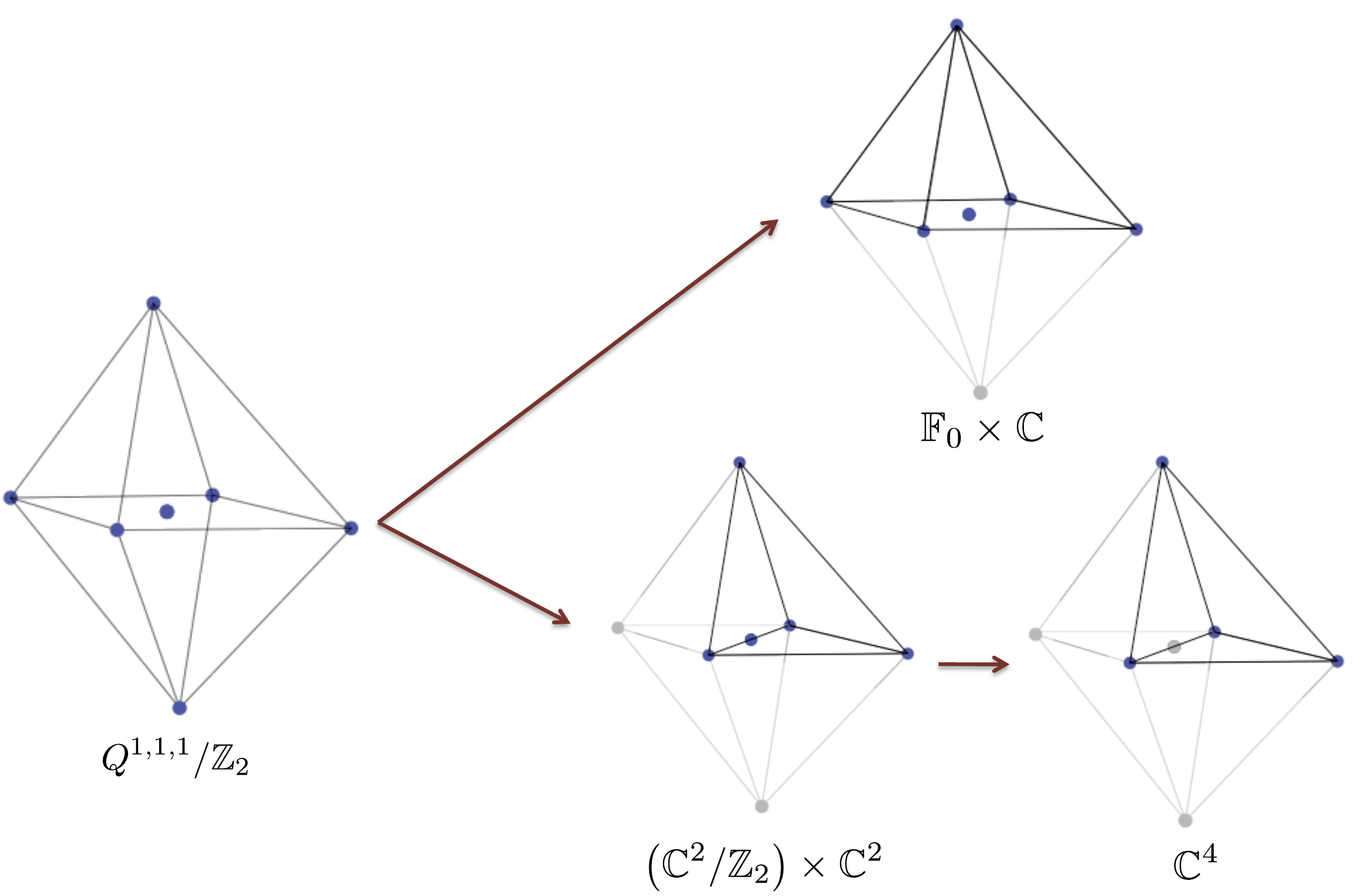}
\caption{Partial resolutions of $Q^{1,1,1}/\BZ_2$.}
\label{f:PRq111z2}
\end{center}
\end{figure}

\begin{figure}[htbp]
\begin{center}
\includegraphics[totalheight=6cm]{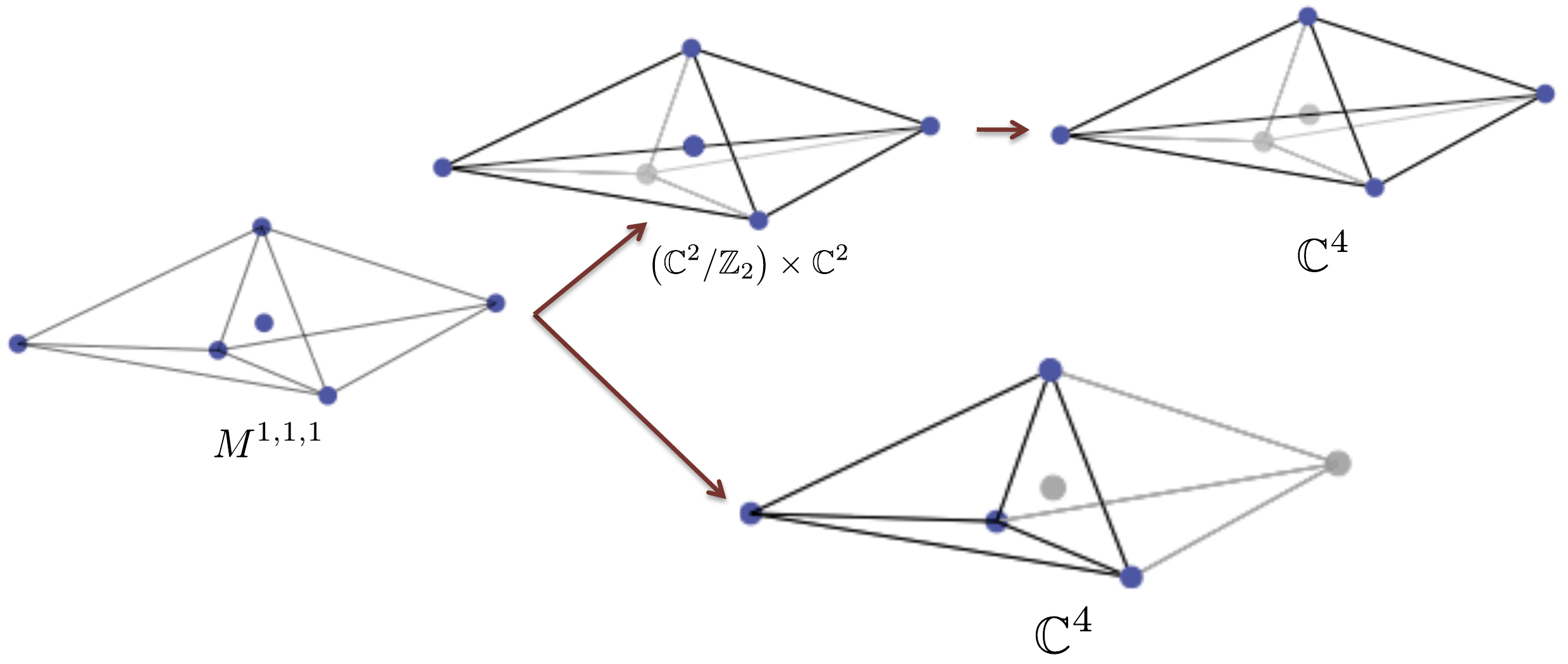}
\caption{Partial resolutions of $M^{1,1,1}$.}
\label{f:PRm111}
\end{center}
\end{figure}

\begin{figure}[htbp]
\begin{center}
\includegraphics[totalheight=4cm]{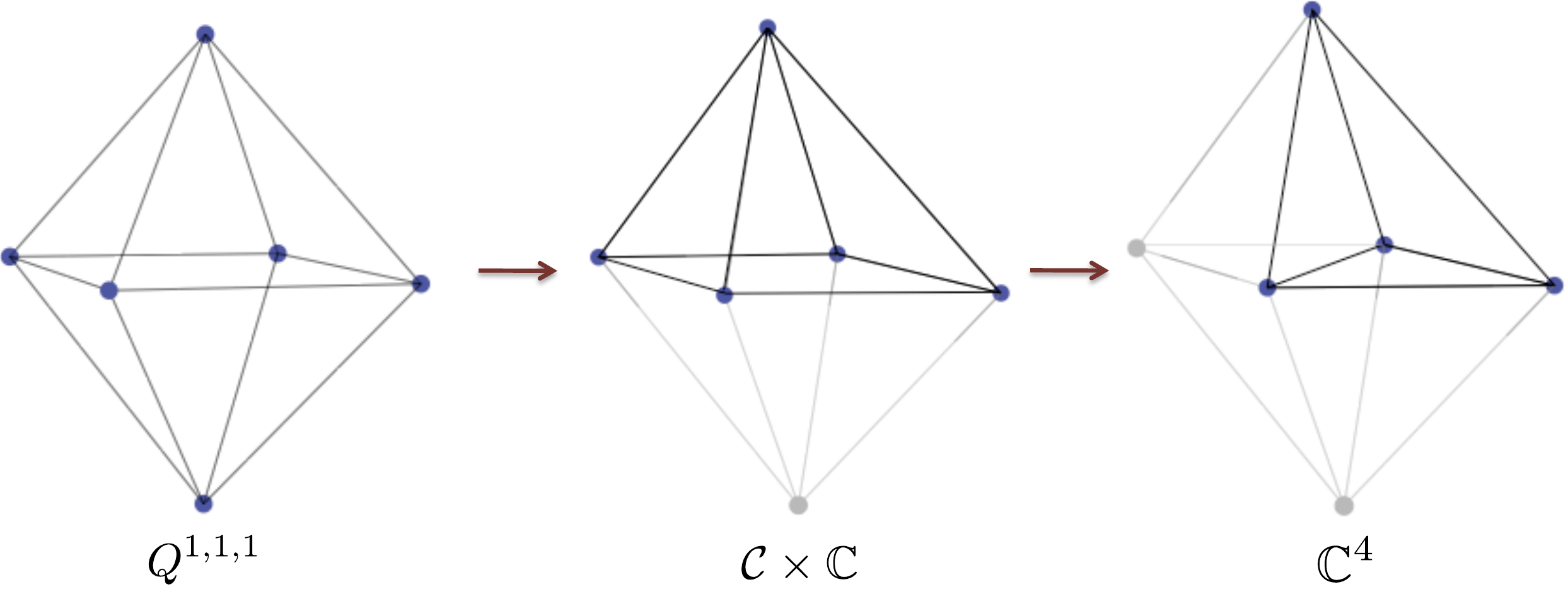}
\caption{Partial resolutions of $Q^{1,1,1}$. Note that this figure has also been discussed in \cite{Franco:2009sp}.}
\label{f:PRq111}
\end{center}
\end{figure}

\begin{figure}[htbp]
\begin{center}
\includegraphics[totalheight=4cm]{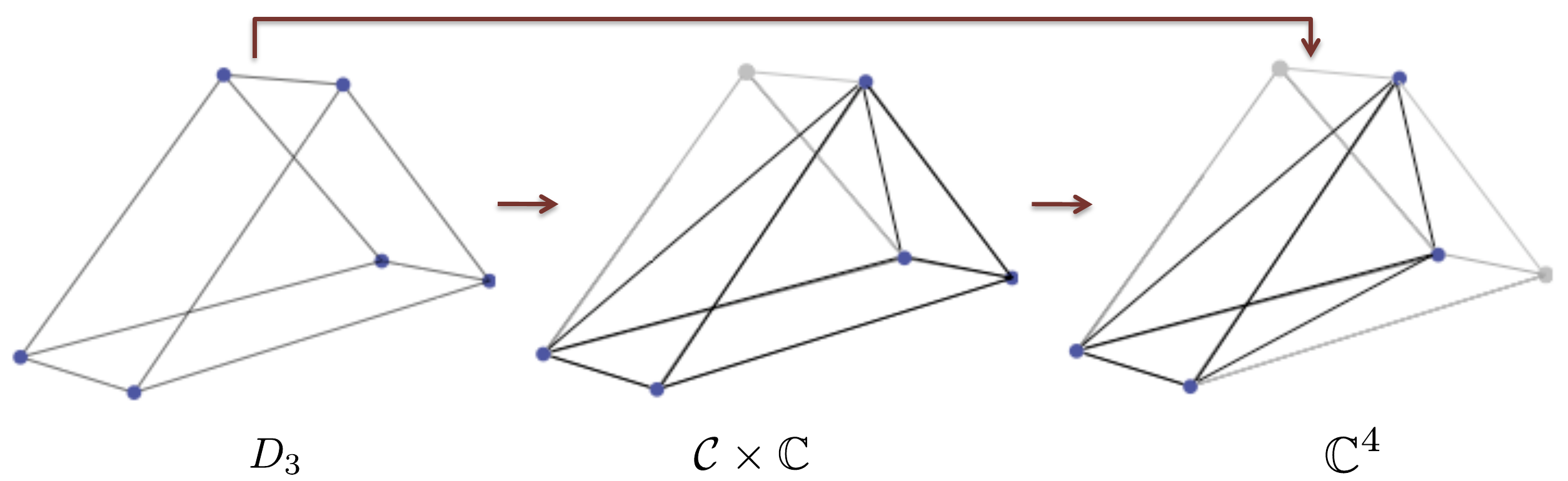}
\caption{Partial resolutions of $D_3$.}
\label{f:PRd3}
\end{center}
\end{figure}

If theory A gets higgsed to a different theory B and the number of points in the toric diagram of B is strictly less than that of the theory A, then it is said that theory B can be obtained by a method of {\it partial resolutions} of theory A.  We summarise how points of the toric diagram are removed as a result of the Higgs mechanism from Figure \ref{f:PRq111z2} to Figure \ref{f:PRd3}.

\comment{
\bea \nn
\xymatrix{
D_3 \ar[r] \ar[dr] 		&  \CC \times \BC  \ar@(r,u)  \ar[d] 		& \ar[l]  Q^{1,1,1} \\
M^{1,1,1} \ar[r] \ar[dr]         &   \BC^4                			&        \\
					&   \BC^2/\BZ_2 \times \BC^2  & \ar[l] \ar[dl] Q^{1,1,1}/\BZ_2 \\     
					&   \BF_0 \times \BC			 &               
}
\eea}


\section{A Summary of The $\mathcal{N}=2$ Supersymmetric Chern--Simons Theory} \label{summary}
Below, we give a brief summary of some useful results on the $(2 + 1)$-dimensional CS theory.
A  review can be found in \cite{phase}.

It is known that field theories on the worldvolume of M2-branes probing Calabi-Yau four-fold ($\mathrm{CY}_4$) singularities are $(2 + 1)$-dimensional $\mathcal{N}=2$ supersymmetric Chern--Simons theories with gauge group $U(N)^G$ and with
bi-fundamental and adjoint matter fields \cite{Martelli:2008si, Ueda:2008hx, Hanany:2008cd}.  The Lagrangian can be written in the superspace notation as
\begin{equation} \label{lagrange}
\mathcal{L}= -\int d^4 \theta\left( \sum\limits_{X_{ab}} X_{ab}^\dagger e^{-V_a} X_{ab} e^{V_b}
-i \sum\limits_{a=1}^G k_a \int\limits_0^1 dt\; V_a \bar{\mathcal{D}}^{\alpha}(e^{t V_a} \mathcal{D}_{\alpha} e^{-tV_a})
\right) + 
\int d^2 \theta \;W(X_{ab}) + \mathrm{c.c.}
\end{equation}
where $a$ indexes the factors in the gauge group ($a=1,\ldots, G$), $X_{ab}$ are the superfields accordingly charged, $V_a$ are the vector multiplets, $\mathcal{D}$ is the superspace derivative, $W$ is the superpotential and $k_a$ are the CS levels which are integers; an overall trace is implicit since all the fields are matrix-valued.
The superpotential $W(X_{ab})$ is assumed to satisfy the {\bf toric condition} \cite{Feng:2002zw}: each chiral multiplet appears precisely twice in the superpotential, once with a positive sign and once with a negative sign. 

The vacuum equations are given by
\begin{eqnarray}
\nn \partial_{X_{ab}} W &=& 0~, \\
\nn \mu_a(X) := \sum\limits_{b=1}^G X_{ab} X_{ab}^\dagger - 
\sum\limits_{c=1}^G  X_{ca}^\dagger X_{ca} + [X_{aa}, X_{aa}^\dagger] &=&  4k_a\sigma_a~, \\
\label{DF} \sigma_a X_{ab} - X_{ab} \sigma_b &=& 0 \ .
\end{eqnarray}
The first set of equations in \eref{DF} is referred to as the \emph{F-term equations}.  
The space of solutions of these equations is called the {\bf Master space} \cite{master}.
The others are in analogy to the  \emph{D-term equations} of $\CN=1$ gauge theories in 3+1 dimensions, with the last equation being a new addition.  
Note that, in the absence of CS terms, this theory can be viewed as a dimensional reduction of a $(3 + 1)$-dimensional ${\CN}=1$ supersymmetric theory. 
In particular, $\sigma_a$, the real scalar in the vector multiplet, arises from the zero mode of the component of the vector field in the reduced direction.
We refer to the space of all solutions for \eref{DF} as the {\bf mesonic moduli space}, and denote it as $\CMm$.   

In \cite{Martelli:2008si, Ueda:2008hx, Hanany:2008cd}, it was shown that
\bea
\label{sumk}
\sum_{a} k_a = 0
\eea 
is a necessary condition for the moduli space to have a branch which is a Calabi--Yau four-fold.  This branch is conjectured to coincide with the mesonic moduli space and is interpreted as the space transverse to the M2-branes.  

Let us focus on the abelian case\footnote{We may as well consider the mesonic moduli space of the non-abelian $U(N)^G$ theory.  This is expected to be the $N$-th symmetric product of the moduli space for the abelian case. The Hilbert series can be obtained using the \emph{plethystic exponential} \cite{pleth, Hanany:2008qc, master, Kim:2009wb}, even though a direct derivation is still evasive.} in which the gauge group is $U(1)^{G}$.
We are interested in the branch in which all the bi-fundamental fields are
generically different from zero.  In this case, the solutions to the
first set of equations in (\ref{DF}) give the {\it irreducible component} of the Master space, $\firr{~}$ \cite{master}.
The third equation of \eref{DF} sets all $\sigma_a$ to a single field, let's say $\sigma$.
The second set of equations in \eref{DF} consists of $G$ equations.  The sum of all the equations is zero, and so there are only $G-1$ linearly independent equations. 
These $G-1$ equations can be divided into one along the direction of the vector $k_a$, and $G-2$ perpendicular to the vector $k_a$.
The former fixes the value of $\sigma$ and leaves a $\mathbb{Z}_k$ action, where $k \equiv  \gcd(\{k_a\})$, by which we need to quotient out in order to obtain the mesonic moduli space. The remaining $G-2$ equations can be imposed by the symplectic quotient of $U(1)^{G-2}$.
Thus, the mesonic moduli space can be written as
\bea
\CMm = \firr{} // \left(U(1)^{G-2} \times \BZ_k \right)~.
\eea
Note that these $G-2$ directions are {\bf baryonic directions} arising from the D-terms\footnote{This does not imply that all possible baryonic directions of the particular Calabi-Yau 4-fold are given by these $G-2$ directions.  It only provides a lower bound. For a given toric phase there are \emph{at least} $G-2$ such baryonic directions and a different phase may give more or less than this number, depending on the number of gauge groups. Such a situation occurs, for example, in Phase II of the $\CC \times \BC$ theory and Phase II of the $D_3$ theory.  The precise number of baryonic charges is equal to the number of external points of the toric diagram minus 4 \cite{phase}.}.
They are in the kernel of the matrix
\begin{equation}\label{C}
C =\left(\begin{matrix}
1 & 1 & 1 & \ldots & 1 \\ k_1 & k_2 & k_3 & \ldots & k_G
\end{matrix}\right) \ .
\end{equation}
For simplicity, in many cases $k =  \gcd(\{k_a\})$ is taken to be 1.  However, it is easy to generalise the result for $k>1$; several explicit examples are given in \cite{Hanany:2008cd, Hanany:2008fj, Hanany:2008qc}.

\subsubsection*{Comments on vanishing CS levels}   
In later sections, we encounter models in which all CS levels are zero, $k_a = 0$.  
These models result from higgsing various $(2 + 1)$-dimensional theories.
A straightforward application of the formalism that is used in this paper 
suggests that the moduli space of each of these models contains a branch which is a Calabi--Yau 3-fold ($\mathrm{CY}_3$).
However, since the field theories studied here live on M2-branes, the mesonic moduli space (Higgs branch) is desired to be a 4-fold. Therefore, we need to add an extra complex degree of freedom. This goes as follows.

When the CS levels vanish, the field $\sigma$ is not constrained by any equation and, therefore, can admit any value. 
In the presence of a gauge kinetic term, which by supersymmetry also adds a kinetic term to $\sigma$, there are two new real degrees of freedom on the moduli space: one is the $\sigma$ fields itself and
the other comes from the gauge field and is given by standard Abelian duality arguments. In particular, a real periodic scalar field $\varphi$ is the dual of the centre of mass gauge field $A_0$.   
See, for example, \S6.2 of \cite{Aganagic:2001ug}. One can start from the action 
\bea
g^2 \int |B|^2 + \int \left( B + \sum_{a=1}^G k_a A_a \right) \wedge \ud A_0,
\eea
containing a dynamical vector field $B$ and a gauge coupling $g$.  
Integrating over $B$ yields the known action for the centre of mass gauge field:
\bea
-\frac{1}{4g^2}\int |\ud A_0|^2 + \int d A_0 \wedge \sum_{a=1}^G k_a A_a~, \label{actiona0}
\eea
whereas integrating over $A_0$ yields 
\bea
B = \ud \varphi - \sum_{a=1}^G k_a A_a~,
\eea
and the action becomes 
\bea
g^2 \int |\ud \varphi - \sum_{a=1}^G k_a A_a|^2~,
\eea
containing a kinetic term for $\varphi$. The $\sigma$ and $\varphi$ fields together can be combined to give a new complex degree of freedom, which implies that the mesonic moduli space is $\mathrm{CY}_3 \times \BC$.

We emphasise that this analysis is inspired by the Type IIA--M theory lift.  In the Type IIA theory, there is a gauge kinetic term for the D2-brane centre of mass gauge field, with gauge coupling $g$.  Note that $g^2 = g_s/l_s$ has a mass dimension 1. From the D2-brane perspective, $g$ is small (with respect to any energy scale) and this corresponds to the UV physics.  Flowing to the IR, $g$ becomes large (with respect to any energy scale) and the gauge kinetic term becomes irrelevant being a dimension 4 operator.  In the IR, the theory is in the large $g_s$ regime, the $l_s$ being kept small in order to avoid the stringy corrections to the gauge theory, and this is equivalent to the M-theory lift. 

\subsection{The Forward Algorithm}
Given the data of the quiver diagram, the superpotential and the CS levels, we can determine the toric diagram of the mesonic moduli space.  These pieces of data are encoded respectively in three matrices: the incidence matrix $d$, the perfect matching matrix $P$, and the CS level matrix $C$.

The incidence matrix $d$ contains the charges of the chiral fields under the $U(1)$ factors of the theory and can be easily obtained from the quiver diagram. The matrix $P$ is a map between the perfect matchings (gauge linear sigma model fields) and the quiver fields; it can be easily obtained from the Kasteleyn matrix of the brane tiling (see \sref{sec:tiling} for more details).

We summarise the process leading to the toric diagram, which is given by the $G_t$ matrix, in the flow chart \eref{forward} \cite{taxonomy}.
Note that the subscripts indicate the sizes of matrices, $G$ is the number of factors of the gauge group, $E$ is the number of quiver fields, $c$ is the number of perfect matchings.

{\footnotesize
\begin{equation}\label{forward}
\begin{array}{lllllll}

\fbox{\mbox{
\begin{tabular}{l}
INPUT 1: \\
~~Superpotential
\end{tabular}
}}
& \rightarrow & P_{E \times c}	& \rightarrow
		& (Q_F)_{(c-G-2)\times c} = \ker (P) \\[-0.3cm] 
&&& \searrow &&&\\

\fbox{
\mbox{
\begin{tabular}{l}
INPUT 2: \\
~~Quiver
\end{tabular}
}}
& \rightarrow & d_{G \times E}	& \rightarrow	&
	(Q_D)_{(G-2) \times c} = 
	 \ker{(C)}_{(G-2) \times G} \cdot \widetilde{Q}_{G \times c}\quad {\small (\text{where} ~d_{G \times E} = \widetilde{Q}_{G \times c} \cdot (P^t)_{c \times E})} \\
&&\\[-0.3cm]
\vspace{-0.5cm}&&& \nearrow &&&\\

\fbox{
\mbox{
\begin{tabular}{l}
INPUT 3: \\
~~CS Levels
\end{tabular}
}}
& \rightarrow & C_{2 \times G}	&&&&\\[-0.3cm]

&&&&~~~~~~\downarrow\\
&&&& 
\hspace{-1in}
(Q_t)_{(c-4) \times c} =
\left( \begin{array}{c}
(Q_D)_{(G-2)\times c} \\
(Q_F)_{(c-G-2) \times c} 
\end{array} \right) \rightarrow
\fbox{\mbox{
\begin{tabular}{l}
OUTPUT:  \\
~~$(G_t)_{4 \times c} = {\ker}(Q_t) $\\
\end{tabular}
}}
\end{array}
\end{equation}}

Because the columns of the $G_t$ matrix have length 4, the Calabi-Yau manifold represented by the toric diagram is a 4-fold.
Since one of the rows of $G_t$, let's say the first, can always be picked to be $(1, \ldots, 1)_{1 \times G}$ \cite{phase, taxonomy}, we can neglect it and consider only a $3 \times c$ matrix that we shall call $G'_t$.
The columns of $G'_t$ give the coordinates of points in the toric diagram, which represent the toric 4-fold by an integer polytope in 3 dimensions.

\subsection{A Summary of Brane Tilings} \label{sec:tiling}
The toric condition \cite{Feng:2002zw} gives rise naturally to a periodic bipartite graph also known as a {\bf brane tiling}.  The smallest unit of repetition is called the {\bf fundamental domain} and is represented in the red frame in the figures of subsequent sections. Each face of the tiling corresponds to a gauge group and each edge corresponds to a bi-fundamental field.  

We will use indices $a, b, \ldots$ for faces, $i, j, \ldots$ for edges, and $\wp, \varrho, \ldots$ for nodes. The field $\Phi_i \equiv X_{ab}$ transforms under $U(1)_a$ and $U(1)_b$ gauge groups, corresponding to the two faces $a$ and $b$ sharing the edge $i$.  The bipartiteness gives rise to a natural orientation of each edge $i$ corresponding to the field $\Phi_i$.  This is indicated by an arrow crossing the edge from the face $a$ to the face $b$:  we adopt the convention that the arrow `circulates' clockwise around the white node and counterclockwise around the black nodes.
The superpotential is given by
\bea
W = \sum_\wp \mathrm{sign}(\wp) \prod_{j_\wp} \Phi_{j_\wp}~,
\eea
where the product is taken over the edges $j_\wp$ around the node $\wp$, and $\mathrm{sign}(\wp)$ is +1 if $\wp$ is a white node and $-1$ if $\wp$ is a black node.

We may assign an integer $n_i$ to the edge $i$ such that the CS level $k_a$ of the gauge group $a$ is given by\footnote{This way of representing $k_a$ is introduced in \cite{Hanany:2008cd} and is also used in \cite{Imamura:2008qs}.}
\bea
k_a = \sum_i d_{ai} n_{i}~, \label{kn}
\eea
where $d$ is the incidence matrix. Due to the bipartiteness of the tiling, we see that the relation $\sum_{a} k_a = 0$ is satisfied as required.   

Many important properties of the tiling are governed by the {\bf Kasteleyn matrix} $K(x, y, z)$, which is defined as follows. The entry $K_{\wp \varrho}$ of the Kasteleyn matrix is zero if there is no connection between the black node $\wp$ and the white node $\varrho$.  Otherwise, $K_{\wp \varrho}$ can be written as
\bea
K_{\wp \varrho} (x,y,z) =  \sum_{ \{ j_{\wp \varrho} \} } \Phi_{j_{\wp \varrho}} z^{n_{j_{\wp \varrho}}} w_{j_{\wp \varrho}}(x,y)~, \label{Kasteleyn}
\eea
where $j_{\wp \varrho}$ represents an edge connecting the black node $\wp$ to the white node $\varrho$, $\Phi_{j_{\wp \varrho}}$ is the field associated with this edge, $w_{j_{\wp \varrho}}(x,y)$ is $x$ or $y$ (or $x^{-1}$ or $y^{-1}$, depending on the orientation of the edge) if the edge ${j_{\wp \varrho}}$ crosses the fundamental domain \cite{Hanany:2005ve,Franco:2005rj} and $w_{j_{\wp \varrho}}(x,y) = 1$ if it does not.

A {\bf perfect matching} is a subset of edges in the tiling or, equivalently, a subset of elementary fields that covers each node exactly once. The coherent component of the Master space of a toric quiver theory is generated by perfect matchings of the associated tiling.  The perfect matchings can be easily obtained from the Kasteleyn matrix as follows: {\bf the quiver fields in the $\alpha$-th term of the permanent\footnote{The permanent is similar to the determinant: the signatures of the permutations are not taken into account and all terms come with a $+$ sign. One can also use the determinant but then certain signs must be introduced \cite{Hanany:2005ve,Franco:2005rj}.} of the Kasteleyn matrix are the elements of the $\alpha$-th perfect matching $p_\alpha$},
\bea \label{permk}
\mathrm{perm}~K = \sum_{\alpha=1}^c p_\alpha ~ x^{u_\alpha} y^{v_\alpha} z^{w_\alpha}~.
\eea
We collect the correspondence between the perfect matchings and the quiver fields in an $E \times c$ matrix (where $E$ is the number of quiver fields and $c$ is the number of perfect matchings), called the {\it perfect matching matrix} $P$.

{\bf The coordinates $(u_\alpha, v_\alpha, w_\alpha)$ of the $\alpha$-th point in the toric diagram are given respectively by the powers of $x, y, z$ in \eref{permk}}.  These coordinates can be collected in the columns of the following matrix:
\beq
G_K = 
\left( \begin{array}{ccccc}  u_1& u_2& u_3&\ldots & u_c \\ v_1& v_2&  v_3&\ldots & v_c \\ w_1& w_2& w_3&\ldots & w_c \end{array} \right)~.
\label{e:gk}
\eeq

\subsubsection*{Remarks on the $G_K$ and $G'_t$ matrices} 
Since we can multiply \eref{permk} by a non-zero variable (with a unit power), we may extend the coordinates of the toric diagram to $(1, u_\alpha, v_\alpha, w_\alpha)$, and so the $G_K$ matrix becomes
\beq
\widetilde{G}_K = 
\left( \begin{array}{ccccc}  1 & 1 & 1& \ldots & 1 \\ u_1& u_2& u_3&\ldots & u_c \\ v_1& v_2&  v_3&\ldots & v_c \\ w_1& w_2& w_3&\ldots & w_c \end{array} \right)~.
\label{e:tildegk}
\eeq
There exists a $GL(4,\BZ)$ transformation $\mathcal{R}$ such that $G_t = \mathcal{R} \cdot \widetilde{G}_K$.  Note that since we are working on the ring $\BZ$, whose only invertible elements are $+1$ and $-1$, it follows that $\det \mathcal{R} = \pm 1$.  Equivalently, we can perform a series of elementary row operations on the $\widetilde{G}_K$ matrix and end up with the $G_t$ matrix and \emph{vice versa}.  It can be seen that row operations can be performed such that the rows $(1~1~ \ldots ~ 1)$ in the $\widetilde{G}_K$ and $G_t$ matrices remain unchanged.  Since $(1~1 \ldots~1)$ does not appear in the coordinates of 3d toric diagram, we may omit it and $\widetilde{G}_K$ and $G_t$ become respectively $G_K$ and $G'_t$. 
In many cases, it suffices to consider simply a $GL(3, \BZ)$ transformation $\mathcal{T}$ such that $G_K = \mathcal{T} \cdot G'_t$; the toric diagram is rotated or reflected under this transformation.

\subsection{The Global Symmetries}  
As can be seen from \emph{all} examples in the subsequent sections, it is possible to perform a series of elementary row operations on the $G_K$ (or $G'_t$) matrix such that the rows of the resulting matrix contain weights of non-abelian factors in the mesonic symmetry.  To illustrate this, let us consider the $M^{1,1,1}$ theory (\sref{sec:summ111}), whose mesonic symmetry is $SU(3) \times SU(2) \times U(1)$.  The $G'_t$ matrix is given by \eref{e:toricdiafano24}:
\bea
G'_t = \left(
\begin{array}{cccccc}
 1 & -1 & 0 &  0 & 0 & 0 \\ 
 0 & 1 & -1 &  0 & 0 & 0 \\
  0 & 0 & 0 & 1 & -1 & 0 
\end{array}
\right)~.
\eea
Note that the first two rows contain weights of $SU(3)$ and the third row contains weights of $SU(2)$. Thus, we arrive at the important observation that \emph{the non-abelian mesonic symmetry is encoded in the coordinates of the toric diagram}. 

The existence of a non-abelian $SU(k)$ factor (with $k>1$) in the mesonic symmetry is also implied by the number $k$ of repetitions of columns in the $Q_t$ matrix. 

Since the mesonic symmetry has total rank 4, we can classify all possible mesonic symmetries according to the partitions of 4 as follows:
\begin{itemize}
\item $SU(4) \times U(1)$~,
\item $SU(3) \times SU(2) \times U(1)$~,
\item $SU(3) \times U(1) \times U(1)$~,
\item $SU(2) \times SU(2) \times SU(2) \times U(1)$~,
\item $SU(2) \times SU(2) \times U(1) \times U(1)$~,
\item $SU(2) \times U(1) \times U(1) \times U(1)$~,
\item $U(1) \times U(1) \times U(1) \times U(1)$~.
\end{itemize}

If there is precisely one $U(1)$ factor in the mesonic symmetry, we can immediately identify it with the R-charge.  Otherwise, there is a minimisation problem to be solved in order to determine which linear combination of these $U(1)$ charges gives the right R-charge in the IR \cite{Hanany:2008fj}.  In some simple cases, we can bypass this calculation using a symmetry argument. 

The precise number of baryonic charges is equal to the number of external points of the toric diagram minus 4 \cite{phase}.
The global symmetry of the theory is a product of mesonic and baryonic symmetries.

\subsection{Notation and Nomenclature}  
We denote the $i$-th bi-fundamental field transforming in the fundamental (anti-fundamental) representation of the gauge group $a$ (gauge group $b$) by $X^{i}_{ab}$ and, similarly, $\phi^i_a$ denotes the $i$-th adjoint field in the gauge group $a$ (when there is only one adjoint field charged under the $a$-th gauge group the $i$-index is dropped).

We adopt the nomenclature of toric phases as in \cite{phase}, \emph{e.g.} Phase I of the $\BC^4$ theory refers to the ABJM theory.  When necessary, a shorthand notation for the features of brane tilings as in \tref{t:nomen} may be used, \emph{e.g.} Phase I of the $\BC^4$ theory is referred to as the `chessboard model' and denoted by $\sC$.  However, it should be noted that this by no means specifies a unique model.  In fact, in some cases we need to further specify the CS levels associated with the tiling; these will be written as subscripts, \emph{e.g.} the shorthand notations for Phases III-A and III-B of $\CC \times \BC$ are respectively ${\mathscr{D}_2 \mathscr{H}_1}_{(0,1,-1)}$ and ${\mathscr{D}_2 \mathscr{H}_1}_{(2,-1,-1)}$.

\begin{table}[h]
 \begin{center} 
   {\small
  \begin{tabular}{|c||c|}
  \hline
  Shorthand notation & Object referred to \\
  \hline
  $\mathscr{C}$ & chessboard \\
  $\mathscr{D}_n$ & $n$ double bonds \\   
  $\mathscr{H}_n$ & $n$ hexagons  \\
  $\mathscr{S}_n$ & $n$ squares  \\   
  $\partial_n$ & $n$ diagonals  \\   
  $\mathscr{O}_n$ & $n$ octagons  \\   
  \hline
  \end{tabular}}
  \end{center}
\caption{Shorthand notation for the nomenclature of the brane tilings used in this paper.}
\label{t:nomen}
\end{table}

\section{Higgsing The $\CC \times \BC$ Theory}
\subsection{Higgsing Phase I of $\CC \times \BC$}
\subsection*{A summary of Phase I of $\CC \times \BC$ (the $\sD_1\sC$  model)}
\begin{figure}[ht]
 \centerline{  \epsfxsize = 5cm \epsfbox{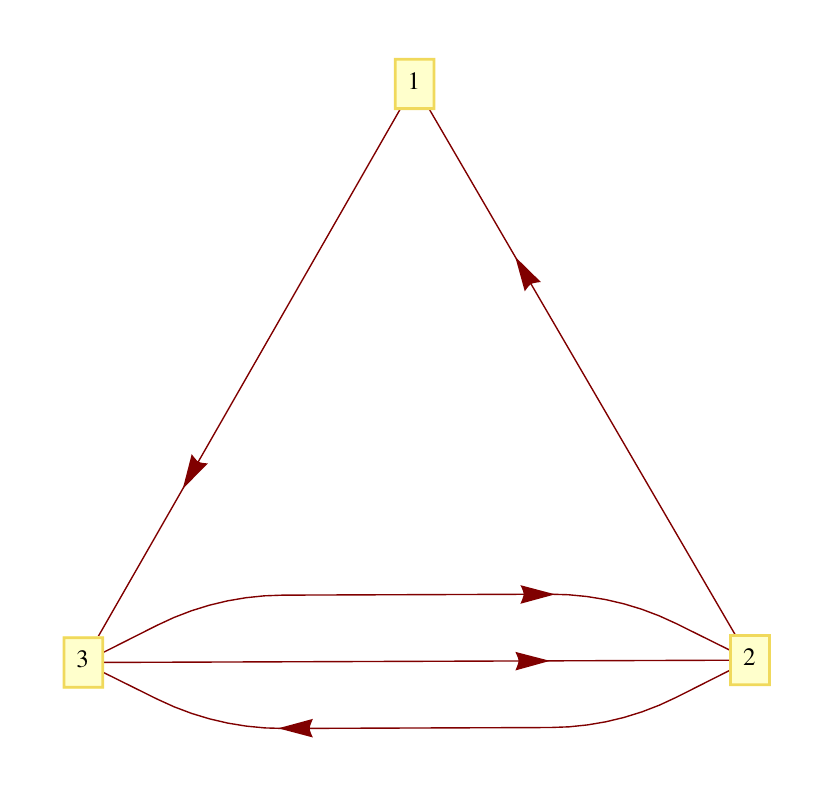}  \hskip 10mm \epsfxsize = 6.5cm \epsfbox{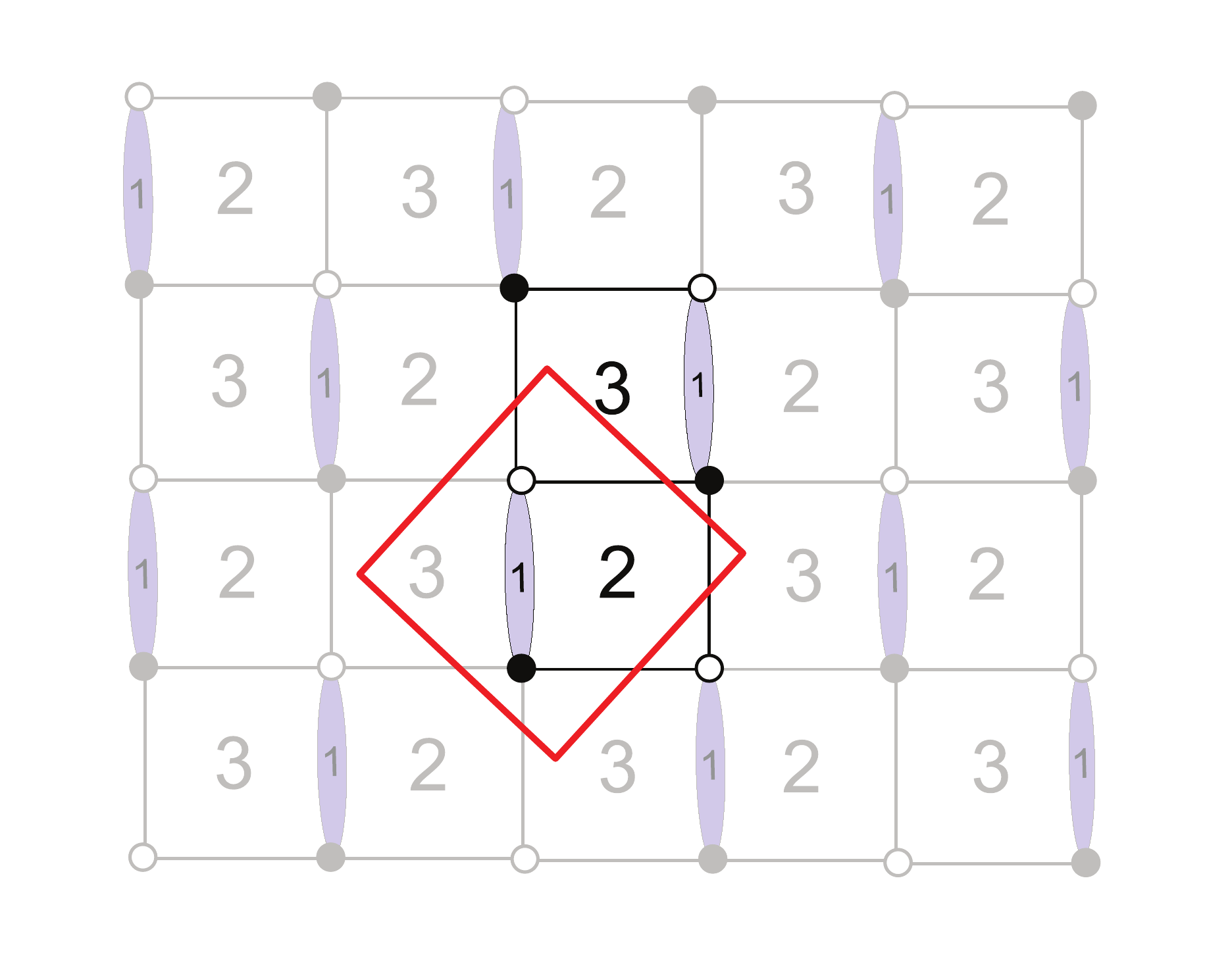}}
\caption{[Phase I of $\CC \times \BC$] (i) Quiver diagram of the $\sD_1\sC$  model.\ (ii) Tiling of the $\sD_1\sC$  model.}
  \label{f:phase1conxc}
\end{figure}

This model has 3 gauge groups and 5 chiral multiplets which are denoted as $X_{13}, X_{23}, X_{21}, X_{32}^1, X_{32}^2$, with a superpotential:
\bea
W = \tr \left(\epsilon_{ij} X_{21} X_{13} X^i_{32} X_{23} X^j_{32}\right)~. 
\eea
The quiver diagram and tiling are given in Figure \ref{f:phase1conxc}.
We choose the CS levels to be 
\bea
k_1 = 1,~k_2 = -1,~ k_3 =0~.
\eea

\begin{figure}[ht]
 \centerline{  \epsfxsize = 9cm  \epsfbox{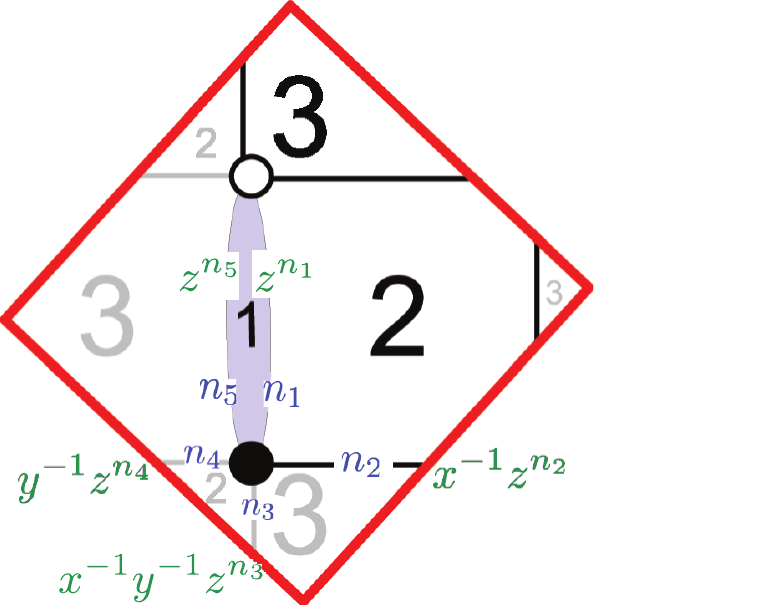} }
 \caption{[Phase I of $\CC \times \BC$]. The fundamental domain of the tiling for the $\sD_1\sC$  model: assignments of the integers $n_i$ to the edges are shown in blue and the weights for these edges are shown in green.}
  \label{f:fdphase1conxc}
\end{figure}

\begin{figure}[h]
\begin{center}
  \includegraphics[totalheight=4cm]{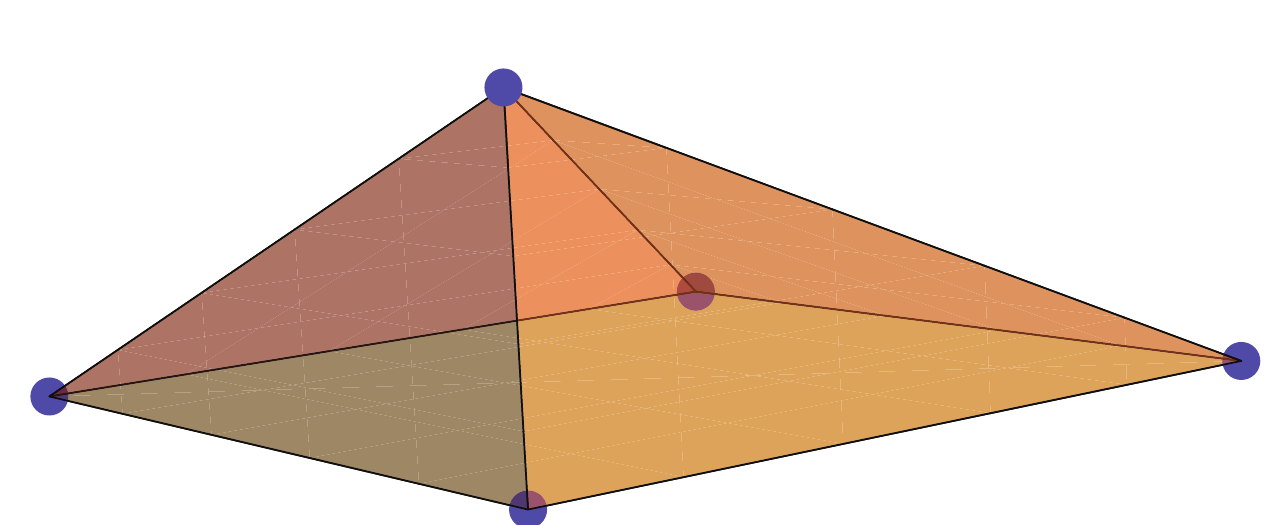}
 \caption{The toric diagram of the $\CC \times \BC$ theory.}
  \label{f:torconxc}
\end{center}
\end{figure}

\paragraph{The Kasteleyn matrix.}   We assign the integers $n_i$ to the edges according to Figure \ref{f:fdphase1conxc}.  We find that 
\bea
\text{Gauge group 1~:} \qquad k_1 &=& -1  = -n_1+n_5 ~, \nn \\
\text{Gauge group 2~:} \qquad k_2 &=& 1 = -n_2 +n_1 - n_4 + n_3 ~,  \nn \\
\text{Gauge group 3~:} \qquad k_3 &=& 0 = -n_3 + n_2 + n_4 -n_5~.
\eea  
We choose
\bea
n_1= 1,\qquad n_i=0 \;\;\text{otherwise}~.
\eea
We can construct the Kasteleyn matrix, which for this case is just a $1\times 1$ matrix and, therefore, coincides with its permanent:
\bea  \label{permKph1conxc}
K &=& X_{13} z^{n_5} + X_{21} z^{n_1} + X^1_{32} x^{-1} z^{n_2} + X_{23} x^{-1} y^{-1} z^{n_3} + X^2_{32} y^{-1} z^{n_4}  \nn \\
&=& X_{13} + X_{21} z + X_{32}^1 x^{-1}  + X_{23} x^{-1} y^{-1}  + X_{32}^2 y^{-1} \qquad \text{(for $n_1 = 1$ and $n_i =0$ otherwise)} ~. \nn\\
\eea
The powers of $x, y, z$ in each term of $K$ give the coordinates of each point in the toric diagram.
We collect these points in the columns of the following $G_K$ matrix:
\bea
G_K = \left(
\begin{array}{ccccc}
 -1 & 0 & -1 & 0 & 0 \\
 -1 & 0 & 0 & -1 & 0 \\
 0 & 0 & 0 & 0 & 1
\end{array}
\right)~. \label{gkconxc}
\eea
The toric diagram is drawn in Figure \ref{f:torconxc}.

\paragraph{The $G_K$ matrix and global symmetry.}  
The extended $G_K$ matrix is given by
\bea
\widetilde{G}_K = \left(
\begin{array}{ccccc}
  1 & 1 &  1 & 1  & 1 \\
 -1 & 0 & -1 & 0 & 0 \\
 -1 & 0 & 0 & -1 & 0 \\
 0 & 0 & 0 & 0 & 1
\end{array}
\right)~.
\eea
We choose a $GL(4, \BZ)$ matrix (which has a determinant $\pm 1$) 
\bea
\mathcal{R}
= \left(
\begin{array}{cccc}
 1 & 0 & 0 & 0 \\
 -1 & 1 & 1 & 1 \\
 0 & 1 & 0 & -1 \\
 0 & 1 & 0 & 0
\end{array}
\right)
\eea
such that the $\widetilde{G}_K$ matrix is transformed into
\bea
G_t = \mathcal{R} \cdot \widetilde{G}_K = \left(
\begin{array}{ccccc}
 1 &  1  & 1 & 1 & 1 \\
 1 & -1 & 0 & 0 & 0 \\
 0 & 0 & 1 & -1 & 0 \\
 1 & 0 & 1 & 0 & 0
\end{array}
\right)~.
\eea
After removing the first row, we obtain 
\bea
G'_t = \mathcal{R} \cdot \widetilde{G}_K = \left(
\begin{array}{ccccc}
 1 & -1 & 0 & 0 & 0 \\
 0 & 0 & 1 & -1 & 0 \\
 1 & 0 & 1 & 0 & 0
\end{array}
\right)~. \label{Gptconxc}
\eea
Observe that the first and the second rows of the $G'_t$ matrix contain weights of $SU(2)$.  These suggest that the non-abelian part of the global symmetry of the $\CC \times \BC$ is given by $SU(2) \times SU(2)$.  Since the total rank of the mesonic symmetry is 4, this is clearly $SU(2) \times SU(2) \times U(1) \times U(1)$ \cite{phase}.

Below, there is a study of the Higgs mechanism of this theory.

\subsubsection{\emph{Phase I of $\BC^4$} from  giving a VEV to $X_{13}$}
\begin{figure}[ht]
 \begin{center} 
   \vskip 2cm
  \hskip -8cm
{ \epsfxsize = 9cm \epsfbox{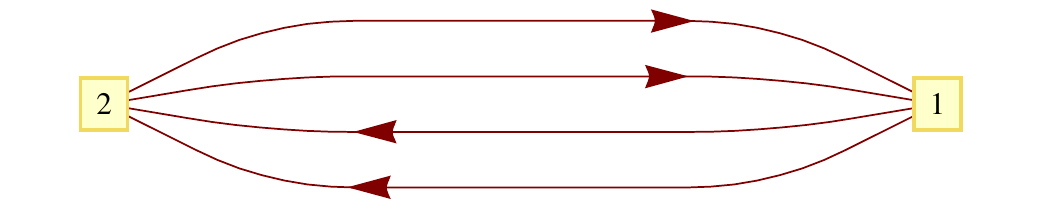}}  
   \vskip -4.0cm
  \hskip 7.5cm
{ \epsfxsize = 7cm \epsfbox{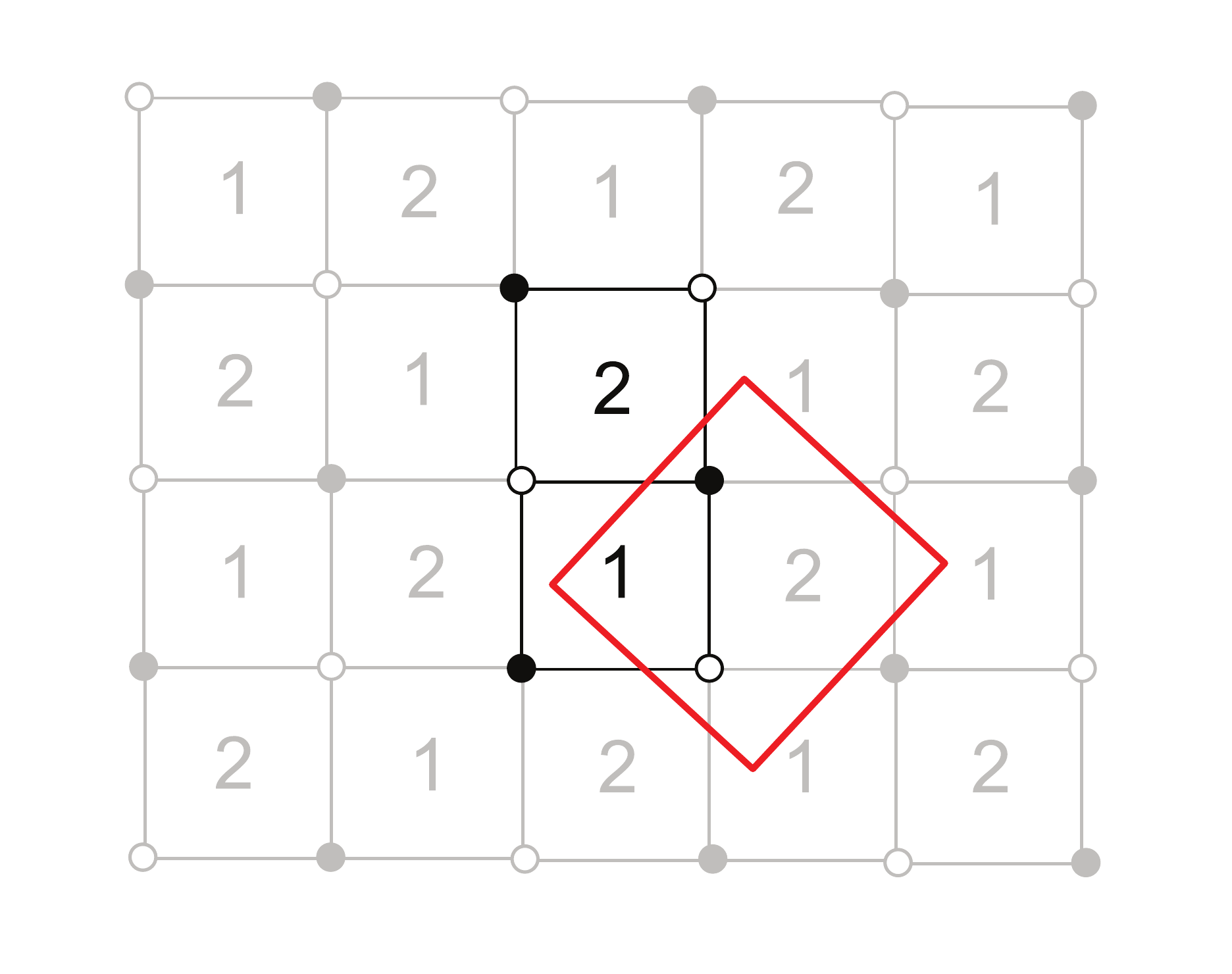}}
  \caption{[Phase I of $\BC^4$] (i) Quiver diagram for the $\mathscr{C}$ model. \ (ii) Tiling for the $\mathscr{C}$ model.}
  \label{f:con}
\end{center}  
\end{figure}

Let us turn on a VEV to $X_{13}$.
Flowing to an energy scale much lower than the scale set by the VEV, we obtain a new field theory resulting from Higgsing gauge groups and integrating out massive fields.  
The quiver nodes 1 and 3 are combined into one node, which is identified as node 1.
The new quiver and tiling are drawn in \fref{f:con}.  
The superpotential is 
\bea
W = \tr \left( \epsilon_{ij} X^2_{12} X^i_{21} X^1_{12} X^j_{21} \right)~. 
\eea
The CS levels associated with the higgsed gauge groups are added, and so the new CS levels are
\bea
k_1 = 1, \quad k_2 = -1~.
\eea
Therefore, the resulting theory is Phase I of $\BC^4$ (the ABJM theory).

\begin{figure}[ht]
 \centerline{  \epsfxsize = 7cm  \epsfbox{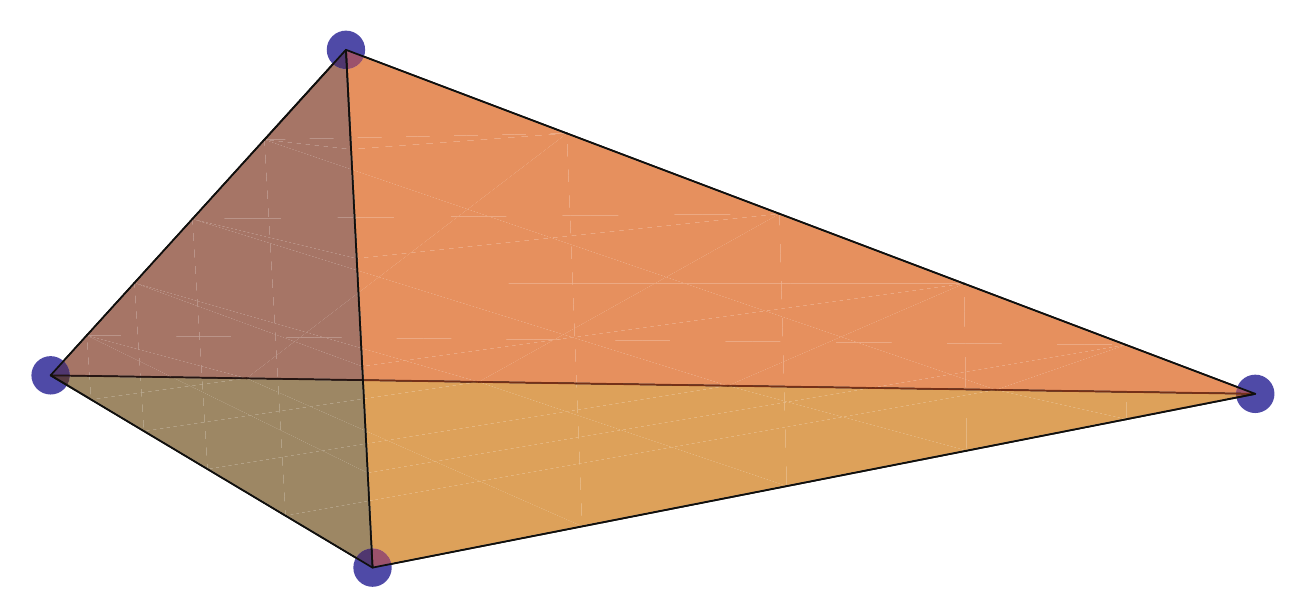} }
 \caption{The toric diagram of the $\BC^4$ theory.}
  \label{f:torabjm}
\end{figure}

\paragraph{The Kasteleyn matrix.} The Kasteleyn matrix can be obtained from \eref{permKph1conxc} by setting $X_{13}$ to zero and identifying subscripts 3 with 1:
\bea 
K = X^1_{21} z + X_{12}^1 x^{-1}  + X^2_{21} x^{-1} y^{-1}  + X_{12}^2 y^{-1}~. \label{permKabjm}
\eea
The powers of $x, y, z$ in each term of $K$ give the coordinates of each point in the toric diagram.
We collect these points in the columns of the following matrix:
\bea
G_K = \ \left(
\begin{array}{cccc}
 -1 & 0 & -1 & 0 \\
 0 & -1 & -1 & 0  \\
 0 & 0 & 0 & 1 
\end{array}
\right)~.
\eea
The toric diagram is drawn in \fref{f:torabjm}.

\subsubsection{\emph{Phase II of $\BC^4$} from giving a VEV to $X_{23}$}
\begin{figure}[ht]
\begin{center}
  \vskip 1cm
  \hskip -8cm
{  \epsfxsize = 7cm  \epsfbox{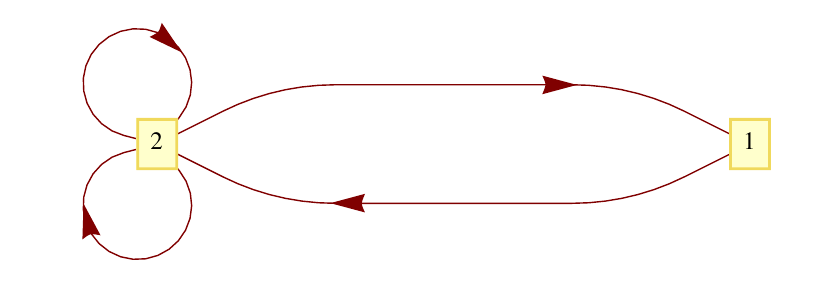} }
  \vskip -3.0cm
  \hskip 8cm
{  \epsfxsize = 6cm \epsfbox{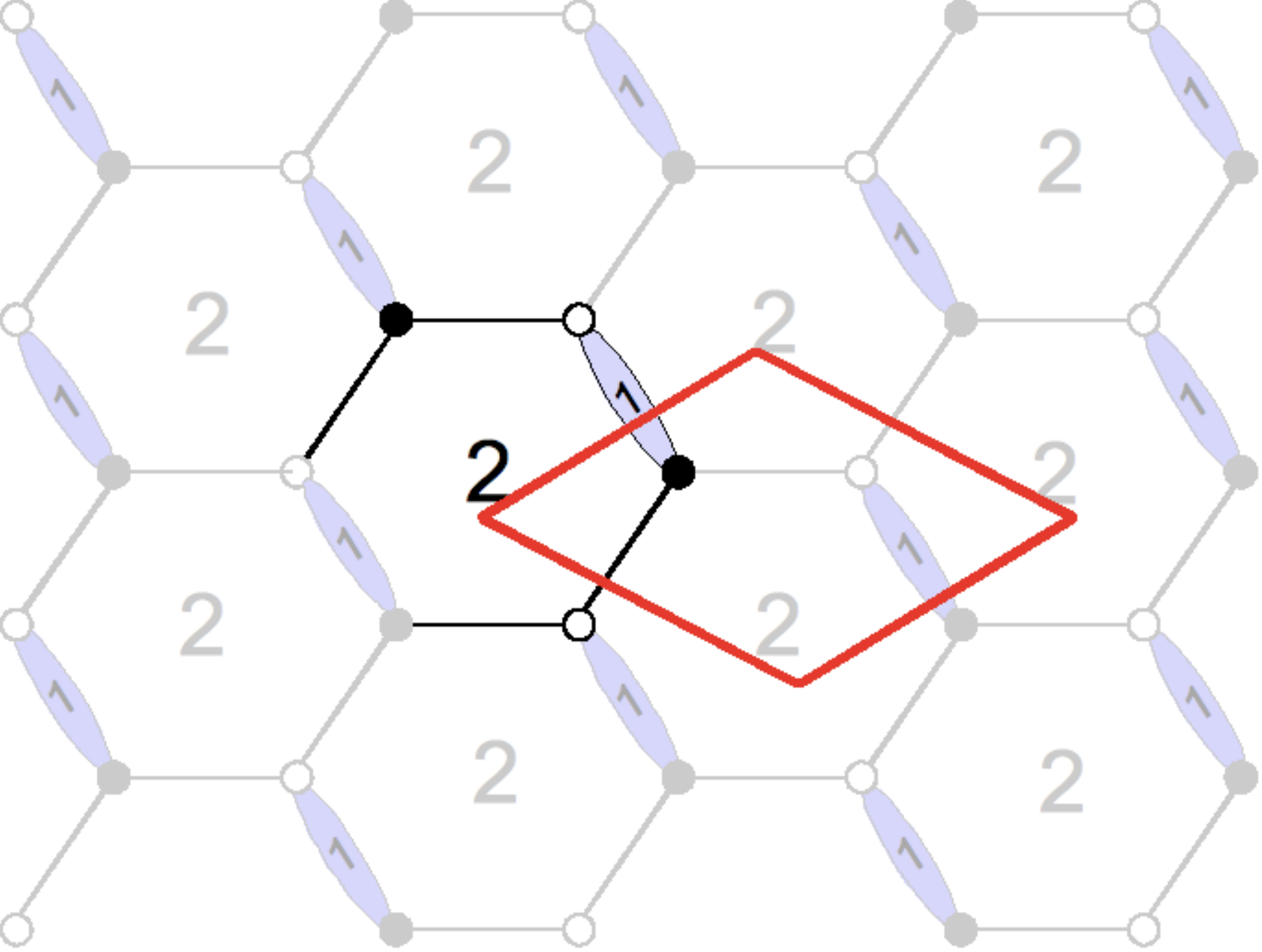}}
 \caption{[Phase II of $\BC^4$] (i) Quiver diagram for the $\sD_1\sH_1$ model. \ (ii) Tiling for the $\sD_1\sH_1$ model.}
  \label{f:phase2c4}
\end{center}
\end{figure}

Let us turn on a VEV to $X_{23}$. 
Nodes 2 and 3 are combined into one node (which is identified as node 2).
The new quiver and tiling are drawn in \fref{f:phase2c4}.  
The superpotential is given by
\bea
W = \tr(X_{21}X_{12}[\phi_2^1,\phi_2^2])~.
\eea

\noindent The CS levels associated with the higgsed gauge groups are added, and so the new CS levels are
\bea
k_1 = 1, \quad k_2 = -1~.
\eea
Therefore, the resulting theory is Phase II of the $\BC^4$ theory. 

\paragraph{The Kasteleyn matrix.} The Kasteleyn matrix can be obtained from \eref{permKph1conxc} by setting $X_{23}$ to zero and identifying subscripts 3 with 2:
\bea \label{Kabjm}
K = X_{12}+X_{21} z + \phi_{2}^1 x^{-1} + \phi_{2}^2 y^{-1}~.
\eea 
The powers of $x, y, z$ in each term of $K$ give the coordinates of each point in the toric diagram.
We end up with the toric diagram drawn in \fref{f:torabjm}.

\paragraph{Higgsing Phase II of $\BC^4$.}  Giving a VEV to $X_{12}$ or $X_{21}$ leads to the one-hexagon tiling with zero CS level.
The tiling suggests that there is a branch of the moduli space which is $\BC^3$.  
As discussed in \sref{summary}, in the presence of a gauge kinetic term, there is an additional complex degree of freedom.
In which case, the mesonic moduli space is $\BC^4$. 

\subsubsection{\emph{The $\CC \times \BC$ theory} from giving a VEV to $X_{21}$} \label{sec:d2onconwithRR}
Let us turn on a VEV to $X_{21}$.
Nodes 1 and 2 are combined into one node (which is identified as node 1).
Therefore, we are left with gauge groups 1 and 3.  
For convenience, let us relabel the gauge group 3 as 2.
The new quiver and tiling are drawn in \fref{f:con} (with gauge groups 1 and 2 interchanged).  
The new superpotential is 
\bea
W = \tr \left( \epsilon_{ij} X^1_{12} X^i_{21} X^2_{12} X^j_{21}\right)~. 
\eea
The CS levels associated with the higgsed gauge groups are added, and so
\bea
k_1 = 0,\quad k_2 = 0~.
\eea
The tiling suggests that there is a branch of the moduli space which is the conifold ($\CC$).
As discussed in \sref{summary}, in the presence of the gauge kinetic term, an additional complex degree of freedom arises.  
Therefore, the mesonic moduli space is $\CC \times \BC$.

\comment{
\paragraph{The Kasteleyn matrix.} The Kasteleyn matrix can be obtained from \eref{permKph1conxc} by setting $X_{21}$ to zero and identify subscripts 1 with 2:
\bea
K = X_{13} + X_{32}^1 x^{-1}  + X_{23} x^{-1} y^{-1}  + X_{32}^2 y^{-1}~. 
\eea
The powers of $x, y$ in each term of $K$ give the coordinates of each point in the toric diagram.
The toric diagram is drawn in \fref{f:torconifold}.
\begin{figure}[ht]
 \centerline{  \epsfxsize = 3cm  \epsfbox{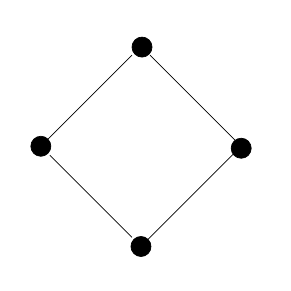} }
 \caption{The toric diagram of the conifold ($\CC$) theory.}
  \label{f:torconifold}
\end{figure}}

\subsection{Higgsing Phase II of $\CC\times \BC$}
\subsection*{A summary of Phase II of $\CC\times \BC$ (the $\sH_2$ Model)}
This model has 2 gauge groups and 6 chiral multiplets denoted as $\phi_1, \phi_2, X_{12}^1, X_{12}^2, X_{21}^1, X_{21}^2$.  The quiver and toric diagrams are drawn in Figure \ref{f:phase2conxc}. Note that in 3+1 dimensions this tiling corresponds to the $\BC^2/\BZ_2 \times \BC$ theory. The superpotential is given by
\begin{equation} \label{spph1conc}
W = \tr \left( \phi_1 (X_{12}^2 X_{21}^1 - X_{12}^1 X_{21}^2 ) + {\phi}_2 (X_{21}^2 X_{12}^1 - X_{21}^1 X_{12}^2) \right) \ .
\end{equation}
We take the Chern--Simons levels to be $k_1=-k_2=1$. \\

\begin{figure}[ht]
\begin{center}
  \vskip 1cm
  \hskip -6cm
  \includegraphics[totalheight=1.2cm]{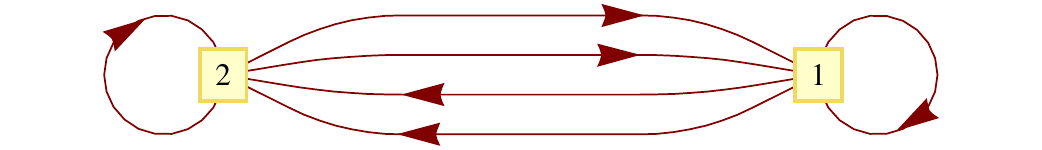}
  \vskip -3.0cm
  \hskip 9cm
  \includegraphics[totalheight=4.5cm]{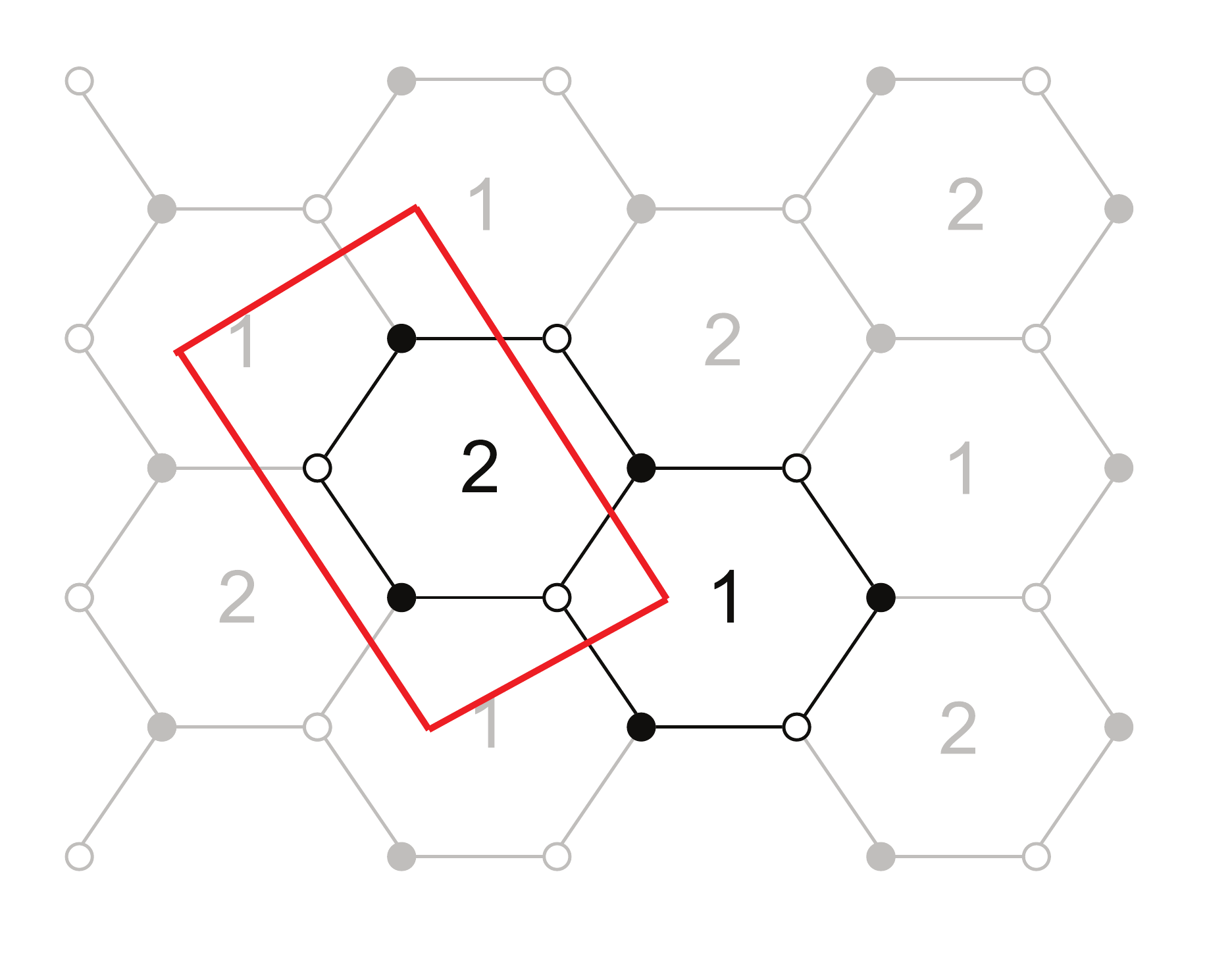}
 \caption{[Phase II of $\CC \times \BC$] (i) Quiver diagram for the $\sH_2$ model. \ (ii) Tiling for the $\sH_2$ model.}
  \label{f:phase2conxc}
\end{center}
\end{figure}

\begin{figure}[ht]
\begin{center}
   \includegraphics[totalheight=8.0cm]{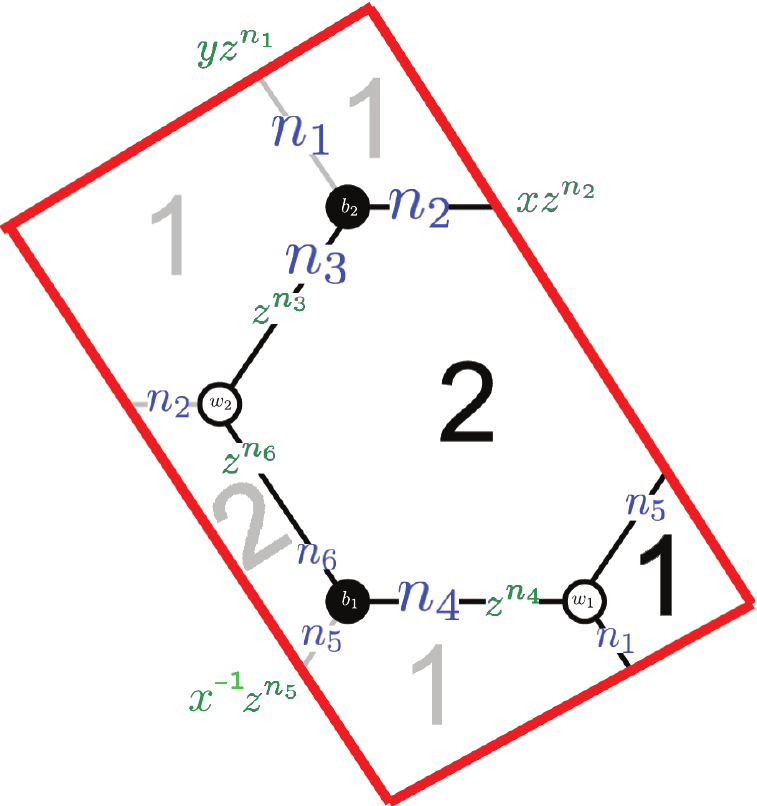}
 \caption{[Phase II of $\CC \times \BC$]. The fundamental domain of tiling for the $\sH_2$ model: assignments of the integers $n_i$ to the edges are shown in blue and the weights for these edges are shown in green.}
  \label{f:fdphase2conxc}
\end{center}
\end{figure}

\paragraph{The Kasteleyn matrix.} We assign the integers $n_i$ to the edges according to Figure \ref{f:fdphase2conxc}.  We find that 
\bea
\text{Gauge group 1~:} \qquad k_1 &=& 1  = -n_2+n_3+n_4 - n_5 ~, \nn \\
\text{Gauge group 2~:} \qquad k_2 &=& -1 = n_2 - n_3 -n_4 + n_5 ~.
\eea  
We choose
\bea
n_3= 1,\quad n_i=0 \; \text{otherwise}~.
\eea
We can now construct the Kasteleyn matrix:
\be
K =   \left(
\begin{array}{c|cc}
\; & w_1 & w_2 \\
\hline
b_1 & X^{1}_{21} x^{-1} z^{n_5}+ X^{2}_{12} z^{n_4}  & \ \phi_2 z^{n_6}  \\
 b_2 & \phi_1 y z^{n_1}      & \  X^{2}_{21} x z^{n_2}+ X^{1}_{12} z^{n_3}
\end{array}
\right) ~.
\ee
The permanent of this matrix is
\bea
\perm~K &=&   X^{1}_{21}X^{2}_{21} z^{n_2+n_5}  + X^{2}_{12} X^{2}_{21} x z^{n_2+n_4}  + X^{1}_{21}X^{1}_{12} x^{-1} z^{n_3+n_5} + X^{1}_{12} X^{2}_{12} z^{n_3+n_4} + \phi_{1}\phi_{2} y z^{n_1+n_6} \nn \\
&=& X^{1}_{21}X^{2}_{21}  + X^{2}_{12} X^{2}_{21} x   + X^{1}_{21}X^{1}_{12} x^{-1} z + X^{1}_{12} X^{2}_{12} z + \phi_{1}\phi_{2} y  \nn \\  
&&\text{(for $n_3 = 1$ and $n_i =0$ otherwise)} ~ ,
\label{permph1conxc}
\eea
where the powers of $x, y, z$ in each term give the coordinates of each point in the toric diagram.  
We collect these points in the columns of the following $G_K$ matrix:
\bea
G_K = \left(
\begin{array}{ccccc}
 0 & 0 & 1 &-1 & 0 \\
 0 & 0 & 0 & 0 & 1 \\
 1 & 0 & 0 & 1 & 0
\end{array}
\right)~.
\label{e:gkconxcph1}
\eea
Note that this $G_K$ matrix can be obtained by multiplying \eref{gkconxc} on the left by the matrix
\bea
\mathcal{T} = \left(
\begin{array}{ccccc}
-1 & 1 & 0\\
 0 & 0 & 1\\
 0 &-1 & 0
\end{array}
\right) \in GL(3, \BZ)~.
\label{e:gkconxcph2}
\eea
Thus, \eref{gkconxc} and \eref{e:gkconxcph2} are the same up to a $GL(3, \BZ)$ transformation.
It is also clear that the $G'_t$ matrix of this phase coincides with \eref{Gptconxc}, and that the mesonic symmetry is $SU(2) \times SU(2) \times U(1) \times U(1)$ \cite{phase}.
The toric diagram is presented in Figure \ref{f:torconxc}. 

Below, there is a study of the Higgs mechanism of this theory.

\subsubsection{\emph{The $\BC^4$ theory} from giving VEV to one of $X^i_{12}$ or $X^i_{21}$}
By symmetry, we see that giving a VEV to either $X^1_{12}$, $X^2_{12}$, $X^1_{21}$ or $X^2_{21}$ yields the same result. 
For definiteness, let us turn on a VEV to $X^1_{12}$.  
This amounts to removing one of the edges that separate the faces corresponding to gauge groups 1 and 2, and collapsing the two vertices adjacent to a bivalent vertex into a single vertex of higher valence  (see $\S5$ of \cite{Franco:2005rj}).
As a result, the gauge groups 1 and 2 are combined into one gauge group, which is identified as 1, and the edges corresponding to $X^1_{21}$ and $\phi_1$ are removed.
Hence, we are left with 3 adjoint fields transforming under gauge group 1.
The resulting theory can be represented by a one-hexagon tiling.
The CS levels associated with the higgsed gauge groups are added, so that the resulting CS level is
\bea
k = 0 ~.
\eea
The tiling suggests that there is a branch of the moduli space which is $\BC^3$.
In the presence of the gauge kinetic term, there is an additional complex degree of freedom, and the mesonic moduli space is $\BC^4$.

\comment{
\paragraph{The Kasteleyn matrix} The Kasteleyn matrix for this model coincides with its permanent and can be written as:
\bea
K &=& \phi^1_1 x + \phi^2_1 x^{-1} + \phi^3_1 y~.
\eea
Accordingly, the toric diagram lies on a 2 dimensional plane as can be seen in Figure \ref{f:tdc3}.
\begin{figure}[ht]
\begin{center}
  \includegraphics[totalheight=3.0cm]{missing.pdf}
 \caption{The toric diagram of Phase II the $\BC^3$}
  \label{f:tdc3}
\end{center}
\end{figure}}

\subsection{Higgsing Phases III-A and III-B of $\CC \times \BC$}
\subsection*{A summary of Phases III-A and III-B of $\CC \times \BC$ (the $\sD_2 \sH_1$ model)}
\begin{figure}[ht]
\begin{center}
 \vskip 1cm
  \hskip -7cm
  \includegraphics[totalheight=1.5cm]{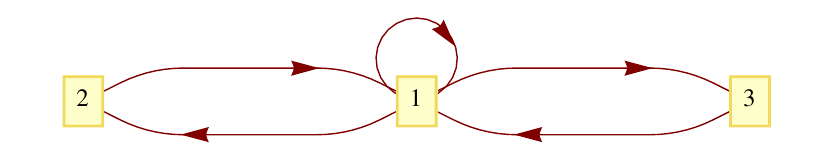}
   \vskip -3cm
  \hskip 8cm
  \includegraphics[totalheight=5.0cm]{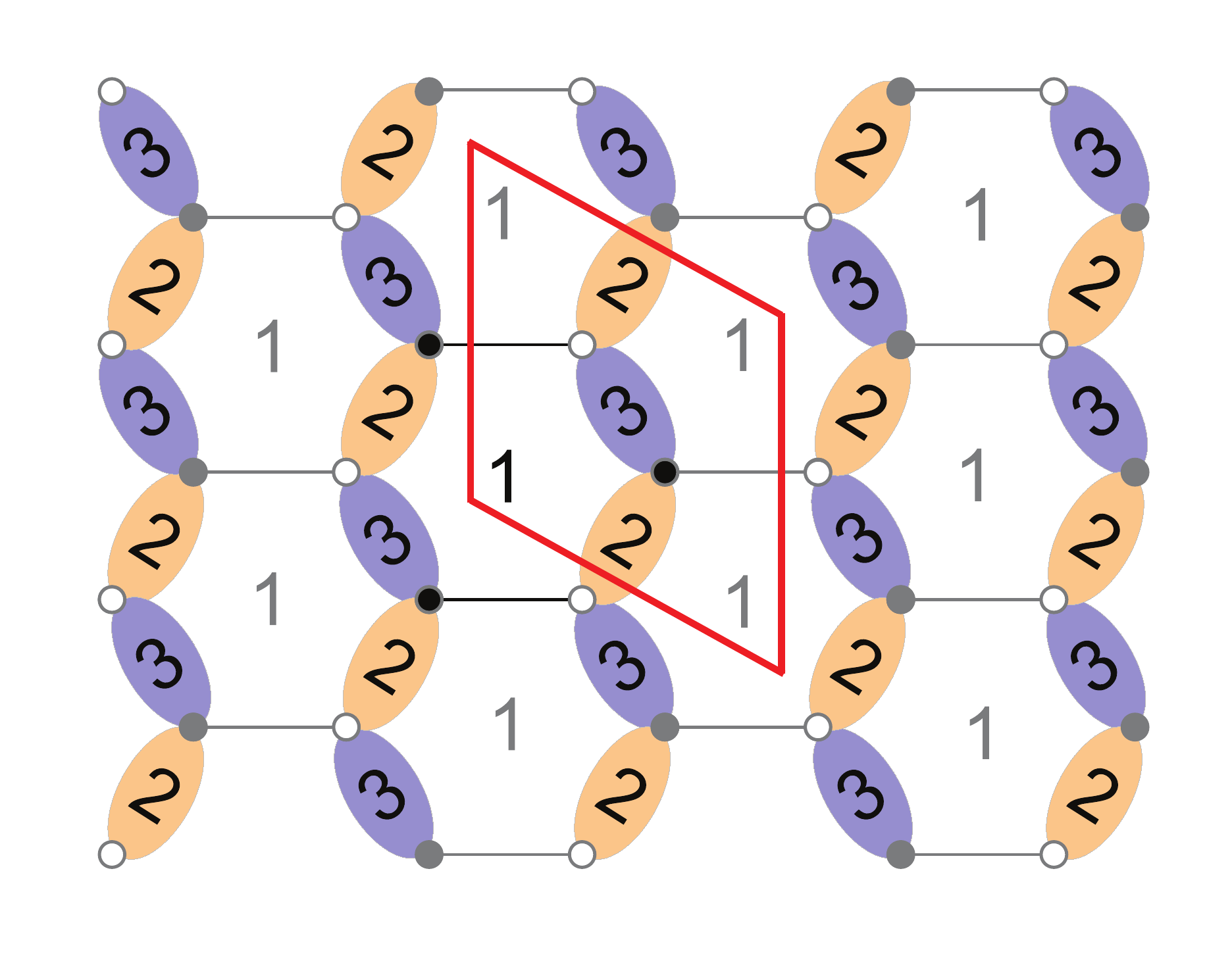}
 \caption{[Phase III of $\CC \times \BC$]  (i) Quiver diagram of the $\sD_2\sH_1$  model.\ (ii) Tiling of the $\sD_2\sH_1$  model.}
  \label{f:phase3conxc}
\end{center}
\end{figure}

This model  has 3 gauge groups and 5 chiral multiplets which are denoted as $X_{12}$, $X_{21}$, $X_{13}$, $X_{31}$ and $\phi_1$. 
The quiver diagram and tiling are drawn in Figure \ref{f:phase3conxc}.
The superpotential is given by
\bea
W= \tr  \left( \phi_1 \left[X_{12} X_{21}, X_{13} X_{31}\right] \right)~. 
\eea
 There are two choices of CS levels that lead to the same toric diagram:
\begin{itemize}
\item  $k_1 = 0,~k_2 = 1,~ k_3 =-1$~;
\item  $k_1 = 2,~k_2 = -1,~ k_3 =-1$~.
\end{itemize}
We refer to the model with the first option as Phase III-A, and to the model with the second option as Phase III-B of $\CC\times \BC$.

\begin{figure}[ht]
\begin{center}
   \includegraphics[totalheight=8.0cm]{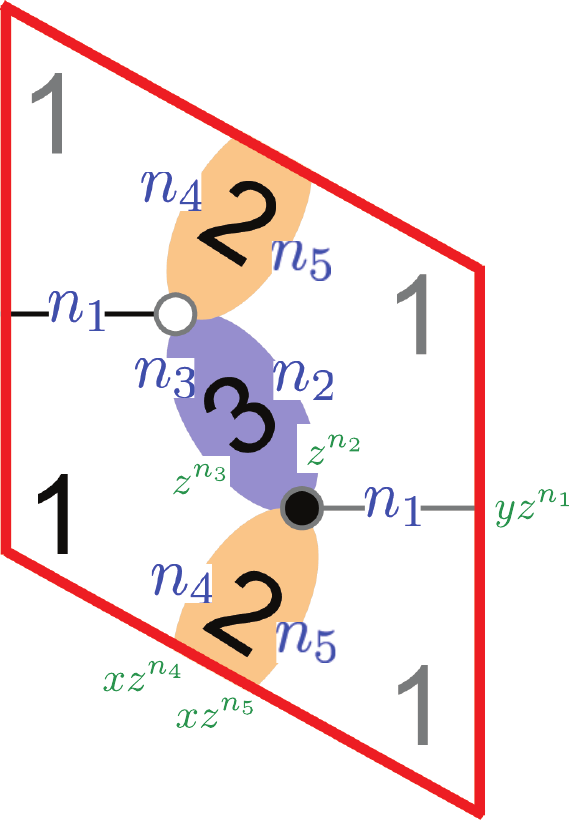}
 \caption{[Phase III of $\CC \times \BC$]  The fundamental domain of tiling for the $\sD_2\sH_1$  model: assignments of the integers $n_i$ to the edges are shown in blue and the weights for these edges are shown in green.}
  \label{f:fdphase3conxc}
\end{center}
\end{figure}

\paragraph {The Kasteleyn matrix for Phase III-A.}  We assign the integers $n_i$ to the edges according to Figure \ref{f:fdphase3conxc}.  We find that 
\bea
\text{Gauge group 1~:} \qquad k_1 &=& 0  = n_2 - n_3 + n_4 -n_5 ~, \nn \\
\text{Gauge group 2~:} \qquad k_2 &=& 1 = -n_4 +n_5 ~,  \nn \\
\text{Gauge group 3~:} \qquad k_3 &=& -1 = -n_2 + n_3~.
\eea  
We choose
\bea
n_2 = n_5 = 1,\qquad n_i=0 \;\;\text{otherwise}~. \label{nph3aconxc}
\eea
We can construct the Kasteleyn matrix, which for this case is just a $1\times 1$ matrix and, therefore, coincides with its permanent:
\bea \label{Kph3genconxc}
K = \phi_1 y z^{n_1} + X_{13} z^{n_2} + X_{31} z^{n_3} + X_{12} x z^{n_4} + X_{21} x z^{n_5}~.
\eea
Thus, from \eref{nph3aconxc}, we find that for Phase III-A, we have
\bea \label{Kph3aconxc}
K_A =  \phi_1 y + X_{13} z  + X_{31} + X_{12} x + X_{21} x z \quad \text{(for $n_2 = n_5 = 1$ and $n_i =0$ otherwise)} ~. \nn\\
\eea
The powers of $x, y, z$ in each term of $K_A$ give the coordinates of each point in the toric diagram.
We collect these points in the columns of the following $G^A_K$ matrix:
\bea
G^A_K =  \left(
\begin{array}{ccccc}
 1 & 0 & 1 & 0 & 0   \\
 0 & 0 & 0 & 0 & 1   \\
 1 & 0 & 0 & 1 & 0 
\end{array}
\right)~.
\eea

\paragraph {The Kasteleyn matrix for Phase III-B.} We now make a different choice of $n_i$'s:
\bea
n_2 = n_4 = 1,\qquad n_i=0 \;\;\text{otherwise}~.
\eea
This leads to the expected Chern-Simons levels:
\bea
\text{Gauge group 1~:} \qquad k_1 &=&  2 =   n_2 - n_3 + n_4 -n_5 ~, \nn \\
\text{Gauge group 2~:} \qquad k_2 &=& -1 = - n_4 + n_5 ~,  \nn \\
\text{Gauge group 3~:} \qquad k_3 &=& -1 = - n_2 + n_3~.
\eea  
Having made this particular choice on the integers $n_i$'s, the permanent of the Kasteleyn matrix written in (\ref{Kph3genconxc}) now becomes:
\bea \label{Kph3bconxc}
K_B &=& \phi_1 y + X_{13} z  + X_{31} + X_{12} x z + X_{21} x  \quad \text{(for $n_2 = n_4 = 1$ and $n_i =0$ otherwise)} ~. \nn\\
\eea
The powers of $x, y, z$ in each term of $K_B$ give the coordinates of each point in the toric diagram.
We collect these points in the columns of the following $G_K$ matrix:
\bea
G^B_K = \left(
\begin{array}{ccccc}
 1 & 0 & 1 & 0 & 0   \\
 0 & 0 & 0 & 0 & 1   \\
 1 & 0 & 0 & 1 & 0 
\end{array}
\right)~.
\eea
The two matrices $G^A_K$ and $G^B_K$ are equal and both of them can be transformed into the matrix \eref{Gptconxc} by interchanging the second and the third row and by multiplying the first and the new third row by $-1$.
Thus, the mesonic symmetry is $SU(2) \times SU(2) \times U(1) \times U(1)$ \cite{phase}.
The toric diagram is drawn in \fref{f:torconxc}.

Below, there is a study of the Higgs mechanism of this theory.

\subsubsection{\emph{Phase II of $\BC^4$} from giving a VEV to one of $X_{12}$, $X_{21}$, $X_{13}$, $X_{31}$}
By symmetry, we see that giving a VEV to any of the bi-fundamental fields leads to the same field theory, up to relabelling gauge groups and fields.
For definiteness, let examine the case in which $X_{13}$ acquires a VEV. From the tiling shown in Figure \ref{f:phase3conxc}, we see that removing the edge corresponding to $X_{13}$ amounts to combining gauge group 1 and 3, so that the double bond corresponding to the gauge group 3 disappears. 
Thus, the resulting tiling is a double-bonded hexagon (\fref{f:phase2c4}).
Since the CS levels associated with the higgsed gauge groups are added, higgsing both Phase III-A and Phase III-B yields to the same CS levels: 
\bea
k_1 = 1, \quad  k_2=-1~. 
\eea
Thus, the resulting theory is indeed Phase II of $\BC^4$.
The toric diagram is drawn in \fref{f:torabjm}.

\section{Higgsing The $M^{1,1,1}$ Theory} \label{sec:summ111}
\subsection*{A summary of the $M^{1,1,1}$ theory} 
\begin{figure}[ht]
 \centerline{  \epsfxsize = 6cm \epsfbox{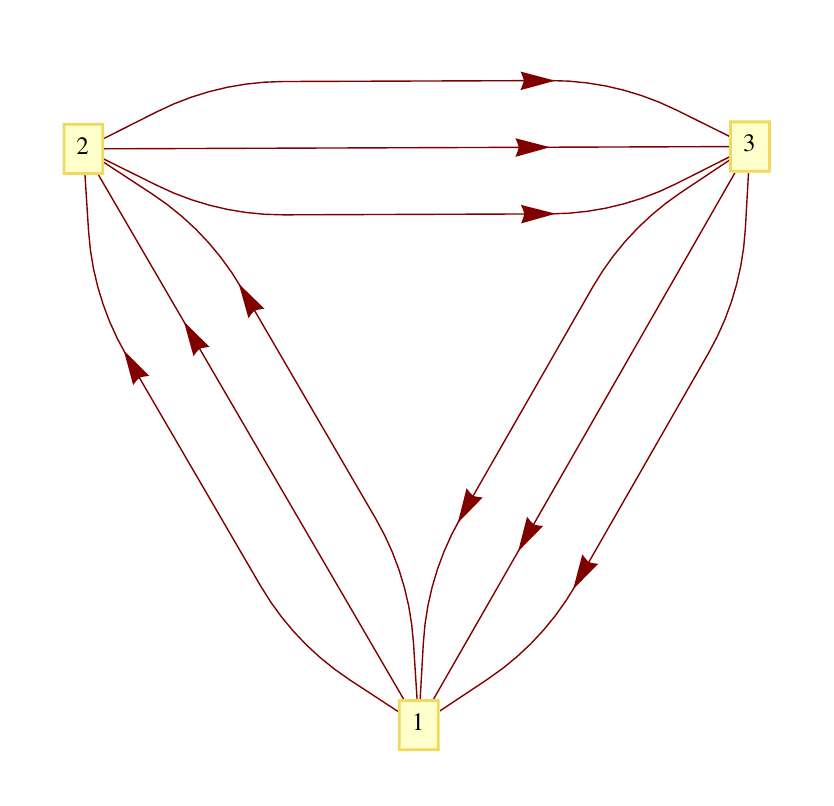}  \hskip 10mm \epsfxsize = 7cm \epsfbox{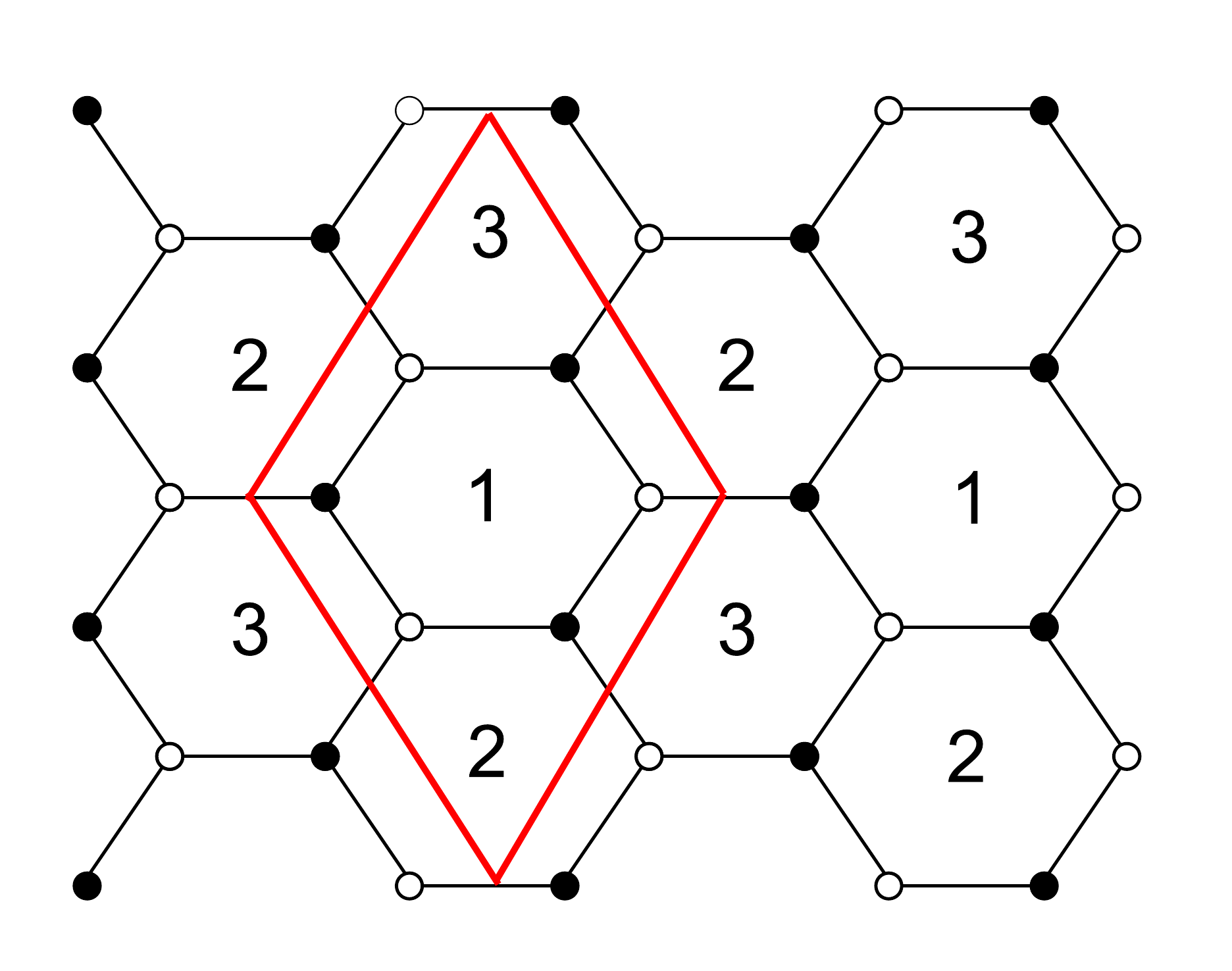}}
\caption{(i) Quiver diagram of the $M^{1,1,1}$  theory.\ (ii) Tiling of the $M^{1,1,1}$  theory.}
  \label{f:m111}
\end{figure}

The $M^{1,1,1}$ theory \cite{Hanany:2008cd, Martelli:2008si, Hanany:2008fj, Hanany:2009vx, Petrini:2009ur, Fabbri:1999hw} has 3 gauge groups and 9 chiral multiplets which are denoted as $X_{12}^i, X_{23}^i$ and $X_{31}^i$ (with $i=1,2,3$) . The quiver diagram and tiling are given in \fref{f:m111}.  
Note that in $3+1$ dimensions, this tiling corresponds to the gauge theory living on D3-branes probing the cone over the $dP_0$ surface. Appendix \ref{app:IIA} discusses how this theory arises on the world volume of a D2-brane which probes this $\mathrm{CY}_3$ with one unit of RR 4-form flux on the $\CP^2$.
The superpotential is given by
\bea
W= \tr \left( \epsilon_{ijk} X^i_{12} X^j_{23} X^k_{31} \right)~. 
\eea
The CS levels are $(k_1,  k_2 , k_3) = (1, -2, 1) $. 

\begin{figure}[ht]
 \centerline{  \epsfxsize = 7cm  \epsfbox{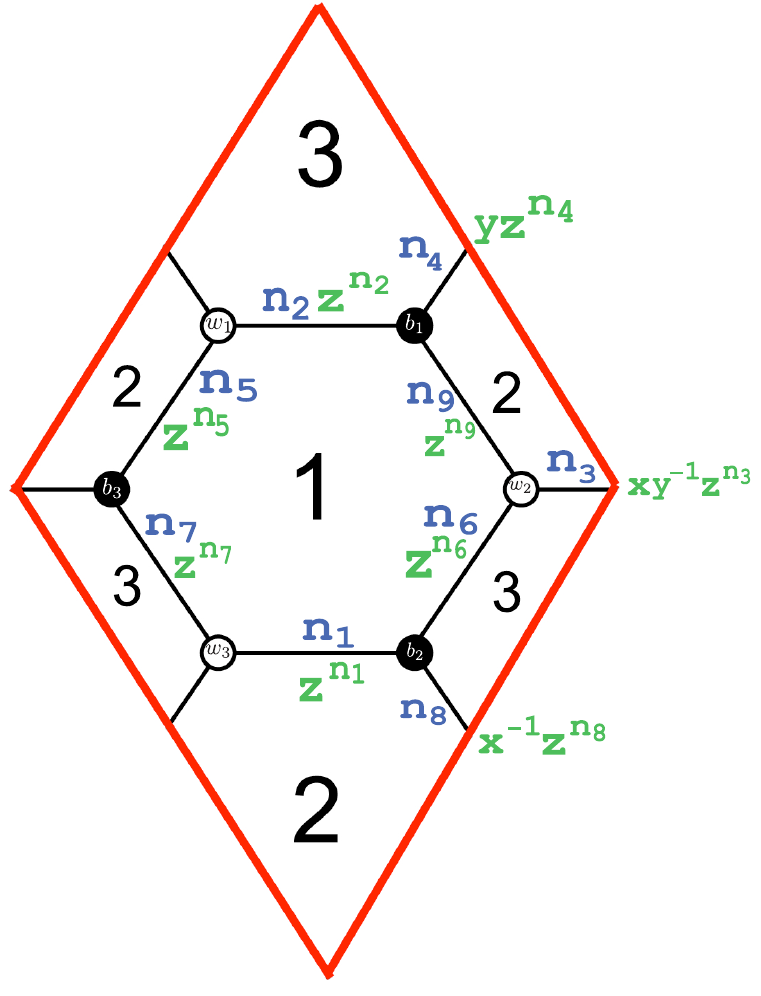} }
 \caption{The fundamental domain of the tiling for the $M^{1,1,1}$ theory: assignments of the integers $n_i$ to the edges are shown in blue and the weights for these edges are shown in green.}
  \label{f:fdm111}
\end{figure}

\paragraph{The Kasteleyn matrix.} We assign the integers $n_i$ to the edges according to Figure \ref{f:fdm111}.  Then, 
\bea
\text{Gauge group 1~:} \qquad k_1 &=& 1  = + n_1 - n_2 + n_5 - n_6 - n_7 + n_9 ~,  \nn \\
\text{Gauge group 2~:} \qquad k_2 &=& -2 = - n_1 + n_3 + n_4 - n_5 + n_8 - n_9  ~,  \nn \\
\text{Gauge group 3~:} \qquad k_3 &=& 1  =   n_2 - n_3 - n_4 + n_6 + n_7 - n_8 ~.  \label{e:kafano24}
\eea  
We choose:
\bea
n_1=-n_3= 1,\qquad n_i=0 \;\;\text{otherwise}~.
\eea
We can now determine the Kasteleyn matrix. Since the fundamental domain contains 3 pairs of black and white nodes, the Kasteleyn matrix is $3 \times 3$: 
\bea
K =   \left(
\begin{array}{c|ccc}
& w_1 & w_2 & w_3\\
\hline
b_1 & X^1_{31} z^{n_2} & X^3_{12} z^{n_9} & X^2_{23} y z^{n_4} \\
b_2 & X^3_{23} \frac{1}{x} z^{n_8} & X^2_{31} z^{n_6} & X^1_{12} z^{n_1} \\
b_3 & X^2_{12} z^{n_5} & X^1_{23} \frac{x}{y} z^{n_3} & X^3_{31} z^{n_7} \end{array}
\right) ~.
\label{e:kastfano24}
\eea
The permanent of the Kasteleyn matrix is given by
\bea
\mathrm{perm}(K) &=&  X^1_{12} X^1_{23} X^1_{31} x y^{-1} z^{n_1 + n_2 + n_3}
+X^2_{12} X^2_{23} X^2_{31} y z^{n_4 + n_5 + n_6}
+X^3_{12} X^3_{23} X^3_{31} x^{-1} z^{n_7 + n_8 +n_9} \nn \\
&+&X^1_{12} X^2_{12} X^3_{12} z^{n_1+n_5+n_9}
+X^1_{23} X^2_{23} X^2_{23} z^{n_3 + n_4 + n_8}
+X^1_{31} X^2_{31} X^3_{31}  z^{n_2 + n_6 +n_7}\nn \\&=&
X^1_{12} X^1_{23} X^1_{31} x y^{-1}
+X^2_{12} X^2_{23} X^2_{31} y
+X^3_{12} X^3_{23} X^3_{31} x^{-1}
X^1_{12} X^2_{12} X^3_{12} z \nn \\
&+&X^1_{23} X^2_{23} X^3_{23} z^{-1}
+X^1_{31} X^2_{31} X^3_{31}\; \text{(for $n_1 = - n_3 = 1,~ n_i=0 \;
\text{otherwise}$)~.}
\label{e:charpolyfano24}
\eea

\paragraph{The perfect matchings.} From (\ref{e:charpolyfano24}), we can take the perfect matchings to be
\bea 
&& p_1 = \left\{X^1_{12}, X^1_{23}, X^1_{31}\right\}, \;\; p_2 = \left\{X^2_{12}, X^2_{23}, X^2_{31}\right\}, \;\; p_3 = \left\{X^3_{12}, X^3_{23}, X^3_{31}\right\}, \nn \\  
&&  r_1 = \left\{X^1_{12}, X^2_{12}, X^3_{12}\right\}, \;\; r_2 = \left\{X^1_{23}, X^2_{23}, X^3_{23}\right\}, \;\; s_1 = \left\{X^1_{31}, X^2_{31}, X^3_{31}\right\}\ . \qquad
\eea
Looking at (\ref{e:charpolyfano24}), we see that the perfect matchings $p_1,~p_2,~p_3,~r_1,~r_2$ correspond to external points in the toric diagram, whereas $s_1$ corresponds to the internal point.
We can also parametrise the chiral fields in terms of perfect matchings as follows:
\bea
&& X^1_{12} = p_1 r_1 , \quad X^1_{23} = p_1 r_2 , \quad X^1_{31} = p_1 s_1 \nn \\
&& X^2_{12} = p_2 r_1 , \quad X^2_{23} = p_2 r_2 , \quad X^2_{31} = p_2 s_1 \nn \\
&& X^3_{12} = p_3 r_1 , \quad X^3_{23} = p_3 r_2 , \quad X^3_{31} = p_3 s_1 ~.
\eea
We can collect all these pieces of information in the perfect matching matrix:
\beq
P=\left(\begin{array} {c|cccccc}
  \;& p_1 & p_2 & p_3 & r_1 & r_2 & s_1\\
  \hline 
  X^{1}_{12}& 1&0&0&1&0&0\\
  X^{1}_{23}& 1&0&0&0&1&0\\
  X^{1}_{31}& 1&0&0&0&0&1\\
  X^{2}_{12}& 0&1&0&1&0&0\\
  X^{2}_{23}& 0&1&0&0&1&0\\
  X^{2}_{31}& 0&1&0&0&0&1\\
  X^{3}_{12}& 0&0&1&1&0&0\\
  X^{3}_{23}& 0&0&1&0&1&0\\
  X^{3}_{31}& 0&0&1&0&0&1\\
  \end{array}
\right).
\eeq
The nullspace of $P$ is 1 dimensional and is spanned by the vector that we write in the row of the following charge matrix:
\be
Q_F =   \left(
\begin{array}{cccccc}
1,&1,&1,&-1,&-1,&-1
\end{array}
\right)~.  \label{e:qffano24}
\ee
Hence, the perfect matchings satisfy the relation:
\bea
p_1 + p_2 + p_3 - r_1 - r_2 - s_1 = 0.
\label{e:relpmfano24}
\eea

\paragraph{The toric diagram.} We construct the toric diagram of this model using two methods:
\begin{itemize}
\item {\bf The charge matrices.}
Because the number of gauge groups of this model is $G = 3$, there is $G-2 =1$ baryonic charge coming from the D-terms. The baryonic charges of the perfect matchings are collected in the $Q_D$ matrix:
\be
Q_D =   \left(
\begin{array}{cccccc}
0, & 0, & 0, & -1,& -1,& 2
\end{array}
\right) \label{e:qdfano24}
\ee
The matrices (\ref{e:qffano24}) and (\ref{e:qdfano24}) can be combined in a single matrix, $Q_t$, that contains all the baryonic charges of the perfect matchings:
\be
Q_t = { Q_D \choose  Q_F} =   \left( 
\begin{array}{cccccc} 
0 & 0& 0& -1& -1& 2 \\
1 & 1& 1& -1& -1& -1
\end{array}
\right) 
\label{e:qtfano24}
\ee
We can now obtain the $G_t$ matrix and, after removing the first row, we get a matrix whose columns represent the coordinates of the toric diagram:
\bea
G'_t = \left(
\begin{array}{cccccc}
 1 & -1 & 0 &  0 & 0 & 0 \\ 
 0 & 1 & -1 &  0 & 0 & 0 \\
 0 & 0 & 0 & 1 & -1 & 0 
\end{array}
\right)~. \label{e:toricdiafano24}
\eea
The toric diagram is presented in Figure \ref{f:tdtoricfano24}.  

\begin{figure}[ht]
\begin{center}
  \includegraphics[totalheight=3.0cm]{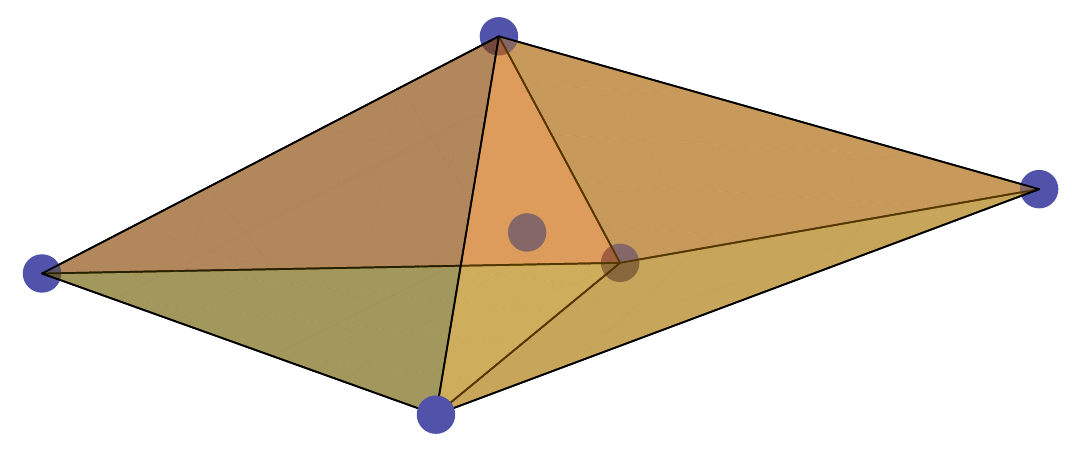}
 \caption{The toric diagram of the $M^{111}$ theory.}
  \label{f:tdtoricfano24}
\end{center}
\end{figure}

\item {\bf The Kasteleyn matrix.} The powers of $x, y, z$ in each term of \eref{e:charpolyfano24} give the coordinates of each point in the toric diagram. We collect these points in the columns of the following $G_K$ matrix: 
\bea
G_K = \left(
\begin{array}{cccccc}
  1 &-1 & 0 & 0 &  0 & 0 \\
  0 & 1 & -1 & 0 &  0 & 0\\
  0 & 0 &  0 & 1 & -1 & 0
\end{array}
\right) = G'_t~.
\eea

\end{itemize}

\paragraph{The baryonic charges.} Since the toric diagram of this model has 5 external points, there is precisely $5-4 = 1$ symmetry, which we shall denote as $U(1)_{B}$. From the discussion on the charge matrices above, we understand that the baryonic charge of the perfect matchings come from the row of the $Q_D$ matrix.

\paragraph{The global symmetry.} We can observe that the $Q_t$ matrix \eref{e:qtfano24} has a pair and a `triplet' of repeated columns. Since the total rank of the mesonic symmetry is 4, the mesonic symmetry of this model is $SU(3)\times SU(2)\times U(1)$.  This can also be seen from the $G'_t$ matrix \eref{e:toricdiafano24} by noticing that the first two rows contain weights of $SU(3)$ and the third row contains weights of $SU(2)$. 
Since there is precisely one factor of $U(1)$, this can be unambiguously identified with the R-symmetry of the theory. 
The global symmetry of this theory is a product of mesonic and baryonic symmetries: $SU(3)\times SU(2)\times U(1)_R \times U(1)_B$.
The R-charge of each perfect matching can be determined as follows \cite{Hanany:2008fj}.  

\paragraph{R-charges of the perfect matchings.}  
In order to determine the R-charge of the perfect matching $p_1$, we must first derive the refined Hilbert series of the mesonic moduli space.
Since the non-abelian fugacities do not play any role in the volume minimization, we may set them to unity. Also, since the R-charge of the internal perfect matching $s_1$ is zero, we may likewise set the corresponding fugacity to unity. We denote by $t_1, t_2, t_3$ the fugacities of $p_1,p_2,p_3$, and by $t_4, t_5$ the fugacities of $r_1, r_2$.  From the $Q_t$ matrix (\ref{e:qtfano24}), the Hilbert series of the mesonic moduli space $\CMm= \BC^6//Q_t$ is given by
\bea
\gm (t_{\alpha}; M^{1,1,1})= \oint \limits_{|z| =1} {\frac{\ud z}{2 \pi i z }}\oint \limits_{|b| =1} {\frac{\ud b}{2 \pi i b }} \frac{1}{\left(1- t_1 z\right)\left(1-t_2 z\right)\left(1-t_3 z\right)\left(1-\frac{t_4}{b z }\right)\left(1-\frac{t_5}{b z}\right)\left(1- \frac{b^2}{z}\right)}~,\nn \\
\label{e:HStrfano24}
\eea
where $z$ is the fugacity associated with the $Q_F$ charges, and $b$ is the fugacity associated with the $Q_D$ charges. The computation shows that the result of the integration depends only on a specific combination of the $t_{\alpha}$'s, namely $t_1 t_2 t_3 t_4 t_5$. Hence, we can define a new fugacity $t$ such that
\bea
t^{18} = t_1 t_2 t_3 t_4 t_5~,
\eea
where the power 18 is introduced for convenience.  The Hilbert series of the mesonic moduli space  can then be rewritten in terms of $t$ as 
\bea
\gm (t; M^{1,1,1}) = \frac{1 + 26 t^{18} + 26 t^{36} +  t^{54}}{(1-t^{18})^4}~. \label{unrefHSM111}
\eea
In fact, it is not a surprise that the mesonic Hilbert series depends on a single variable, as there is only one $U(1)$ symmetry, which is identified as the R-symmetry.

Each term in the superpotential is the product of all the external perfect matchings.  
Therefore, it scales like $t^{18}$.  
Since the R-charge of the superpotential is 2,  it follows that the R-charge associated with $t$ is $1/9$. 
In other words, we may write
\bea
t = e^{- \mu/9}~,
\eea 
where $\mu$ is the chemical potential of the R-charge associated with $t$.

Next \cite{Hanany:2008fj}, let us compute the Hilbert series of the divisor corresponding to $p_1$, which we will refer to as $D_1$. This would be the integral over the baryonic fugacities of the Hilbert series of the space of perfect matchings multiplied by the inverse of the fugacity relative to $p_1$:
\bea
g (D_1; t_{\alpha}; M^{1,1,1}) &=& \oint \limits_{|z| =1} {\frac{\ud z}{2 \pi i z }}\oint \limits_{|b| =1} {\frac{\ud b}{2 \pi i b }} \frac{(t_1 z)^{-1}}{\left(1- t_1 z\right)\left(1-t_2 z\right)\left(1-t_3 z\right)\left(1-\frac{t_4}{b z }\right)\left(1-\frac{t_5}{b z}\right)\left(1- \frac{b^2}{z}\right)},\nn \\
\label{e:HSD1fano24}
\eea
where, again, we have set the non-abelian fugacities to unity as they do not matter in the computation of volumes. As before, the result of the integration depends only on the product of $t_\alpha$'s and, therefore, it can be rewritten in terms of $t$:
\bea
g (D_1; t ; M^{1,1,1}) &=& \frac{3t^{18} (6 + 11t^{18} + t^{36})}{(1-t^{18})^4}~.
\eea
Thus, the $R$-charge of the perfect matching $p_1$ is given by
\bea
R_{1} &=& \lim_{\mu\rightarrow0}\frac{1}{\mu} \left[ \frac{g(D_1; e^{- \mu/9 }; M^{1,1,1}) }{\gm(e^{-\mu/9};M^{1,1,1})}- 1 \right]= \frac{4}{9}~.
\eea
The computations for the other perfect matchings can be done in a similar way.  The results, as well as the charges under the other global symmetries,  are presented in Table \ref{t:chargefano24}:
\begin{table}[h]
 \begin{center}  
  \begin{tabular}{|c||c|c|c|c|c|c|}
  \hline
  \;& $SU(3)$&$SU(2)$&$U(1)_R$&$U(1)_B$&fugacity\\
  \hline  \hline 
  $p_1$&$(1,0)$&$0$&4/9&0&$t^4 y_1  $\\
  \hline
  $p_2$&$(-1,1)$&$0$&4/9&0&$t^4 y_2 / y_1 $\\
  \hline
  $p_3$&$(0,-1)$&$0$&4/9&0&$t^4 / y_2 $\\
  \hline
  $r_1$&(0,0)&1&1/3&$-1$&$t^3 x / b$ \\
  \hline
  $r_2$&(0,0)&$-1$&1/3&$-1$&$t^3/(x b)$\\
  \hline
  $s_1$&(0,0)&0&0&2&$  b^2$\\
  \hline
  \end{tabular}
  \end{center}
\caption{Charges of the perfect matchings under the global symmetry of the $M^{1,1,1}$ theory. Here $t$ is the fugacity of the R-charge (in the unit of $1/9$), $y_1,y_2$ are the fugacities of the $SU(3)$ symmetry, $x$ is the fugacity of the $SU(2)$ symmetry and $b$ is the fugacity of the $U(1)_B$ symmetry.  We have used the notation $(a,b)$ to represent a weight of $SU(3)$.}
\label{t:chargefano24}
\end{table}

\begin{table}[h]
 \begin{center}  
  \begin{tabular}{|c||c|}
  \hline
  \; Generators &R-charge\\
  \hline  \hline 
  $X^{i}_{12}$ &  $7/9$\\
  \hline
  $X^{i}_{23}$ &  $7/9$\\
  \hline
  $X^{i}_{31}$ &  $4/9$\\
  \hline
  \end{tabular}
  \end{center}
\caption{R-charges of the quiver fields for the $M^{1,1,1}$ theory.}
\label{t:Rgenfano24}
\end{table}

\paragraph{The Hilbert series.} The coherent component of the Master space is generated by the perfect matchings, which are subject to the relation (\ref{e:relpmfano24}):
\bea
\firr{M^{1,1,1}} = \BC^6//Q_F~.  \label{firrfano24}
\eea
It follows that the Hilbert series of the coherent component of the Master space of this model can be obtained by integrating the Hilbert series of the space of perfect matchings over the fugacity $z$:
\bea
g^{\firr{}} (t, x, y_1, y_2, b; M^{1,1,1}) &=& \oint \limits_{|z| =1} {\frac{\ud z}{2 \pi i z }} \frac{1}{\left(1- t^4 y_1 z\right)\left(1-\frac{t^4 y_2 z}{y_1}\right)\left(1-\frac{t^4 z}{y_2}\right)\left(1-\frac{t^3 x}{b z }\right)\left(1-\frac{t^3}{x b z}\right)\left(1- \frac{b^2}{z}\right)}~. \nn \\
\label{e:HSmasterfano24}
\eea
The unrefined version of the result of the integration can be written as:
\bea
g^{\firr{}} (t, 1, 1, 1, 1; M^{1,1,1}) &=& \frac{1 - 6 t^{11} - 3 t^{14} + 2 t^{15} + 12 t^{18} + 2 t^{21} - 3 t^{22} - 
 6 t^{25} + t^{36}}{\left(1-t^4\right)^3\left(1-t^7\right)^6}~. \nn \\
\eea
Integrating (\ref{e:HSmasterfano24}) over the baryonic charge $b$ gives the Hilbert series of the mesonic moduli space:
\bea
\gm (t,x,y_1,y_2; M^{1,1,1}) &=& \oint_{|b|=1} \frac{\ud b}{2\pi i b}\;\; g^{\firr{}} (t, x, y_1, y_2, b; M^{1,1,1}) \nn \\
&& =\frac{P_{M^{1,1,1}}(t,x,y_1,y_2)}{\left(1-\frac{t^{18} y^3_1}{x^2}\right)\left(1-t^{18} x^2 y^3_1\right)\left(1-\frac{t^{18} x^2}{y^3_2}\right)\left(1-	\frac{t^{18}}{x^2 y^3_2}\right)\left(1-\frac{t^{18} y^2_2}{x^2 y^3_1}\right)\left(1-\frac{t^{18} x^2 y^3_2}{y^3_1}\right)} \nn \\ 
&& =\sum^{\infty}_{n=0}\left[3n,0;2n\right]t^{18n}~,
\label{e:HSmesfano24}
\eea
where $P_{M^{1,1,1}}(t,x,y_1,y_2)$ is a polynomial of degree 90, too long to be presented here. 
Instead, the last expression in \eref{e:HSmesfano24} gives a more convenient representation of this Hilbert series. It is a sum over all irreducible representations of the form $[3n, 0; 2n]$, where the first two numbers are highest weights of an $SU(3)$ representation (totally symmetric $3n$ tensor), and the last number is the highest weight of an $SU(2)$ representation (of spin $n$). Indeed, this result confirms the known KK spectrum on $M^{1,1,1}$, see for example \cite{Fabbri:1999hw}.

The totally unrefined mesonic Hilbert series is given by \eref{e:HStrfano24}.
The generators of the mesonic moduli space can be determined from the plethystic exponential of (\ref{e:HSmesfano24}):
\bea
\PL[\gm (t,x,y_1,y_2,M^{1,1,1})] &=& \left[3,0;2\right]t^{18} - ( \left[6,0;0\right] + \left[4,1;2\right] + \left[2,2;4\right] \nn \\
&& \left[2,2;0\right] + \left[0,3;2\right] )t^{36} + O(t^{54})~,   \label{plm111}
\eea
where the transformation laws of the relations can be computed by subtracting $[6,0;4]$ from the symmetric product of 2 copies of $[3,0;2]$.
The 30 generators can be written in terms of perfect matchings as:
\bea
p_i ~p_j ~p_k ~r_l ~r_m ~s_1~,  \label{genm111}
\eea
where $i,j,k=1,2,3$ and $l,m=1,2$. 
As a check, we note that $p_i p_j p_k$ has $\frac{3 \times 4 \times 5}{3!} = 10$ independent components and $r_l r_{m}$ has $\frac{2 \times 3}{2!}  = 3$ independent components, so that there are indeed 30 generators.

\begin{figure}[ht]
 \centerline{  \epsfxsize = 4cm \epsfbox{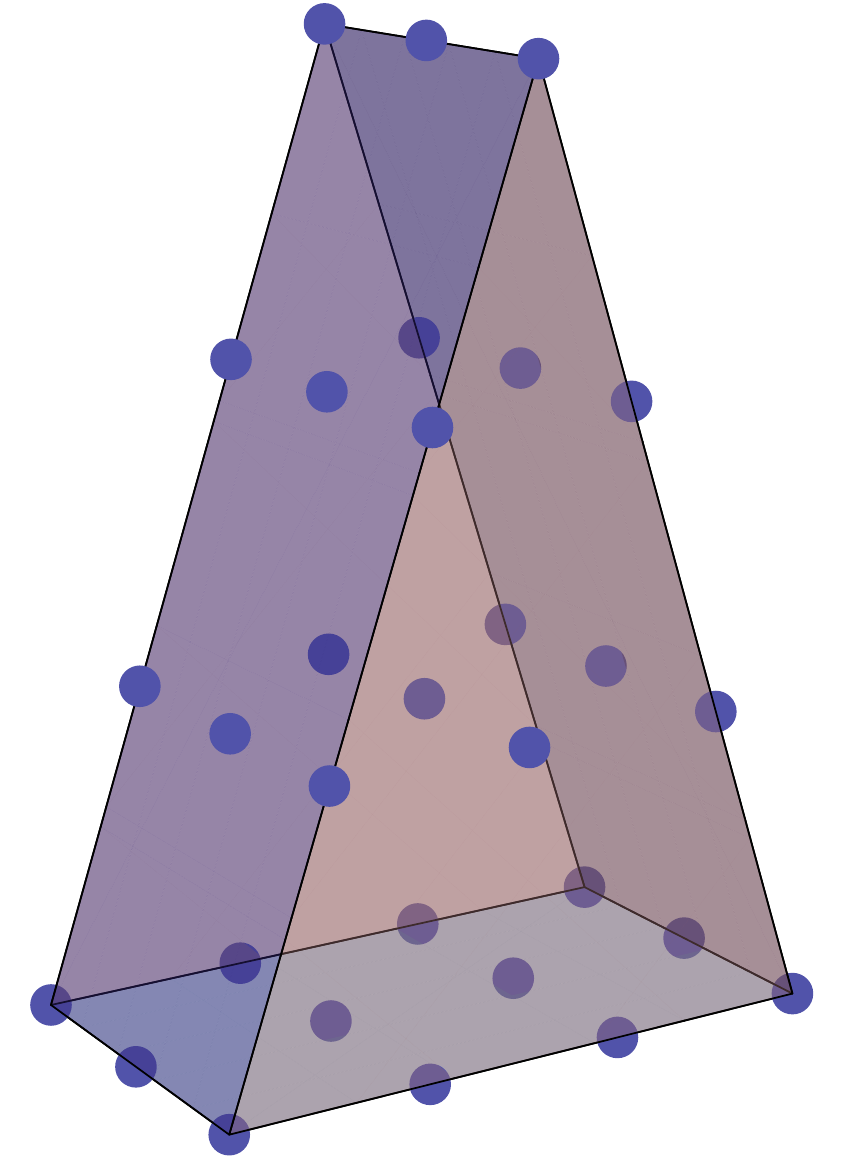}}
\caption{The lattice of generators of the $M^{1,1,1}$ theory}
  \label{f:lattm111}
\end{figure}

\paragraph{The lattice of generators.} We can represent the generators \eref{genm111} in a lattice (\fref{f:lattm111}) by plotting the powers of each monomial in the characters of the first term of \eref{plm111}.
Note that the lattice of generators is the dual of 
the toric diagram (nodes are dual to faces and edges are dual to edges): the toric diagram 
has 5 nodes, which are the external points of the polytope, 9 edges and 6 faces, whereas the generators form a convex polytope that has 
6 nodes, which are the corners of the polytope, 9 edges and 5 faces. 

\subsection{\emph{Phase II of $\BC^4$} from giving a VEV to one of $X^i_{12}$} 
Let us turn on a VEV to one of the $X^i_{12}$ fields.  
This amounts to removing one of the edges that separate the hexagons corresponding to gauge groups 1 and 2, and collapsing the two vertices adjacent to 
a bivalent vertex into a single vertex of higher valence.
As a result, the gauge groups 1 and 2 are combined into one gauge group (which is identified as  2), and the edges corresponding to $X_{31}^1,X_{31}^2,X_{23}^1,X_{23}^2$ are removed.
Then, the hexagon corresponding to gauge group 3 becomes a double bond.  
For convenience, let us relabel the gauge group 3 as 1.
The quiver diagram and tiling are drawn in \fref{f:phase2c4}. 
The CS levels associated with the higgsed gauge groups are added, and so the new CS levels are
\bea
k_1 = -1,\quad k_2=1~.
\eea
The resulting theory is indeed the one double-bonded one-hexagon ($\sD_1\sH_1$) model (Phase II of $\BC^4$) whose toric diagram is given by \fref{f:torabjm}.

\subsection{\emph{Phase II of $\BC^4$} from giving a VEV to one of $X^i_{23}$} 
Let us turn on a VEV to one of the $X^i_{23}$ fields.  This amounts to removing one of the edges that separate the hexagons corresponding to gauge groups 2 and 3, and collapsing the two vertices adjacent to a bivalent vertex into a single vertex of higher valence. The resulting quiver diagram and tiling are drawn in \fref{f:phase2c4}.  The CS levels associated with the higgsed gauge groups are added, therefore the new CS levels are
\bea
k_1 = 1,\quad k_2 = -1~, 
\eea
This model is actually a parity dual of the previous model, so that it can be identified with Phase II of $\BC^4$.

\subsection{\emph{Phase I of $(\BC^2/\BZ_2) \times \BC^2$} from giving a VEV to one of $X^i_{31}$}  \label{sec:ph1c2z2xc2}
Let us turn on a VEV to one of the $X^i_{31}$ fields.  
This amounts to removing one of the edges that separate the hexagons corresponding to gauge groups 1 and 3, and collapsing the two vertices adjacent to a bivalent vertex into a single vertex of higher valence.
The quiver diagram and tiling are given by \fref{f:phase2c4}.
The new CS levels are 
\bea
k_1 = 2,\quad k_2 = -2~.
\eea
Since $\mathrm{gcd}(k_1,k_2) =2$, the mesonic moduli space of this new theory is a $\BZ_2$ quotient of that of the $\sD_1\sH_1$ model.  
The $\BZ_2$ orbifold acts on the generators as one of the gauge groups, let's say $(1, -1, 0, 0)$ on $\left( X_{12}, X_{21}, \phi^1_2, \phi^2_2\right)$.  
The mesonic moduli space of this new theory can be written as
\bea
\CMm =  \left( \BC^2/\BZ_2 \right) \times \BC^2~.
\eea
Henceforth, we shall refer to this model as {\bf Phase I of the $\left( \BC^2/\BZ_2 \right) \times \BC^2$ theory}.
The Hilbert series of this moduli space is given by
\bea
\gm(t) = \frac{1}{2} \left[ \frac{1}{(1-t)^2(1-t)^2} + \frac{1}{(1+t)^2(1-t)^2} \right]  
= \frac{1+t^2}{(1-t)^2 (1-t^2)^2}~, \label{gmesph1c2z2xc2}
\eea
where $t$ is the fugacity counting the scaling dimensions.

\section{Higgsing The $D_3$ Theory}
\subsection{Higgsing Phase I of $D_3$}
\subsection*{A summary of Phase I of $D_3$ (the $\sD_2 \sC$ model)}
\begin{figure}[h]
\begin{center}
  \vskip 1cm
  \hskip -7cm
  \includegraphics[totalheight=4cm]{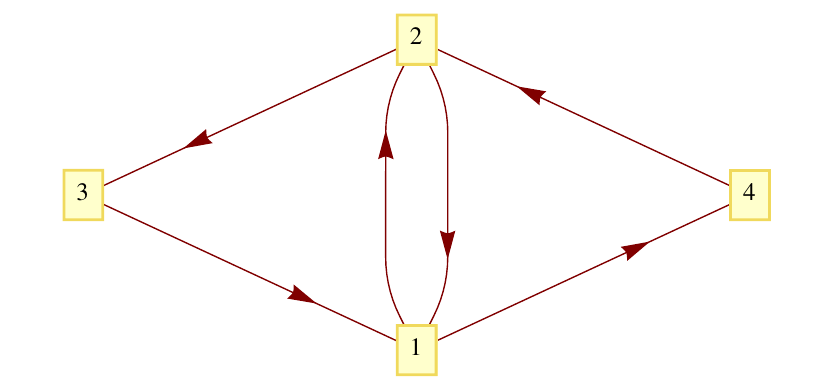}
    \vskip -4.5cm
  \hskip 8.4cm
  \includegraphics[totalheight=5cm]{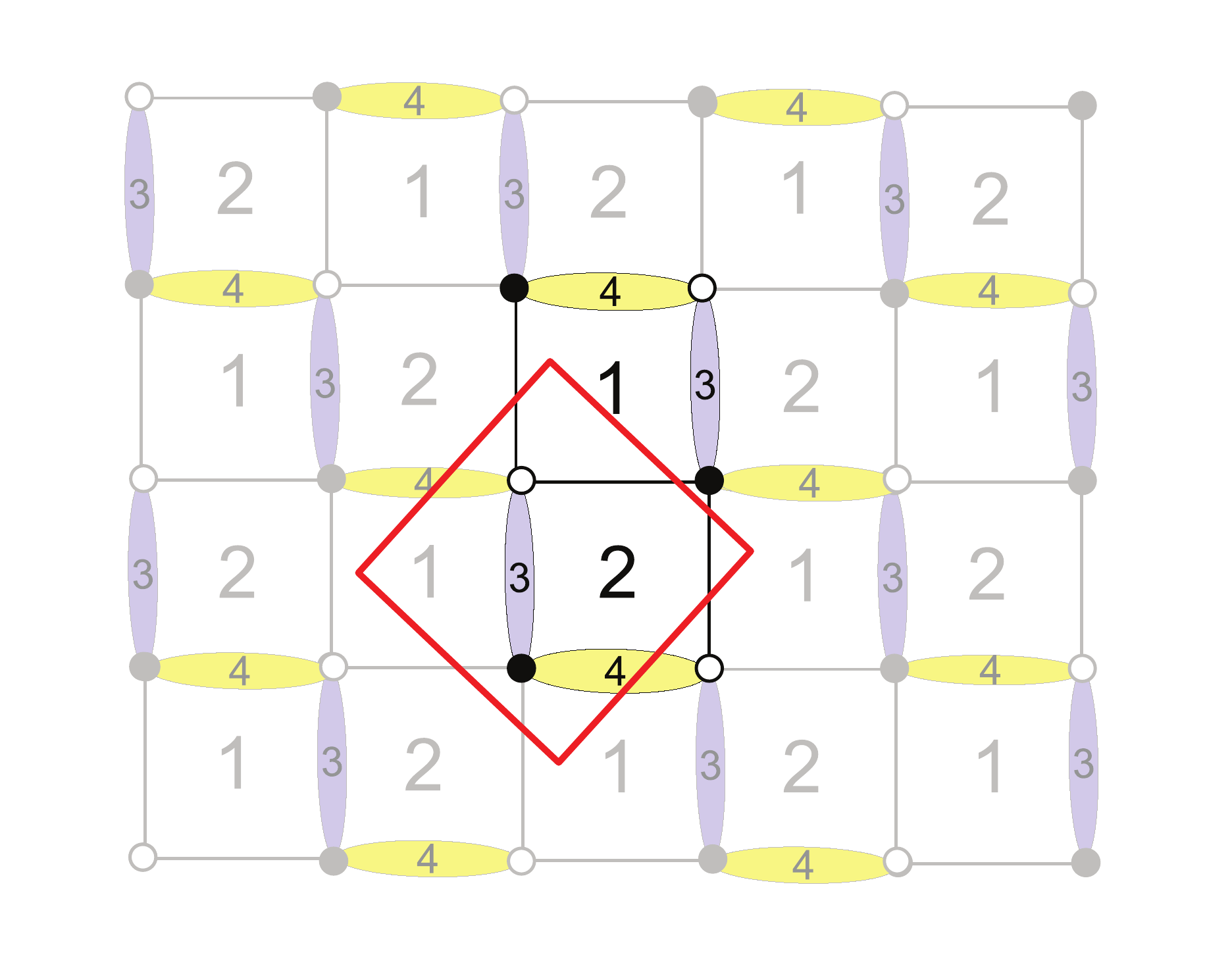}
 \caption{[Phase I of the $D_3$ theory] (i) Quiver diagram of the $\sD_2 \sC$ model.\ (ii) Tiling of the $\sD_2 \sC$ model.}
  \label{f:phase1D3}
\end{center}
\end{figure}

The quiver diagram and tiling of this model are drawn in Figure \ref{f:phase1D3}.
The superpotential is given by
\bea
W = \tr \left( X_{14}X_{42}X_{21}X_{12}X_{23}X_{31} - X_{14}X_{42}X_{23}X_{31}X_{12}X_{21} \right)~.
\eea
We choose the CS levels to be $(k_1,k_2,k_3,k_4) = (1,1,-1,-1)$.

\begin{figure}[h]
\begin{center}
   \includegraphics[totalheight=7cm]{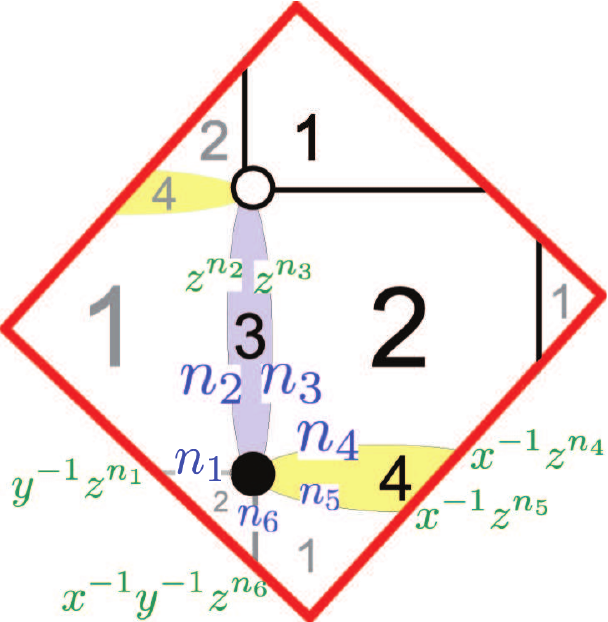}
 \caption{[Phase I of the $D_3$ theory] The fundamental domain of tiling for the $\sD_2 \sC$ model: assignments of the integers $n_i$ to the edges are shown in blue and the weights for these edges are shown in green.}
  \label{f:fdph1d3}
\end{center}
\end{figure} 

\paragraph{The Kasteleyn matrix.}
We assign the integers $n_i$ to the edges according to Figure \ref{f:fdph1d3}.  
Then, we find that 
\bea
\text{Gauge group 1~:} \qquad k_1 &=& 1  = n_1 - n_2 +n_5 - n_6 ~, \nn \\
\text{Gauge group 2~:} \qquad k_2 &=& 1 = - n_1 + n_3 - n_4 +n_6  ~, \nn \\
\text{Gauge group 3~:} \qquad k_3 &=& -1 = n_2 - n_3~, \nn \\
\text{Gauge group 4~:} \qquad k_4 &=& -1 = n_4 - n_5~.
\eea  
We choose
\bea
n_3 = n_5 = 1,\quad n_i=0 \; \text{otherwise}~.
\eea
Since the fundamental domain contains only one white node and one black node, the Kasteleyn matrix is $1\times 1$ and, therefore, coincides with its permanent:
\bea \label{Kph2d3}
K &=& X_{31}z^{n_2} + X_{23}z^{n_3} +  X_{42} x^{-1} z^{n_4} + X_{14} x^{-1} z^{n_5} + X_{21} x^{-1} y^{-1} z^{n_6} + X_{12} y^{-1} z^{n_1} \nn \\
&=& X_{31} + X_{23}z +  X_{42} x^{-1}  + X_{14} x^{-1} z +  X_{21} x^{-1} y^{-1}  + X_{12} y^{-1} \nn \\ 
&& \text{(for $n_3 = n_5 =1$ and $n_i=0$)}~,
\eea

\paragraph{The toric diagram.}  We construct the toric diagram of this model using two methods
\begin{itemize}
\item {\bf The charge matrices.}
From \eref{Kph2d3}, we can take the perfect matchings to be
\bea
p_1= X_{23}, \quad p_2 = X_{42}, \quad p_3 = X_{12}, \quad p_4 = X_{21}, \quad p_5 = X_{31}, \quad p_6 = X_{14}~.
\eea
Since there is a one-to-one correspondence between the perfect matchings and the quiver fields,
\bea
Q_F = 0~.
\eea
Since the number of gauge groups is $G=4$, there are $G-2 = 2$ baryonic charges coming from the D-terms.  We find that the $Q_D$ matrix is given by
\bea
Q_D = \left(
\begin{array}{cccccc}
 1 & 0 & -1 & 1 & 0 & -1\\
 1 & 1 & 0 & 0 & -1 & -1 
\end{array}
\right)~. \label{qdph1d3}
\eea
Thefore, the total charge matrix $Q_t$ coincides with $Q_D$:
\bea
Q_t = \left(
\begin{array}{cccccc}
 1 & 0 & -1 & 1 & 0 & -1\\
 1 & 1 & 0 & 0 & -1 & -1 
\end{array}
\right)~. \label{qtph1d3}
\eea 
Hence, the $G'_t$ matrix is given by
\bea
G'_t =\left(
\begin{array}{cccccc}
  0 & 1 & 0 & 1 & 1 & 0\\
 0 & 0 & 1 & 1 & 0 & 0 \\
1 & 0 & 0 & 0 & 1 & 0
\end{array}
\right)~. \label{gtpph1d3}
\eea
Thus, we arrive at the toric diagram in Figure \ref{f:tord3}.

\begin{figure}[h]
\begin{center}
  \includegraphics[totalheight=4cm]{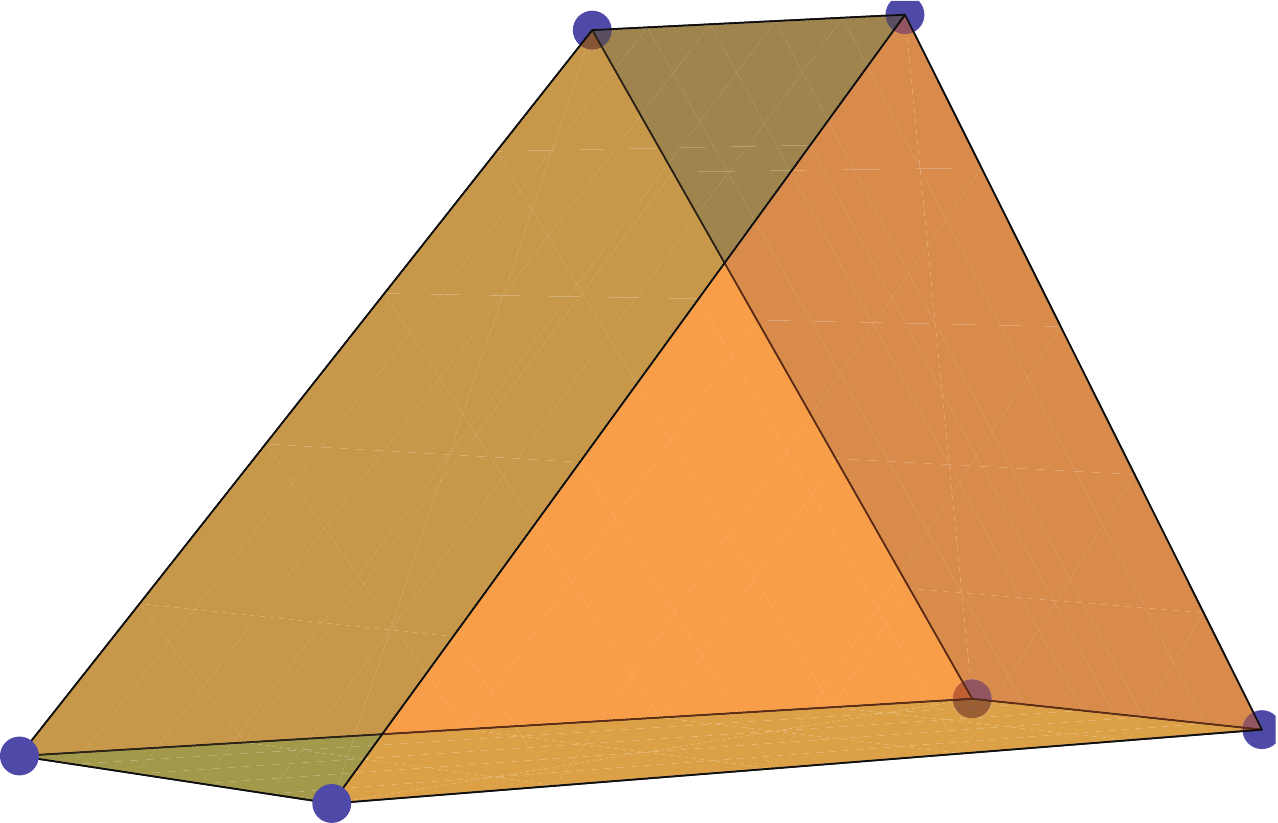}
 \caption{The toric diagram of the $D_3$ theory.}
  \label{f:tord3}
\end{center}
\end{figure}

\item {\bf The Kasteleyn matrix.} The powers of $x, y, z$ in each term of \eref{Kph2d3} give the coordinates of each point in the toric diagram. We collect these points in the columns of the following $G_K$ matrix: 
\bea
G_K = \left(
\begin{array}{cccccc}
 0 & -1 & 0 & -1 & -1 & 0 \\
 0 & 0 & -1 & -1 & 0 & 0 \\
 1 & 0 & 0 & 0 & 1 & 0
\end{array}
\right)~.
\eea
Note that the toric diagrams constructed from $G_K$ and $G'_t$ are the same up to a transformation
{\footnotesize $\CT =\left( \begin{array}{ccc} -1&0&0\\ 0&-1&0 \\ 0&0&1 \end{array} \right)  \in GL(3, \BZ)$},
where we have $G_K = \CT \cdot G'_t$.
\end{itemize}

\paragraph{The global symmetry.}  
Since all columns of the $Q_t$ matrix are distinct, the symmetry of the mesonic moduli space is expected to be $U(1)^3 \times U(1)_R$. 
Since there are 6 external points in the toric diagram, we have $6-4 =2$ baryonic symmetries, under which the perfect matchings are charged according to the $Q_D$ matrix.  
The global symmetry of this model is the product of mesonic and baryonic symmetries:  $U(1)^3 \times U(1)_R \times U(1)_{B_1} \times U(1)_{B_2}$.  

Below, there is a study of the Higgs mechanism of this theory.

\subsubsection{\emph{Phase I of $\CC \times \BC$} from giving a VEV to one of $X_{14}$, $X_{42}$, $X_{23}$, $X_{31}$}
By symmetry, giving a VEV to any of these fields leads to the same theory.
For definiteness, let us examine the case of $X_{14}$.
This amounts to removing one of the edges that separate the faces corresponding to gauge groups 1 and 4. As a result, these gauge groups are combined into one gauge group, which is identified as 1.
The quiver diagram and tiling are drawn in \fref{f:fdphase1conxc} (up to relabelling of gauge groups). 
The CS levels associated with the higgsed gauge groups are added, and so the new CS levels are
\bea
k_1 = 0,\quad k_2=1,\quad k_3=-1~.
\eea
Therefore, the resulting theory is the $\sD_1\sC$ model (Phase I of $\CC \times \BC$).


\subsubsection{\emph{Phase III-B of $\CC \times \BC$} from giving a VEV to one of $X_{12}$, $X_{21}$}
Let us first give a VEV to $X_{12}$.     
This amounts to removing one of the edges that separate the faces corresponding to gauge groups 1 and 2.  As a result, the gauge groups 1 and 2 are combined into one gauge group, which is identified as 1.  For convenience, we relabel gauge groups 3, 4 as 2, 3. 
The quiver diagram and tiling are drawn in \fref{f:phase3conxc}. 
The CS levels associated with the higgsed gauge groups are added, and so the new CS levels are
\bea
k_1 = 2,\quad k_2=-1,\quad k_3=-1~.
\eea
Thefore, the resulting theory is Phase III-B of $\CC \times \BC$.

\subsection{Higgsing Phase II of $D_3$}
\subsection*{A summary of Phase II of $D_3$ (the $\sH_{2} \partial_1$ model)}
\begin{figure}[ht]
\begin{center}
  \vskip -0.5cm
  \hskip -7cm
  \includegraphics[totalheight=6.2cm]{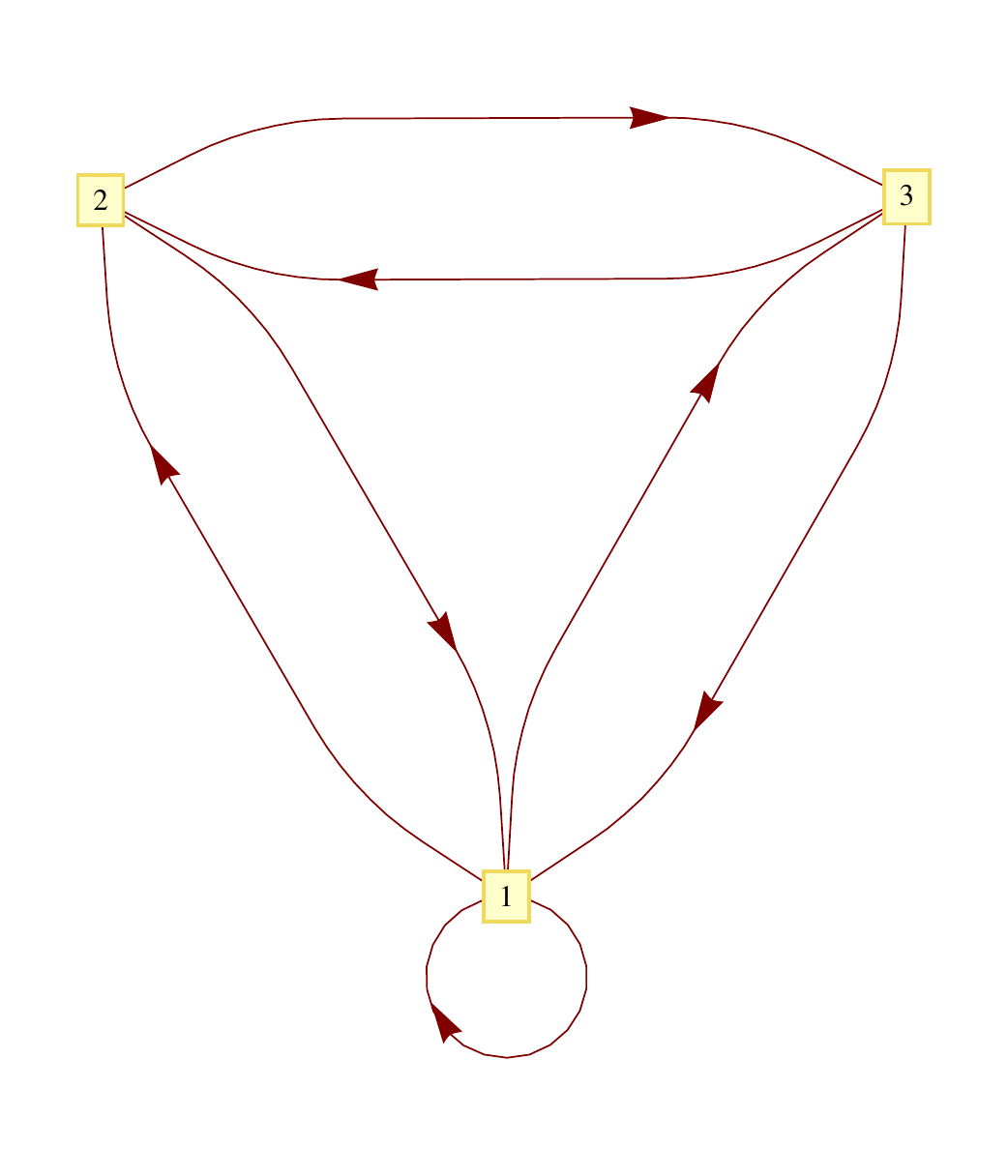}
    \vskip -5.5cm
  \hskip 8cm
  \includegraphics[totalheight=4.9cm]{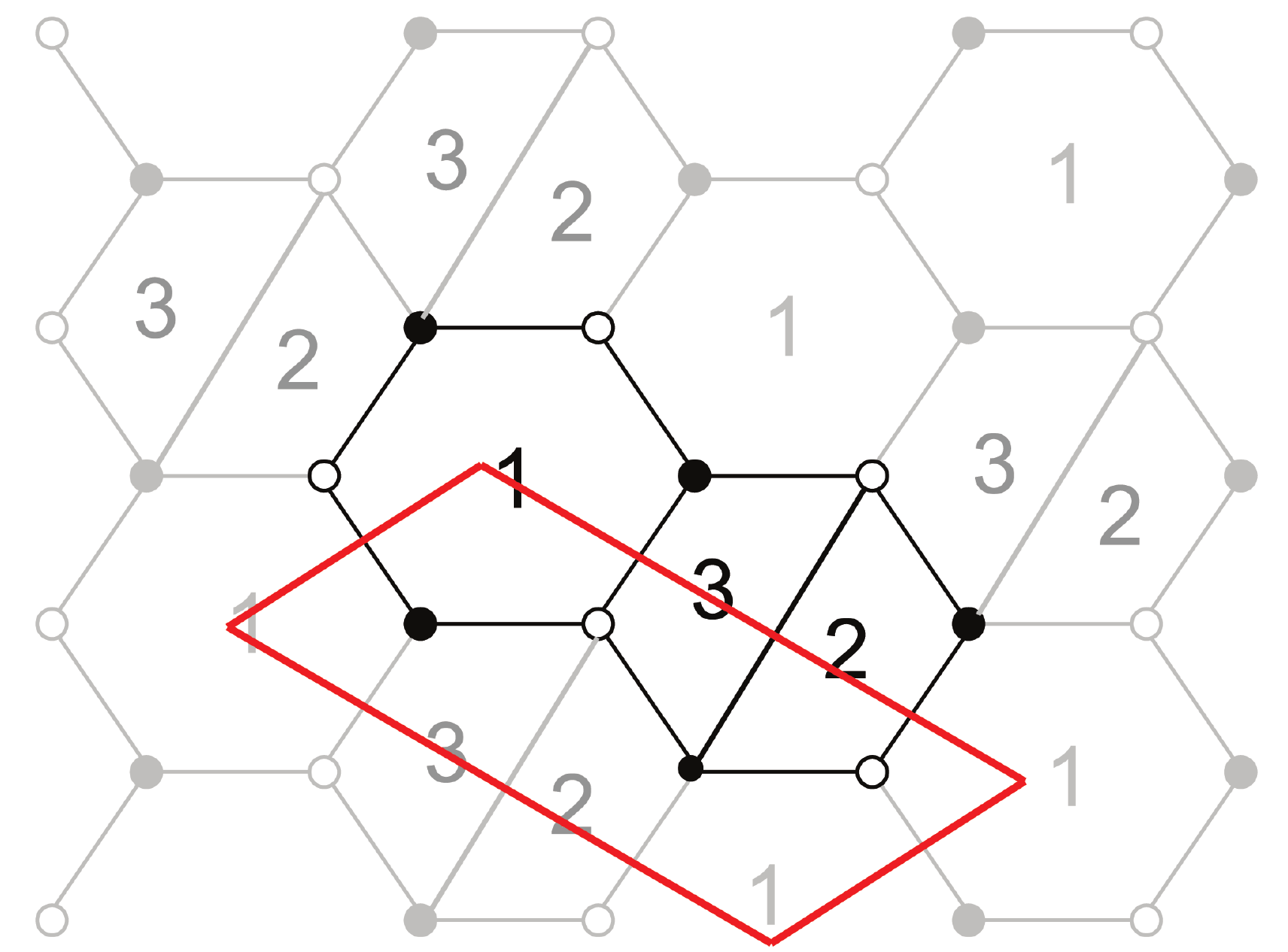}
 \caption{[Phase II of the $D_3$ theory]  (i) Quiver diagram for the $\sH_{2} \partial_1$ model.\ (ii) Tiling of the $\sH_{2} \partial_1$ model.}
  \label{f:phase2D3}
\end{center}
\end{figure}

The quiver diagram and tiling of this model are given in Figure \ref{f:phase2D3}.  The superpotential is given by
\bea
W = \tr \left( X_{32}X_{23}X_{31}X_{13} - X_{23}X_{32}X_{21}X_{12}  - \phi_1 \left(X_{13}X_{31} - X_{12}X_{21}\right) \right) \ .
\eea
We choose the CS levels to be $k_1= 1,~k_2= -1,~k_3 = 0$.

\begin{figure}[ht]
\begin{center}
   \includegraphics[totalheight=6.0cm]{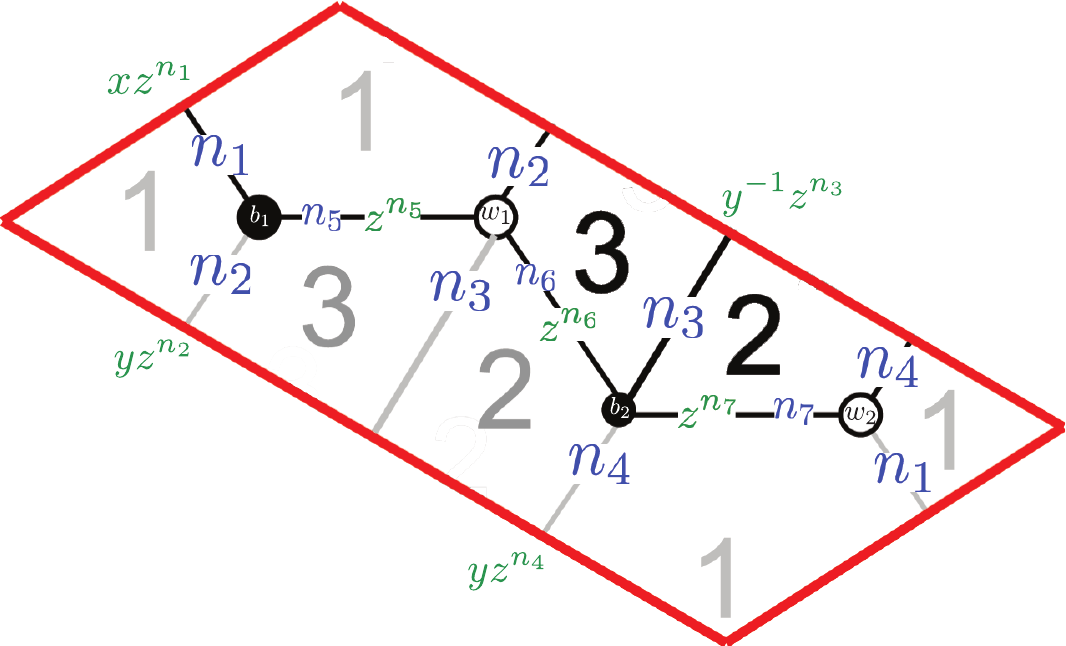}
 \caption{[Phase II of the $D_3$ theory]  The fundamental domain of tiling for the $\sH_{2} \partial_1$ model: assignments of the integers $n_i$ to the edges are shown in blue and the weights for these edges are shown in green.}
  \label{f:fdphase2d3}
\end{center}
\end{figure}

\paragraph{The Kasteleyn matrix.}
We assign the integers $n_i$ to the edges according to Figure \ref{f:fdphase2d3}.  We find that 
\bea
\text{Gauge group 1~:} \qquad k_1 &=& 1  = n_2 - n_4 - n_5 + n_7  ~, \nn \\
\text{Gauge group 2~:} \qquad k_2 &=& -1 =  n_3 + n_4 - n_6  - n_7 ~,  \nn \\
\text{Gauge group 3~:} \qquad k_3 &=& 0 =  - n_2 - n_3  + n_5 + n_6  ~.
\eea  
We choose
\bea
n_4= -1,\quad n_i=0 \; \text{otherwise}~.
\eea
The Kasteleyn matrix for this theory is
\be
K =   \left(
\begin{array}{c|cc}
& w_1 & w_2 \\
\hline
b_1& X_{13} y z^{n_2}+ X_{31} z^{n_5}  &   \  \phi_1 x z^{n_1}  \\
b_2 & X_{32} z^{n_6}+ X_{23} y^{-1}  z^{n_3}  & \  X_{21} y z^{n_4}+ X_{12} z^{n_7} 
\end{array}
\right)~.
\ee
The permanent of this matrix is given by
\bea  \label{permKph2d3}
\perm~ K &=& X_{31} X_{12} z^{n_5+n_7} + X_{13} X_{12} y z^{n_2+n_7} + X_{31} X_{21} y z^{n_4+n_5} +  \nn \\
&& + X_{13} X_{21} y^{2} z^{n_2+n_4} + \phi_1 X_{32} x z^{n_1+n_6} + \phi_1 X_{23} x y^{-1} z^{n_1+n_3} \nn \\
&=& X_{31} X_{12}  + X_{13} X_{12} y  + X_{31} X_{21} y z^{-1}  + X_{13} X_{21} y^{2} z^{-1}  + \phi_1 X_{32} x  + \nn\\
&& + \phi_1 X_{23} x y^{-1}  \qquad \text{(for $n_4= -1$ and $n_i= 0$ otherwise)} \  .
\eea

\paragraph{The perfect matchings.} From the permanent of the Kasteleyn matrix, we can write the perfect matchings as collections of fields as follows:
\bea
&& p_1 = \{X_{31}, X_{12}\},~ p_2 =  \{ X_{21}, X_{13} \} ,~p_3 = \{X_{23}, \phi_1 \}, \nn \\
&& p_4 = \{ X_{32},  \phi_1 \},~p_5 = \{X_{31} , X_{21} \},~ p_6 = \{ X_{12}, X_{13} \} \ . 
\eea
In turn, we find the parameterisation of fields in terms of perfect matchings: 
\bea
&& X_{31} = p_1 p_5, \quad X_{12} = p_1 p_6, \quad X_{21} = p_2 p_5, \nn \\
&& X_{13} = p_2 p_6, \quad X_{23} = p_3, \quad \phi_1 = p_3 p_4, \quad X_{32} = p_4~. \label{quivfieldph2d3}
\eea
This is summarised in the perfect matching matrix:
\beq
P=\left(\begin{array} {c|cccccc}
  \;& p_1&p_2&p_3&p_4&p_5&p_6\\
  \hline
  X_{31}&1&0&0&0&1&0 \\
  X_{12}&1&0&0&0&0&1\\
    X_{21}&0&1&0&0&1&0\\
  X_{13}&0&1&0&0&0&1\\
  X_{23}&0&0&1&0&0&0\\
    \phi_{1}&0&0&1&1&0&0 \\
  X_{32}&0&0&0&1&0&0
 \end{array}
\right).
\eeq
Basis vectors of the null space of $P$ are given in the row of the charge matrix:
\bea
Q_F = (1,1,0, 0 -1,-1)~. \label{qfph2d3}
\eea
Hence, we see that the relation between the perfect matchings is given by
\bea
p_1+p_2 - p_5 - p_6 = 0 \ . \label{relph2d3}
\eea
Since the coherent component of the Master space is generated by the perfect matchings, subject to the relation \eref{relph2d3}, it follows that 
\bea
\firr{\sH_{2} \partial_1} = \BC^6//Q_F = \BC^6//(1,1,0, 0 -1,-1)~.  \label{quot2a}
\eea
Since the quotient $\BC^4//(1,1,-1,-1)$ is known to be the conifold ($\CC$) and $\BC^2$ is parametrised by the remaining perfect matchings with charge 0, it follows that
\bea
\firr{\sH_{2} \partial_1}  = \CC \times \BC^2~.
\eea

\paragraph{The toric diagram.} We demonstrate two methods of constructing the toric diagram. 
\begin{itemize}
\item{\bf The charge matrices.}  
Since the number of gauge groups is $G=3$, there is $G-2 = 1$ baryonic symmetry, which we shall denote as $U(1)_{B_1}$, coming from the D-terms.  We collect the baryonic charges of the perfect matchings in the $Q_D$ matrix:
\bea
Q_D = (1, 0, -1, 1, 0, -1)~.  \label{qdph2d3}
\eea 
Note that since the CS coefficient $k_3=0$, the $Q_D$ matrix \eref{qdph2d3} has been chosen such that the baryonic charge of each quiver field in \eref{quivfieldph2d3} coincides with the quiver charge under gauge group 3.
From \eref{qfph2d3} and \eref{qdph2d3}, the total charge matrix is given by
\bea
Q_t = \left(
\begin{array}{cccccc}
1& 0 & -1 & 1 & 0 & -1 \\
 1 & 1 & 0 & 0 & -1 & -1 
\end{array}
\right)~. 
\eea
Note that this is precisely the same as the $Q_t$ matrix \eref{qtph1d3} for Phase I. 
Thus, we obtain the same matrix $G'_t$ as for Phase I \eref{gtpph1d3}.
Therefore, toric diagram is given by Figure \ref{f:tord3}.   

\item {\bf The Kasteleyn matrix.}
The powers of $x, y, z$ in each term of \eref{permKph2d3} give the coordinates of each point in the toric diagram.
We collect these points in the columns of the following $G_K$ matrix:
\bea
G_K = \left(
\begin{array}{cccccc}
 1 & 0 & 0 & 0 & 1 & 0 \\
 -1 & 1 & 1 & 2 & 0 & 0 \\
 0 & 0 & -1 & -1 & 0 & 0
\end{array}
\right)~.
\eea
Note that the toric diagrams constructed from $G_K$ and $G'_t$ are the same up to a transformation
{\footnotesize $\CT =\left( \begin{array}{ccc} 1&0&0\\-1&1&1\\0&-1&0 \end{array} \right)  \in GL(3, \BZ)$},
where we have $G_K = \CT \cdot G'_t$.
\end{itemize}

Below, there is a study of the Higgs mechanism of this theory.

\subsubsection{\emph{Phase II of $\CC \times \BC$} from giving a VEV to one of $X_{23}$, $X_{32}$}
Let us give a VEV to either $X_{23}$ or $X_{32}$.   
This amounts to removing one of the edges that separate the faces corresponding to gauge groups 2 and 3.  As a result, these two gauge groups are combined into one gauge group, which is identified as gauge group 2.
The quiver diagram and tiling are drawn in \fref{f:phase2conxc}. 
The CS levels associated with the higgsed gauge groups are added, and so the new CS levels are
\bea
k_1 = 1,\quad k_2=-1 ~.
\eea
The resulting theory is Phase II of $\CC \times \BC$.
The toric diagram is drawn in \fref{f:torconxc}. 

\comment{
\begin{figure}[ht]
\begin{center}
  \vskip 2cm
  \hskip -6cm
  \includegraphics[totalheight=1.2cm]{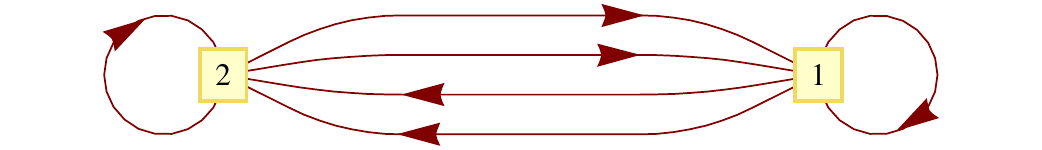}
  \vskip -3.0cm
  \hskip 9cm
  \includegraphics[totalheight=4.5cm]{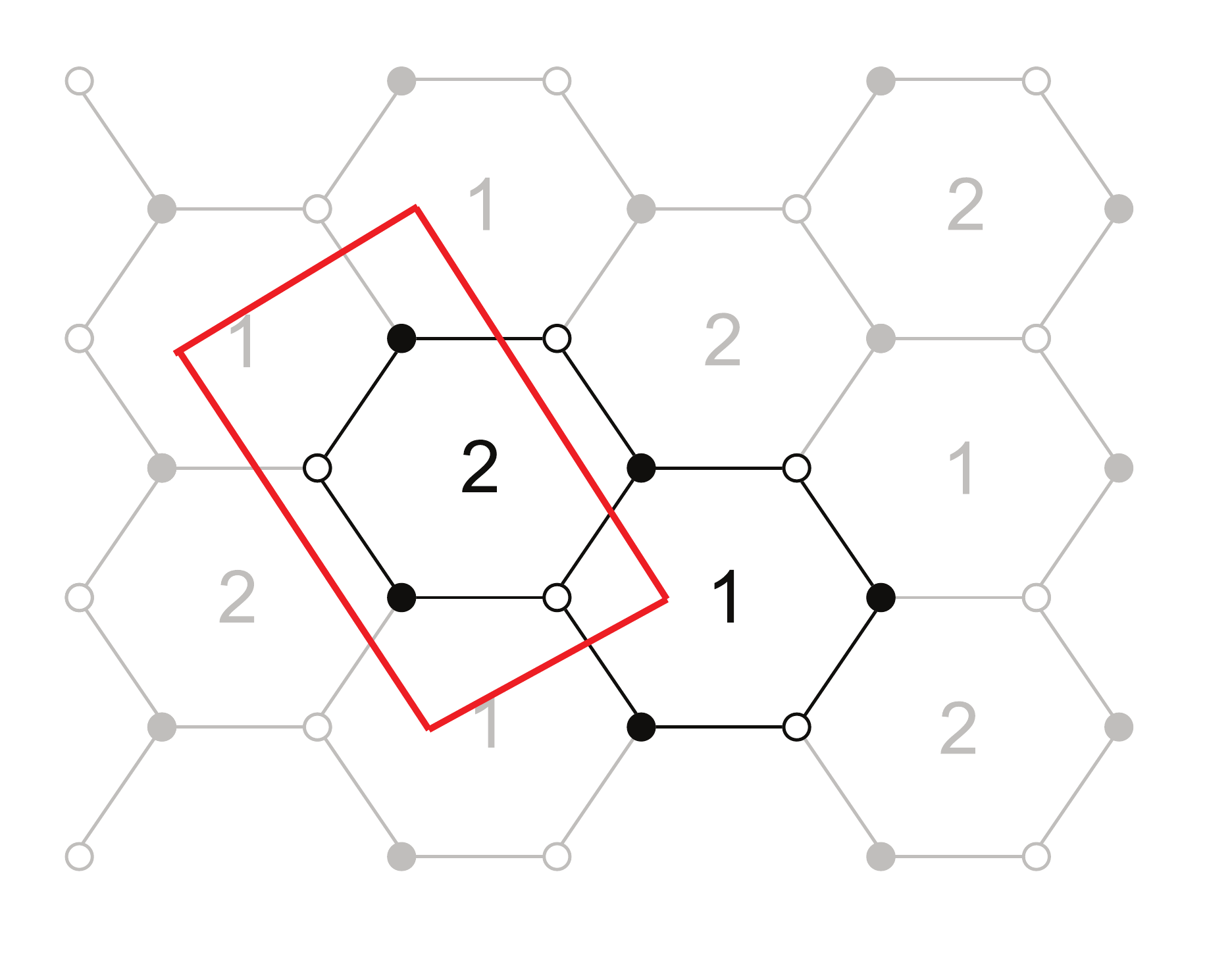}
 \caption{[Phase II of $\CC \times \BC$] (i) Quiver diagram for the $\sH_2$ model. \ (ii) Tiling for the $\sH_2$ model.}
  \label{f:phase2conxc}
\end{center}
\end{figure}} 

\subsubsection{\emph{Phase I of $\BC^4$} from giving a VEV to one of $X_{13}$, $X_{31}$}
For definiteness, let us turn on a VEV to $X_{31}$ (the case of $X_{13}$ can be treated in a similar way).
This amounts to removing one of the edges that separate the faces corresponding to gauge groups 1 and 3, and collapsing the two vertices adjacent to 
a bivalent vertex into a single vertex of higher valence.
As a result, the gauge groups 1 and 3 are combined into one gauge group, which is identified as 1, and the edges corresponding to $\phi_1, X_{13}$ are removed. 
The quiver diagram and tiling are drawn in \fref{f:con}. 
The CS levels associated with the higgsed gauge groups are added, and so the new CS levels are
\bea
k_1 = 1,\quad k_2=-1 ~.
\eea
The resulting theory is Phase I of the $\BC^4$ theory.
The toric diagram is drawn in \fref{f:torabjm}. 

\subsubsection{\emph{The $\CC \times \BC$ theory} from giving a VEV to one of $X_{12}$, $X_{21}$}
This is similar to the previous case.  The quiver diagram and tiling are drawn in \fref{f:con} (with the gauge group 2 being relabelled as 3). 
The CS levels associated with the higgsed gauge groups are added, and so the new CS levels are
\bea
k_1= k_3 =0~.
\eea
The tiling suggests that there is a branch of the moduli space which is the conifold $\CC$.
As discussed in \sref{summary}, in the presence of the gauge kinetic term, there is an additional complex degree of freedom. 
It follows that the mesonic moduli space is $\CC \times \BC$.

\subsection{Higgsing Phase III of $D_3$}
\subsection*{A summary of Phase III of $D_3$ (the $\sD_3 \sH_1$ model)}
\begin{figure}[h]
\begin{center}
  \hskip -7cm
  \includegraphics[totalheight=5cm]{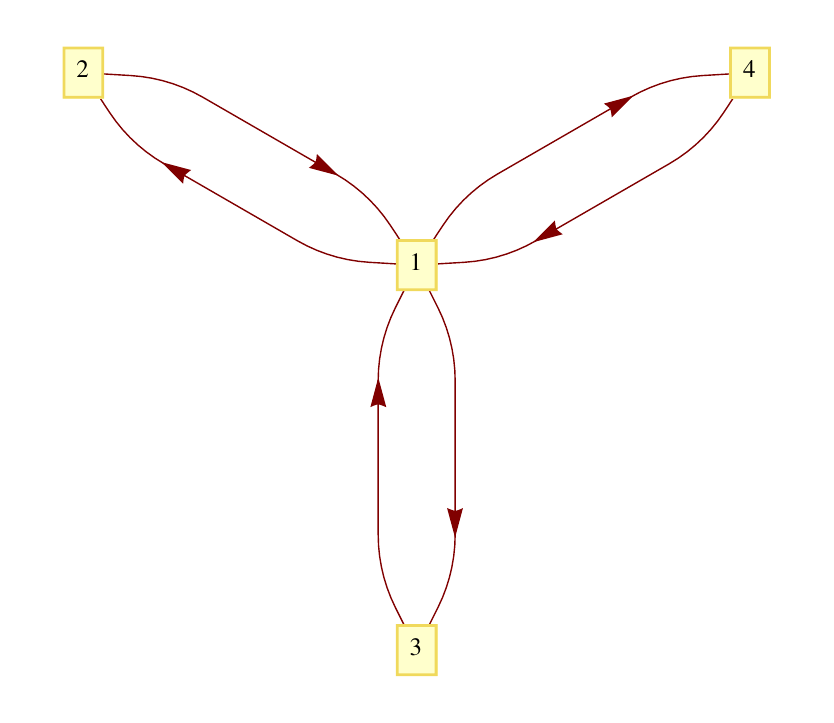}
    \vskip -5.0cm
  \hskip 8.4cm
  \includegraphics[totalheight=5cm]{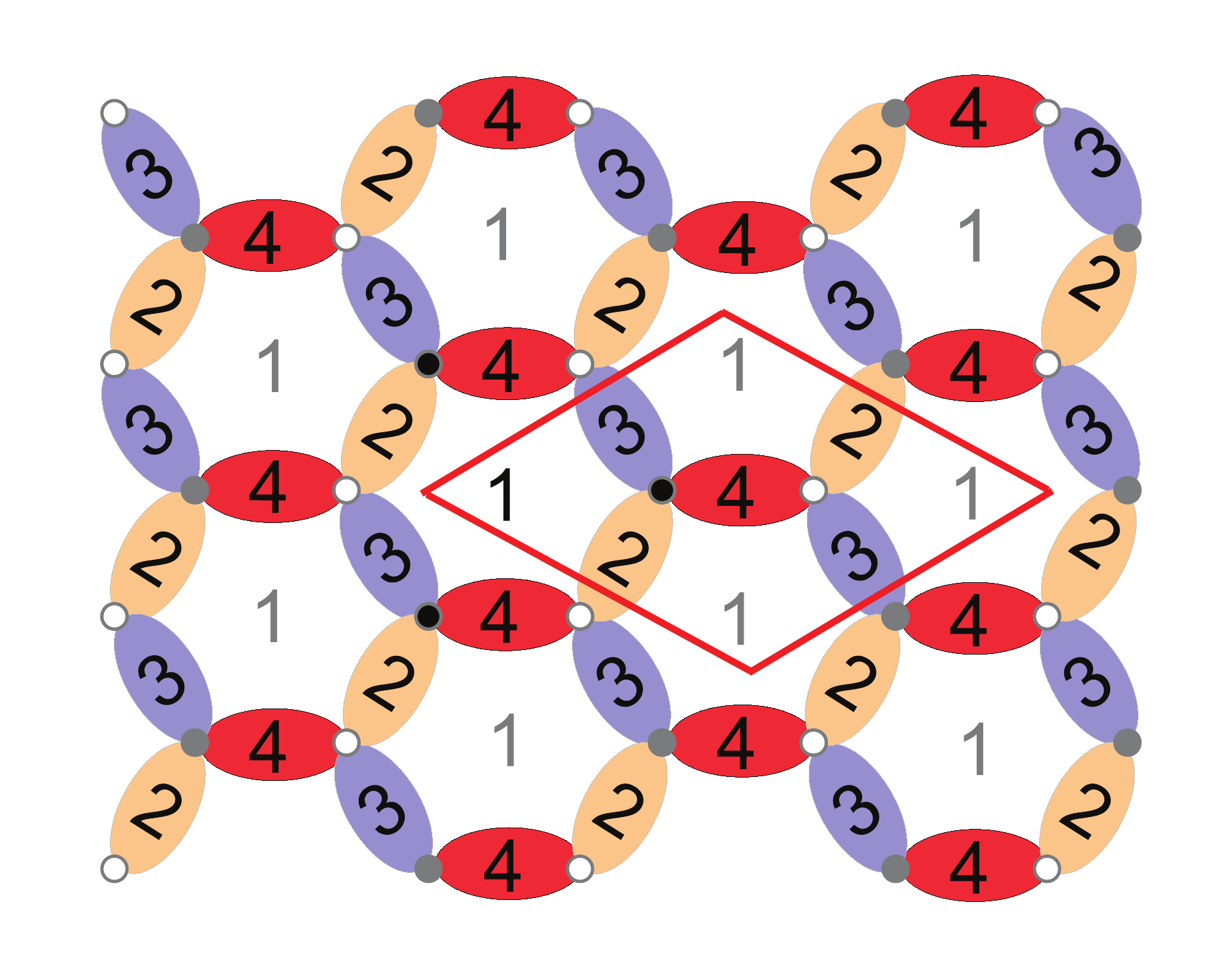}
 \caption{[Phase III of the $D_3$ theory]  (i) Quiver diagram of the $\sD_3\sH_1$ model.\qquad (ii) Tiling of the $\sD_3\sH_1$ model.}
  \label{f:ph3d3}
\end{center}
\end{figure}

The quiver diagram and tiling of this model are drawn in Figure \ref{f:ph3d3}.
The superpotential is given by
\bea
W = \tr \left( X_{13}X_{31}[X_{14}X_{41},X_{12}X_{21}] \right)~.
\eea
We choose the CS levels to be $(k_1,k_2,k_3,k_4) = (1,-1,1,-1)$.

\begin{figure}[h]
\begin{center}
   \includegraphics[totalheight=6.0cm]{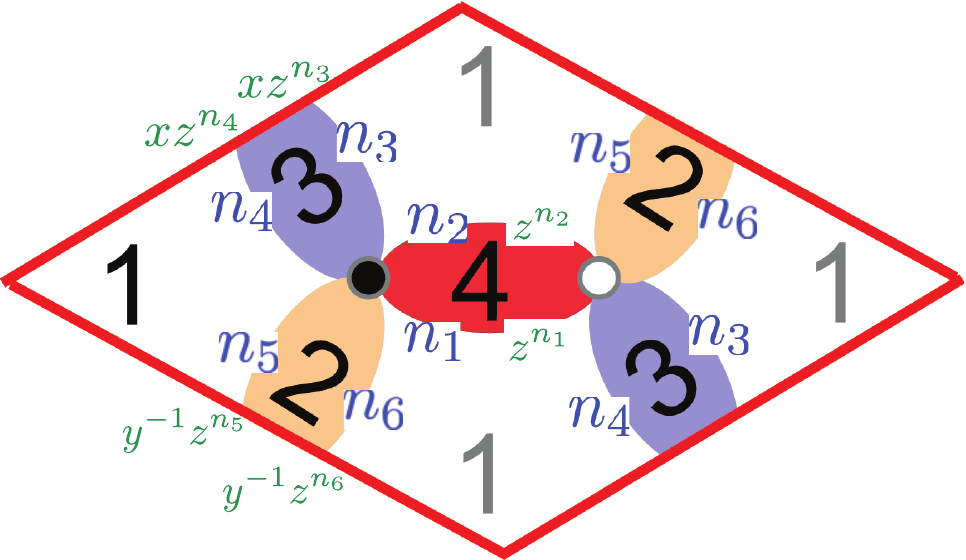}
 \caption{[Phase III of the $D_3$ theory]  The fundamental domain of tiling for the $\sD_3\sH_1$ model: assignments of the integers $n_i$ to the edges are shown in blue and the weights for these edges are shown in green.}
  \label{f:fdph3d3}
\end{center}
\end{figure}

\paragraph{The Kasteleyn matrix.}
We assign the integers $n_i$ to the edges according to Figure \ref{f:fdph3d3}.  We find that 
\bea
\text{Gauge group 1~:} \qquad k_1 &=& 1  = n_1 - n_2 + n_3 - n_4 + n_5 - n_6 ~, \nn \\
\text{Gauge group 2~:} \qquad k_2 &=& -1 = - n_5 +n_6  ~, \nn \\
\text{Gauge group 3~:} \qquad k_3 &=& 1 = - n_3 + n_4 ~, \nn \\
\text{Gauge group 4~:} \qquad k_4 &=& -1 = -n_1 + n_2~.
\eea  
We choose
\bea
n_1 = n_4 = n_5 = 1,\quad n_i=0 \; \text{otherwise}~.
\eea
Since the fundamental domain contains only one white node and one black node, the Kasteleyn matrix is $1\times 1$ and, therefore, coincides with its permanent:
\bea \label{Kph3d3}
K &=& X_{14}z^{n_1} + X_{41}z^{n_2} +  X_{13} x z^{n_3} + X_{31} x z^{n_4} + X_{12} y^{-1} z^{n_5} + X_{21} y^{-1} z^{n_6} \nn \\
&=& X_{14}z + X_{41} +  X_{13} x  + X_{31} x z + X_{12} y^{-1} z + X_{21} y^{-1}  \nn \\ 
&& \text{(for $n_1 = n_4 = n_5 = 1$ and $n_i=0$)}~,
\eea
where the powers of $x, y, z$ in each term give the coordinates of each point in the toric diagram. 
We collect these points in the columns of the following $G_K$ matrix:
\bea
G_K = \left(
\begin{array}{cccccc}
 1 & 0 & 0 & 0 & 1 & 0 \\
 0 & 0 & -1 & -1 & 0 & 0 \\
 0 & 1 & 0 & 1 & 1 & 0
\end{array}
\right)~.
\eea
The toric diagram in drawn in Figure \ref{f:tord3}.

Below, there is a study of the Higgs mechanism of this theory.

\subsubsection{\emph{Phase III-A of $\CC \times \BC$} from giving a VEV to one of $X_{12}$, $X_{21}$, $X_{14}$, $X_{41}$}
For definiteness, let us turn on a VEV to $X_{41}$. 
This amounts to removing one of the edges that separate the faces corresponding to gauge groups 1 and 4.
As a result, the gauge groups 1 and 4 are combined into one gauge group, which is identified as 1.
The quiver diagram and tiling are drawn in \fref{f:phase3conxc}. 
The CS levels associated with the higgsed gauge groups are added, and so the new CS levels are
\bea
k_1 = 0, \quad k_2 = -1, \quad k_3 = 1~.
\eea
Therefore, the resulting theory is Phase III-A of the $\CC \times \BC$ theory.

From symmetries of the quiver diagram and tiling, we see that turning on a VEV to either $X_{12}$, $X_{21}$, or $X_{14}$ yields  the same quiver and tiling as in \fref{f:phase3conxc}.
The CS levels are respectively $(0,1,-1)$, $(0,1,-1)$ and $(0,-1,1)$.
Hence, it can be seen that the resulting theory is Phase III-A of the $\CC \times \BC$ theory, as above.

\subsubsection{\emph{Phase III-B of $\CC \times \BC$} from giving a VEV to one of $X_{13}$, $X_{31}$}
This is similar to the previous case. The resulting quiver diagram and tiling are \fref{f:phase3conxc}, and the new CS levels are
\bea
k_1 = 2, \quad k_2 = -1, \quad k_3 = -1~.
\eea
The resulting theory is Phase III-B of the $\CC \times \BC$ theory.

\section{Higgsing The $Q^{1,1,1}$ Theory}
\subsection*{A summary of the $Q^{1,1,1}$ theory}
\begin{figure}[ht]
  \begin{center} 
   \vskip 0.5cm
  \hskip -8cm
  {  \epsfxsize = 8cm \epsfbox{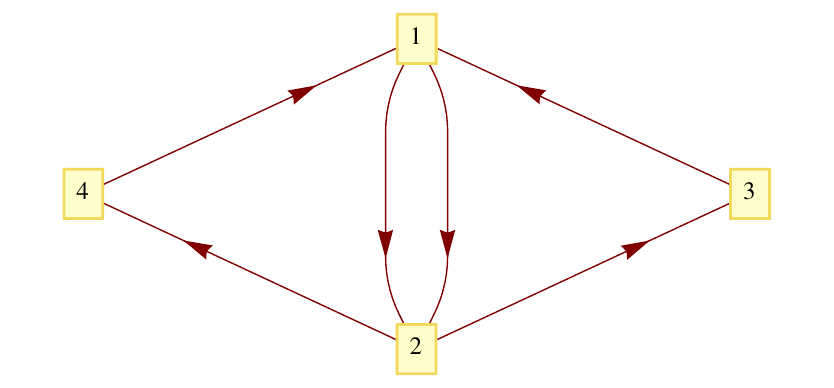}}
    \vskip -4.5cm
  \hskip 7.5cm
  { \epsfxsize = 7cm \epsfbox{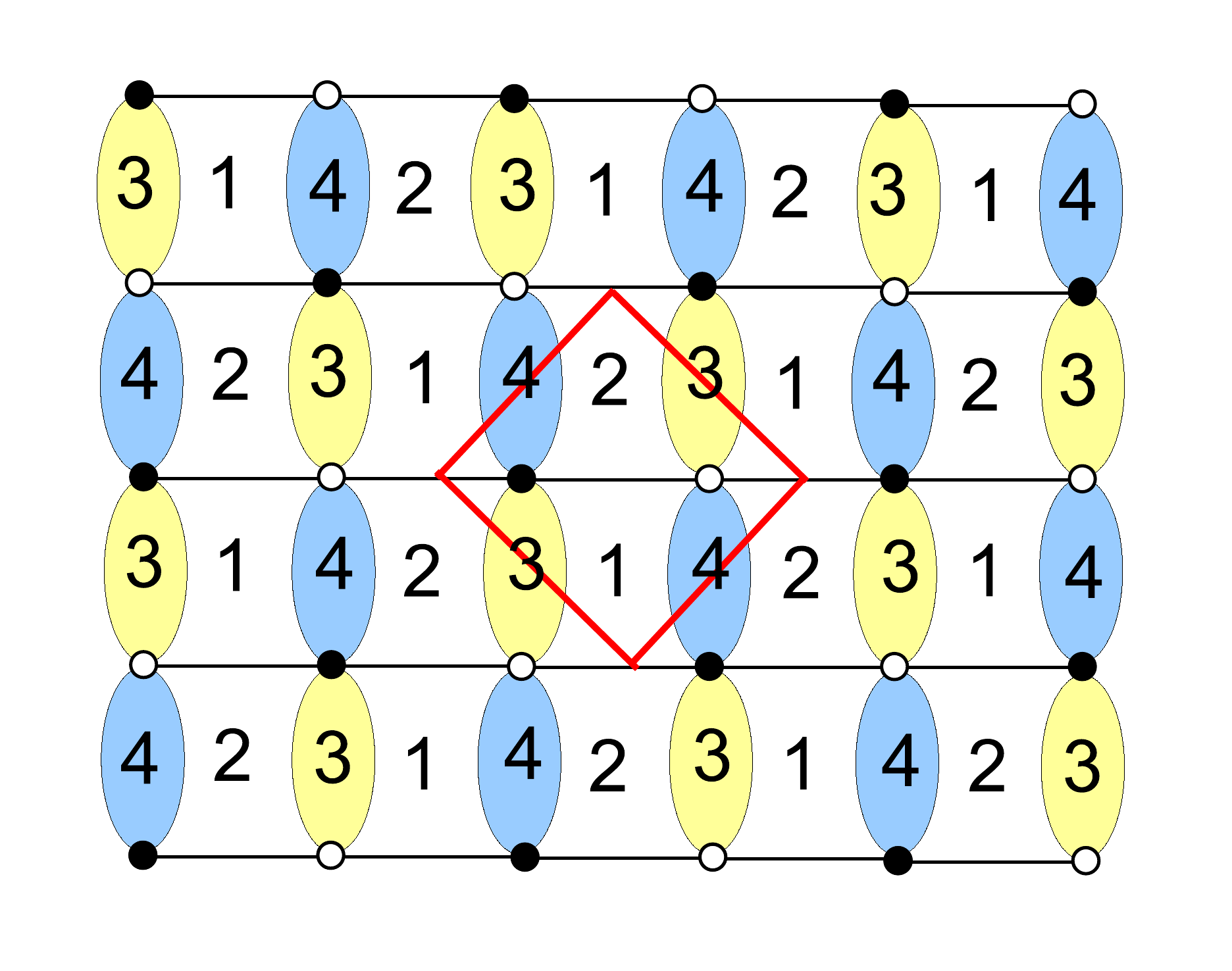}}
\caption{(i) Quiver diagram of the $Q^{1,1,1}$  theory.\ (ii) Tiling of the $Q^{1,1,1}$  theory.}
  \label{f:qutq111}
  \end{center}  
\end{figure}

The $Q^{1,1,1}$ theory \cite{Fabbri:1999hw, Franco:2008um, Franco:2009sp} has 4 gauge groups and 6 chiral fields: $X^1_{12}$, $X^2_{12}$, $X_{23}$, $X_{24}$, $X_{31}$ and $X_{41}$. The quiver diagram and tiling are presented in Figure \ref{f:qutq111}. The superpotential is
\bea
W = \epsilon_{ij} \tr\left(X^i_{12}X_{23}X_{31}X^j_{12}X_{24}X_{41}\right).
\eea
We choose the CS levels to be $(k_1, k_2, k_3, k_4)=(1,1,-1,-1)$.

\begin{figure}[ht]
 \centerline{  \epsfxsize = 9cm  \epsfbox{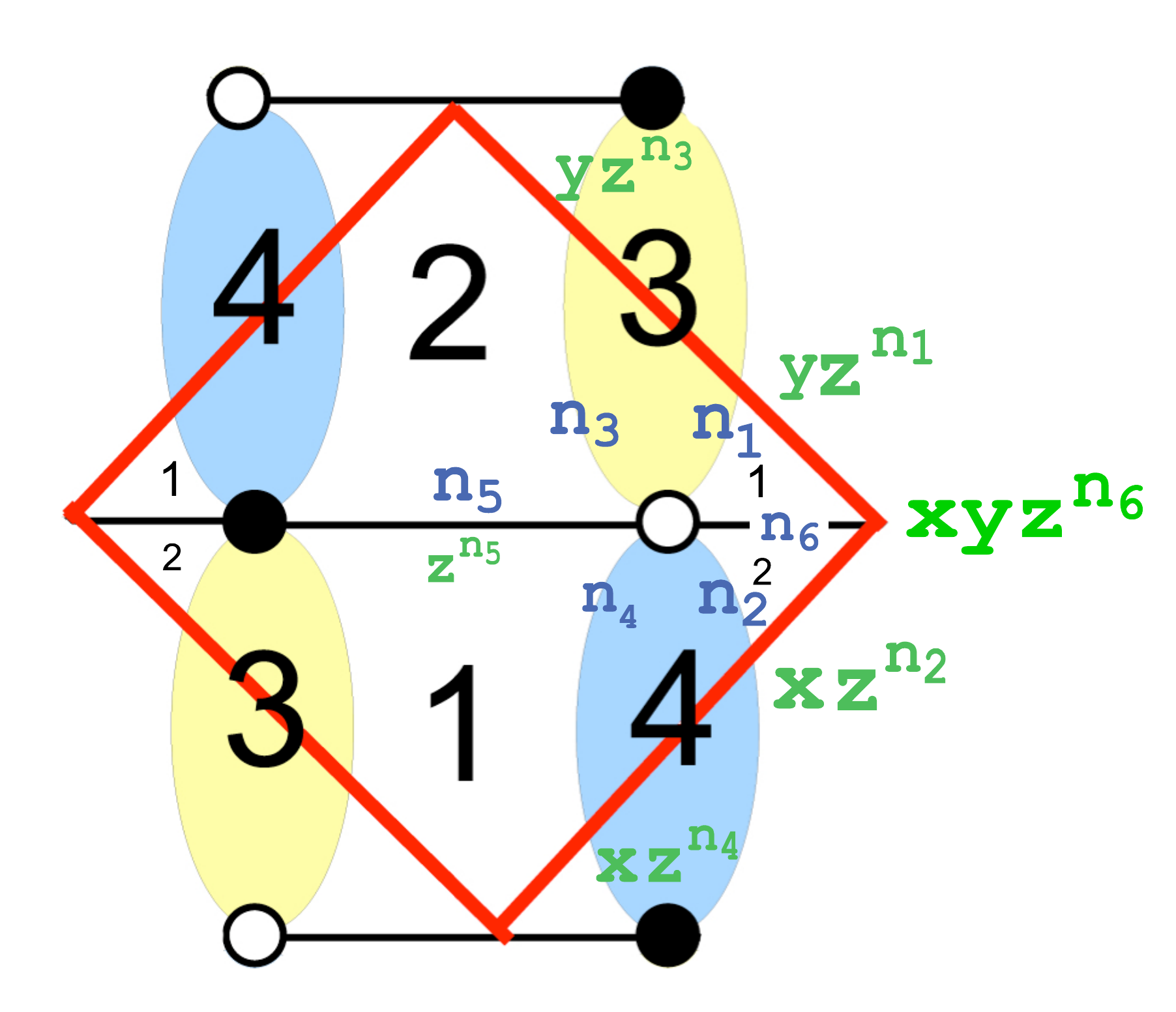} }
 \caption{The fundamental domain of tiling for the $Q^{1,1,1}$ theory: assignments of the integers $n_i$ to the edges are shown in blue and the weights for these edges are shown in green.}
  \label{f:fdq111}
\end{figure}

\paragraph{The Kasteleyn matrix.}
We assign integers $n_i$'s to the edges of the tiling as shown in Figure \ref{f:fdq111}. Accordingly, the Chern-Simons levels can be written as:
\bea
\text{Gauge group 1~:} \qquad k_1 &=&  1  =   n_1 + n_4 - n_5 - n_6 ~,  \nn \\
\text{Gauge group 2~:} \qquad k_2 &=&  1  = - n_2 - n_3 + n_5  + n_6 ~,  \nn \\
\text{Gauge group 3~:} \qquad k_2 &=& -1  =   n_3 - n_1~,  \nn \\
\text{Gauge group 4~:} \qquad k_3 &=& -1  =   n_2 - n_4~.  
\label{e:kaq111}
\eea  
We choose:
\bea
n_1 = - n_2 = 1,\qquad n_i=0 \;\;\text{otherwise}~.
\eea
The fundamental domain contains only two nodes and, therefore, the Kasteleyn matrix coincides with its permanent:
\bea
K &=& X_{31}  y z^{n_1} + X_{24} x z^{n_2} + X_{23}  y z^{n_3} + X_{41} x z^{n_4} + X^{1}_{12} z^{n_5} +   X^2_{12} x y z^{n_6} \nn \\
&=& X_{31} y z + X_{24} x z^{-1} + X_{23} y + X_{41} x  + X^{1}_{12} + X^2_{12} x y\nn \\ 
&& \text{(for $n_1 = - n_2 = 1,~ n_i=0 \; \text{otherwise}$)~.}  \label{kasq111}
\eea

\paragraph{The perfect matchings.}  From \eref{kasq111}, we can take the perfect matchings to be
\bea
p_1 = X_{31}, \quad p_2 = X_{24}, \quad r_1 = X_{23}, \quad r_2 = X_{41}, \quad s_1 = X^1_{12}, \quad s_2 = X^2_{12}~.
\eea
Since there is a one-to-one correspondence between the perfect matchings and the quiver fields, it follows that 
\bea
Q_F = 0~.
\eea

\begin{figure}[ht]
\begin{center}
  \includegraphics[totalheight=7.0cm]{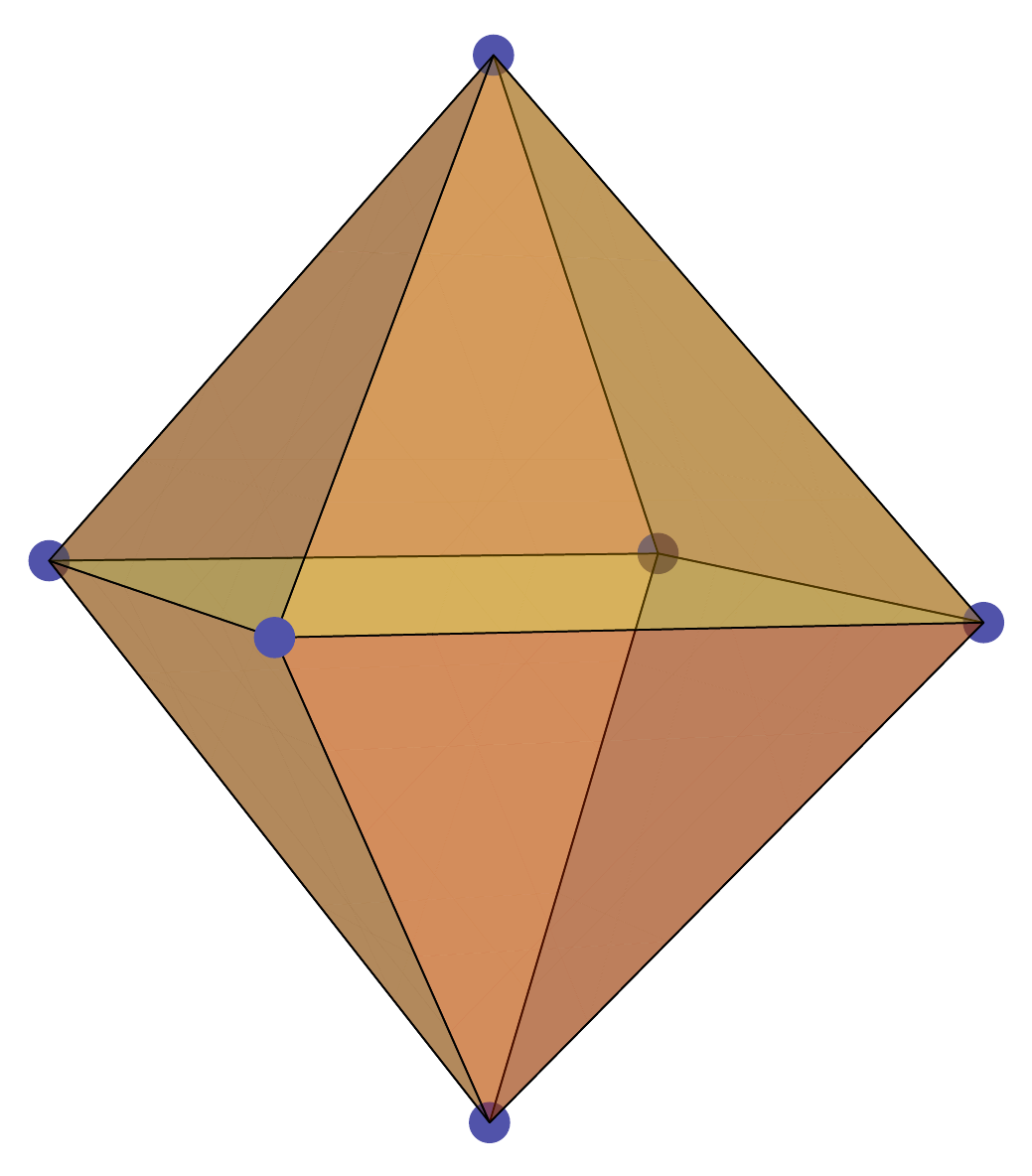}
 \caption{The toric diagram of the $Q^{1,1,1}$ theory.}
  \label{f:tdq111}
\end{center}
\end{figure}

\paragraph{The toric diagram.} We construct the toric diagram of this model using two methods:
\bi
\item {\bf Charge matrices.}
Since the number of gauge groups is $G=4$, there are $G-2=2$ baryonic charges coming from the $D$-terms. These can be collected in the rows of the $Q_D$ matrix, which also coincides with the total charge matrix $Q_t$:
\be
Q_t = Q_D =   \left(
\begin{array}{cccccc}\Blue
1 & 1 & -1 & -1 &  0 &  0 \\
0 & 0 &  1 &  1 & -1 & -1 \Black
\end{array}
\right)~. \label{e:qdq111}
\ee
Taking the kernel of (\ref{e:qdq111}), and deleting its first row, we obtain the $G'_t$, whose columns the coordinates of the toric diagram:
\bea
G'_t = \left(
\begin{array}{cccccc}
 1 & -1 & 0 & 0 & 0 & 0\\
 0 & 0 & 1 & -1 & 0& 0 \\
 0 & 0 & 0 & 0 & 1 & -1 
\end{array}
\right)~. \label{e:gtq111}
\eea
The toric diagram is presented in Figure \ref{f:tdq111}. 

\item {\bf The Kasteleyn matrix.}
The powers of $x,y$ and $z$ in \eref{kasq111} give the coordinates of the toric diagram which are collected in the columns of this matrix:
\bea
G_K = \left(
\begin{array}{cccccc}
 0 & 1 & 1 & 0 & 0 & 1 \\
 1 & 0 & 1 & 0 & 1 & 0 \\
 1 & -1 & 0 & 0 & 0 & 0
\end{array}
\right)~.
\eea

\ei

\paragraph{The baryonic charge.} From Figure \ref{f:tdq111}, we can see that the toric diagram of this model has 6 external points and, accordingly, the number of baryonic symmetries is $6-4=2$. The charges of the perfect matchings under these two symmetries, which we shall refer to as $U(1)_{B_1}$ and $U(1)_{B_2}$, can be found in the rows of the $Q_D$ matrix.

\paragraph{The global symmetry.} From (\ref{e:qdq111}), we observe that the $Q_t$ matrix has three pairs of repeated columns. Since the total rank of the mesonic symmetry is 4, this can be identified with $SU(2)^3 \times U(1)_R$.  Since there is precisely one abelian factor, it can be unambiguously identified with the R-symmetry. 
This mesonic symmetry can also be seen from the $G'_t$ matrix by noticing that the three rows contain weights of $SU(2)$.
The R-charge of each perfect matching can be computed using a symmetry argument: the perfect matchings are completely symmetric and the Calabi-Yau condition simply divides 2 equally among them. Therefore, each perfect matching has R-charge $1/3$. In Table \ref{t:chq111}, we present the global charges of each perfect matching. 

\begin{table}[h]
 \begin{center}  
  \begin{tabular}{|c||c|c|c|c|c|c|c|c|}
  \hline
  \;& $SU(2)_1$&$SU(2)_2$&$SU(2)_3$&$U(1)_R$&$U(1)_{B_1}$&$U(1)_{B_2}$&fugacity\\
  \hline  \hline 
  $p_1$&   1&   0&   0&   1/3&   1&   0&   $t x_1 b_1$\\
  \hline
  $p_2$&  $-1$&   0&   0&   1/3&   1&   0&   $t b_1 / x_1$\\
  \hline
  $r_1$&   0&   1&   0&   1/3& $ -1$&   1&   $t x_2 b_2 / b_1$\\
  \hline
  $r_2$&   0&  $-1$&   0&   1/3&  $-1$&   1&   $t b_2 / (x_2 b_1)$\\
  \hline
  $s_1$&   0&   0&   1&   1/3&   0&  $-1$&   $t x_3 / b_2$\\
  \hline
  $s_2$&   0&   0&  $-1$&   1/3&   0&  $-1$&   $t / (x_3 b_2)$\\
  \hline
  \end{tabular}
  \end{center}
\caption{Charges of the perfect matchings under the global symmetry of the $Q^{1,1,1}$ theory. Here $t$ is the fugacity of the R-charge, $x_1, x_2$ and $x_3$ are the weights of the $SU(2)$ symmetries, and $b_1, b_2$ are the fugacities associated with the baryonic symmetries $U(1)_{B_1}$ and $U(1)_{B_2}$.}
\label{t:chq111}
\end{table}

\paragraph{The Hilbert series.} 
Since the $Q_F$ matrix is zero, the Master space is simply
\bea
\f_{Q^{1,1,1}} = \BC^6~.
\eea
The mesonic moduli space is given by
\bea
\CMm_{Q^{1,1,1}} = \BC^6 // Q_D~.
\eea
Therefore, in order to compute the Hilbert series of the mesonic moduli space, we need to integrate the Hilbert series of the Master space over the two baryonic fugacities $b_1$ and $b_2$:
\bea
\gm (t, x_1, x_2, x_3; Q^{1,1,1}) &=& \oint \limits_{|b_1| =1} {\frac{\ud b_1}{2 \pi i b_1 }}\oint \limits_{|b_2| =1} {\frac{\ud b_2}{2 \pi i b_2 }} \frac{1}{\left(1- t x_1 b_1\right)\left(1- \frac{t b_1}{x_1}\right)\left(1- \frac{t x_2 b_2}{b_1}\right)}\times \nn \\
&&\times \frac{1}{\left(1- \frac{t b_2}{x_2 b_1}\right)\left(1- \frac{t x_3}{b_2}\right)\left(1- \frac{t}{x_3 b_2}\right)}\nn \\
&=& \frac{P(t,x_1, x_2, x_3; Q^{1,1,1})}{\left(1- t^3 x_1 x_2 x_3\right)\left(1-\frac{t^3 x_1 x_2}{x_3}\right)\left(1-\frac{t^3 x_1 x_3}{x_2}\right)\left(1-\frac{t^3 x_2 x_3}{x_1}\right)}\times \nn \\
&& \times \frac{1}{\left(1-\frac{t^3 x_1}{x_2 x_3}\right)\left(1-\frac{t^3 x_2}{x_1 x_3}\right)\left(1-\frac{t^3 x_3}{x_1 x_2}\right)\left(1-\frac{t^3}{x_1 x_2 x_3}\right)}\nn \\
&=& \sum^{\infty}_{n=0} [n; n; n] t^{3n}~,
\label{e:HSmesq111}
\eea
where $P(t,x_1, x_2, x_3; Q^{1,1,1})$ is a polynomial of degree 12 that we do not write here because of its length. The completely unrefined Hilbert series of the mesonic moduli space can be written as:
\bea \label{HSunref1111}
\gm (t, 1, 1, 1; Q^{1,1,1}) = \frac{1 + 4t^3 + t^6}{\left(1-t^3\right)^4} = \sum^{\infty}_{n=0} (n+1)^3 t^{3n}~.
\eea
The plethystic logarithm of (\ref{e:HSmesq111}) can be written as:
\bea
\PL[\gm (t, x_1, x_2, x_3; Q^{1,1,1})] &=& [1;1;1] t^3 - ([1;0;0] + [0;1;0] + [0;0;1])t^6 + O(t^9)~. \qquad
\label{e:plq111}
\eea
Therefore, the 8 generators of the mesonic moduli space can be identified with
\bea
p_i\, r_j\, s_k~,
\eea
where $i,j,k =1,2$.

\begin{figure}[ht]
\begin{center}
  \includegraphics[totalheight=5.0cm]{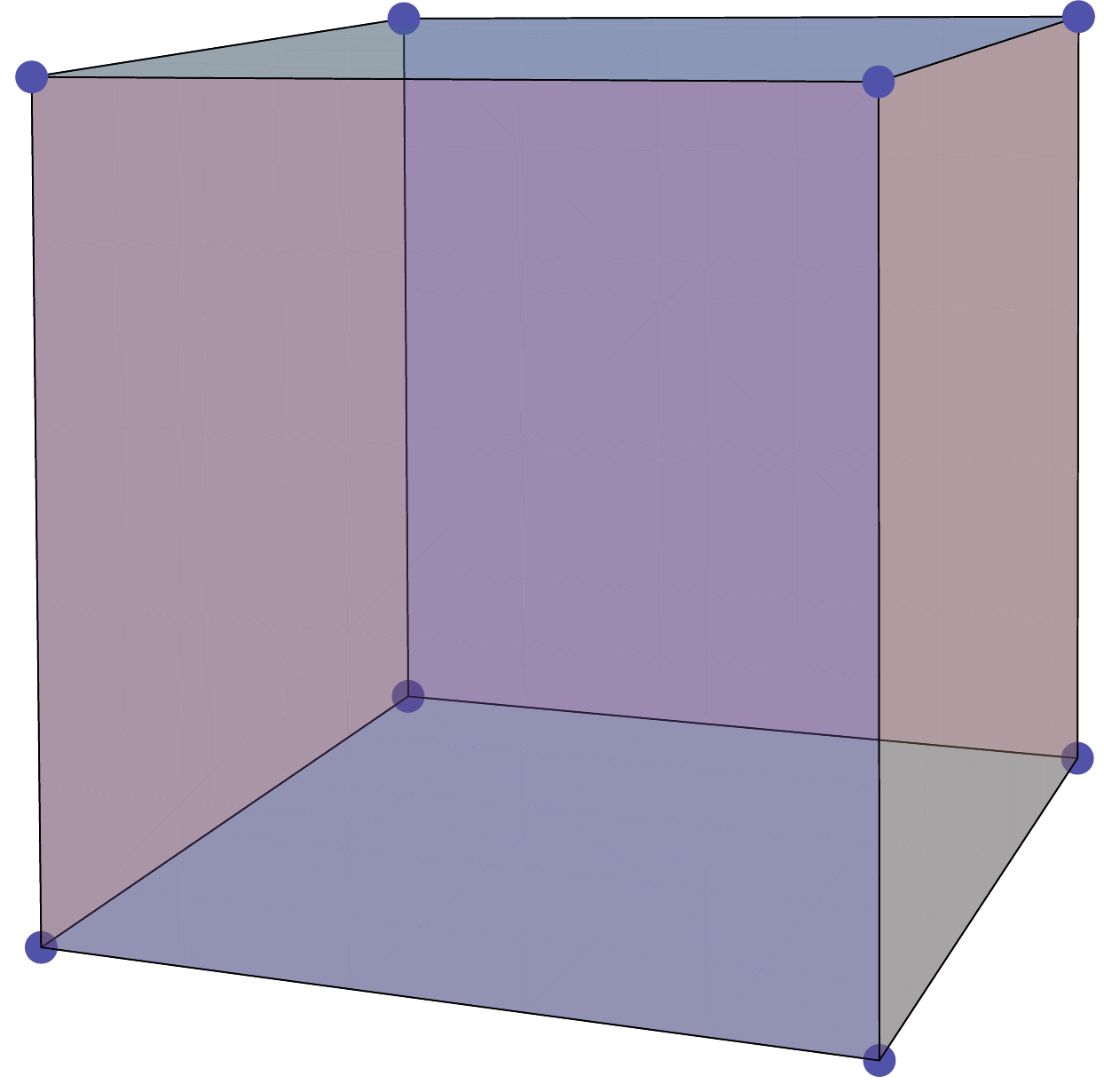}
 \caption{The lattice of generators of the $Q^{1,1,1}$ theory.}
  \label{f:latq111}
\end{center}
\end{figure}

\paragraph{The lattice of generators.} We can represent the generators in a lattice (\fref{f:latq111}) by plotting the powers of each monomial in the character of $[1;1;1]$.
Note that the lattice of generators is the dual of the toric diagram (nodes are dual to faces and edges are dual to edges):
the toric diagram has 6 nodes, 12 edges and 8 faces, whereas the generators form a convex polytope which has 8 nodes, 12 edges and 6 faces.

\paragraph{The Hilbert series for higher $k$.} 
The $Q_D$ matrix \eref{e:qdq111} contains the charges under $U(1)_3-U(1)_4$ and $U(1)_2+U(1)_4$. To construct the orbifold by a $\BZ_k$ action it is enough to choose an action under one of the gauge groups which has a non-zero D term. A possible choice is $U(1)_1$.

The Hilbert series for $Q^{1,1,1}$ with the CS levels $(k, k ,- k, -k)$ is given by
{\small \bea \label{higherkq111}
\gm (t, Q^{1,1,1}/\BZ_k) =  \frac{1}{k} \sum_{j=0}^{k-1}  \frac{1}{(2\pi i)^2} \oint \limits_{|b_1|=1} \frac{db_1}{b_1} \oint \limits_{|b_2|=1} \frac{db_2}{b_2} \frac{1}{ \left(1-t b_1\right) \left(1-t b_1 \omega ^{-j} \right)  \left(1-\frac{t b_2}{b_1}\right) \left(1-\frac{t b_2}{b_1} \omega ^{-j} \right)\left(1-\frac{t }{b_2}\omega ^j \right)^2}~. \nn \\
\eea}
This expression can be written in a closed form as
\bea
\gm (t, Q^{1,1,1}/\BZ_k) = \frac{1+t^3+2 k t^{3 k}-t^{6 k}-2 k t^{3+3 k}-t^{3+6 k}}{\left(1-t^3\right)^3 \left(1-t^{3 k}\right)^2}~.
\eea
Note that, setting $k=1$, we recover \eref{HSunref1111}. This particular $\BZ_k$ action for $k>1$ breaks the mesonic global symmetry down to $SU(2)\times U(1)^2\times U(1)_R$, as can be seen from the charge assignments in \eref{higherkq111}.
By taking the Plethystic Logarithm one finds the generators to be 4 at order $t^3$ carrying $R$ charge 1 and $2k+2$ at order $t^{3k}$ carrying $R$ charge $k$ and transforming as two copies of the $[k]$ representation (of dimension $k+1$) of the $SU(2)$ factor in the mesonic global symmetry.

\subsection{\emph{Phase III-B of $\CC \times \BC$} from giving a VEV to one of the $X^i_{12}$ fields}
Giving a VEV to one of the fields $X^i_{12}$ amounts to removing one of the edges that separate the squares corresponding to gauge groups 1 and 2. As can be seen from Figure \ref{f:qutq111}, the removal of such an edge merges the two squared tiles into an hexagonal tile. Thus, the tiling of the resulting field theory is a two double-bonded one hexagon ($\sD_2 \sH_1$). For convenience, we relabel the gauge groups so that gauge group 1 corresponds to the hexagon and gauge groups 2 and 3 correspond to the double bonds. 
The CS levels associated with the higgsed gauge groups are added, and so the new CS levels are
\bea
k_1 = 2, \quad k_2 = -1, \quad k_3 =  -1~.
\eea
Hence, the resulting theory can be identified with Phase III-B of the $\CC \times \BC$ theory. 
The toric diagram is drawn in \fref{f:torconxc}.

\subsection{\emph{Phase I of $\CC \times \BC$} from giving a VEV to one of $X_{23}$, $X_{31}$, $X_{24}$, $X_{41}$}
By symmetry, turning on a VEV to one of these four fields yields the same result.
For definiteness, let us give a VEV to the field $X_{24}$.  
This amounts to removing the edge corresponding to $X_{24}$ from the tiling in \fref{f:qutq111}.
As a result, the double bond corresponding to the gauge group 4 disappears.
Thus, the resulting tiling is a one double-bonded chessboard tiling ($\sD_1 \sS_2$). 
The CS levels associated with the higgsed gauge groups are added, and so the new CS levels are
\bea
k_1 = 1, \quad k_2 = 0, \quad k_3 = -1~. 
\eea
Hence, the resulting theory can be identified with Phase I of the $\CC \times \BC$ theory.

\section{Higgsing The $Q^{1,1,1}/ \BZ_2$ Theory}
\subsection{Higgsing Phase I of $Q^{1,1,1}/ \BZ_2$}
\subsection*{A summary of Phase I of $Q^{1,1,1}/ \BZ_2$ (the $\mathscr{S}_4$ model)}
\begin{figure}[ht]
\begin{center}
  \hskip -7cm
  \includegraphics[totalheight=6.2cm]{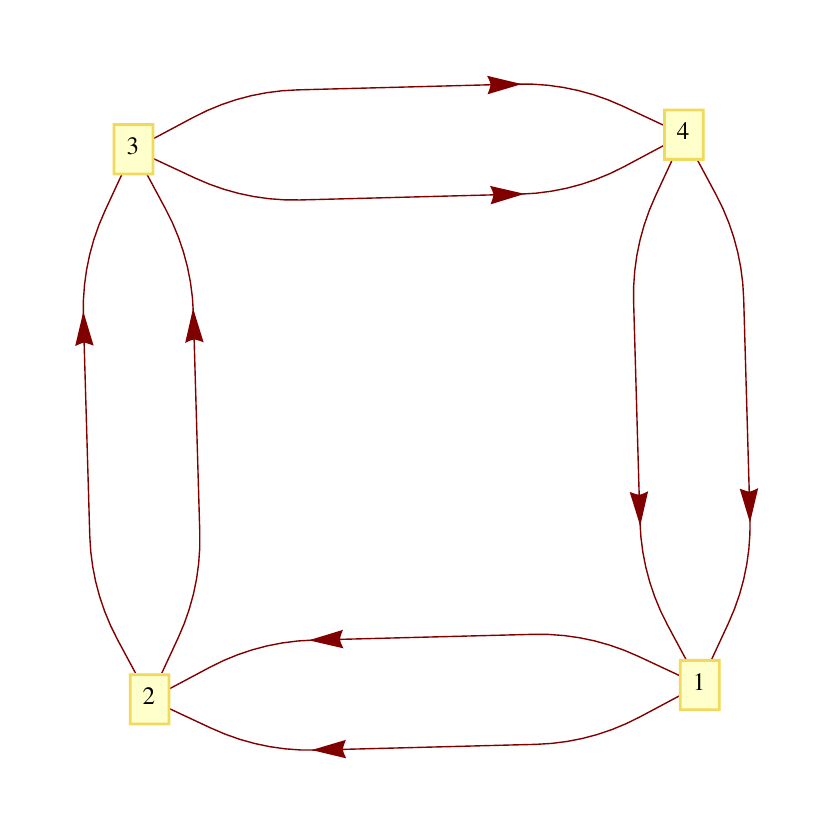}
    \vskip -5.5cm
  \hskip 8.9cm
  \includegraphics[totalheight=5cm]{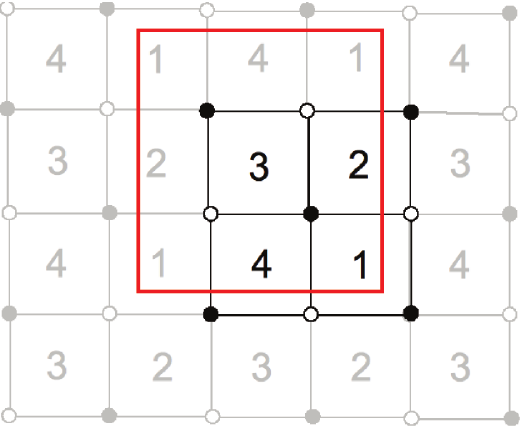}
 \caption{[Phase I of $Q^{1,1,1}/\BZ_2$]  (i) Quiver diagram for the $\mathscr{S}_4$ model. \ (ii) Tiling for the $\mathscr{S}_4$ model.}
  \label{f:phase1f0}
\end{center}
\end{figure}

This model has 4 gauge groups and bi-fundamental fields $X_{12}^i$, $X_{23}^i$, $X_{34}^i$ and $X_{41}^i$ (with $i=1,2$). The quiver diagram and tiling are drawn in Figure \ref{f:phase1f0}.
The superpotential is given by
\bea
W = \epsilon_{ij} \epsilon_{pq} \tr(X_{12}^i X_{23}^p X_{34}^j X_{41}^q)~.
\eea
We choose the CS levels to be $k_1 = -k_2 = -k_3 = k_4=1$.

\begin{figure}
\begin{center}
   \includegraphics[totalheight=8.0cm]{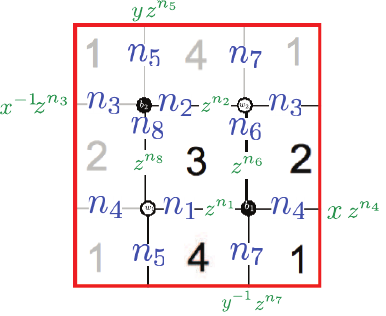}
 \caption{[Phase I of $Q^{1,1,1}/\BZ_2$] The fundamental domain of tiling for the $\mathscr{S}_4$ model: assignments of the integers $n_i$ to the edges are shown in blue and the weights for these edges are shown in green.}
  \label{f:fdph1f0}
\end{center}
\end{figure}

\paragraph{The Kasteleyn matrix.} We assign the integers $n_i$ to the edges according to Figure \ref{f:fdph1f0}.  We find that 
\bea
\text{Gauge group 1~:} \qquad k_1 &=& 1  =   n_3 + n_4 - n_5 - n_7 ~, \nn \\
\text{Gauge group 2~:} \qquad k_2 &=& -1 =    n_6 + n_8 - n_3 - n_4 ~, \nn \\
\text{Gauge group 3~:} \qquad k_3 &=& -1 =   n_1 + n_2 - n_6 - n_8 ~, \nn \\
\text{Gauge group 4~:} \qquad k_4 &=& 1 = - n_1 - n_2 + n_5 + n_7 ~.
\eea  
We choose
\bea
n_3 = - n_1 = 1,\quad n_i=0 \; \text{otherwise}~.
\eea

\noindent We can now construct the Kasteleyn matrix. The fundamental domain contains two black nodes and two white nodes and, therefore, the Kasteleyn matrix is a $2\times 2$ matrix:
\be
K =   \left(
\begin{array}{c|cc}
& w_1 & w_2 \\
\hline
b_1 & X^{1}_{34} z^{n_1} + X^{2}_{12} x z^{n_4} &\ X^{1}_{23} z^{n_6} + X^{2}_{41} y^{-1} z^{n_7}   \\
b_2 & X^{2}_{23} z^{n_8} + X^{1}_{41} y z^{n_5} &\ X^{2}_{34} z^{n_2} + X^{1}_{12} x^{-1} z^{n_3}  
\end{array}
\right) ~.
\ee
The permanent of this matrix is given by
\bea
\perm~K &=& X^{1}_{34} X^{2}_{34}  {z^{n_1 + n_2}} +  X^{1}_{12} X^{2}_{12} z^{n_3 + n_4}  + X^{1}_{34} X^{1}_{12} x^{-1} z^{n_1 + n_3} +  X^{2}_{34} X^{2}_{12} x z^{n_2 + n_4}\nn \\
&& X^{1}_{41} X^{1}_{23} y z^{n_5 + n_6} + X^{2}_{41} X^{2}_{23} y^{-1} z^{n_7 + n_8} +  X^{1}_{41} X^{2}_{41} z^{n_5 + n_7} + X^{1}_{23} X^{2}_{23} z^{n_6 + n_8} \nn \\
&=& X^{1}_{34} X^{2}_{34} z^{-1}+  X^{1}_{12} X^{2}_{12} z + X^{1}_{34} X^{1}_{12} x^{-1}  +  X^{2}_{34} X^{2}_{12} x +  X^{1}_{41} X^{1}_{23} y  + X^{2}_{41} X^{2}_{23}y^{-1} \nn \\
&& +  X^{1}_{41} X^{2}_{41}  + X^{1}_{23} X^{2}_{23} \qquad \text{(for $n_3 = - n_1 = 1,~ n_i=0 \; \text{otherwise}$)} ~. \label{permKph1f0}
\eea

\paragraph{The perfect matchings.} From \eref{permKph1f0}, we write the perfect matchings as collections of fields as follows:
\bea 
&& p_1 = \{ X^1_{34},  X^2_{34} \},\;\; p_2 = \{X^{1}_{12}, X^{2}_{12} \}, \;\; q_1 = \{X^{1}_{34}, X^{1}_{12} \}, \;\; q_2 = \{ X^{2}_{34},  X^2_{12} \}, \nn \\  
&& r_1=  \{ X^{1}_{41}, X^1_{23} \}, \;\; r_2 = \{ X^2_{41}, X^2_{23} \}, \;\; 
s_1 = \{  X^1_{41},  X^2_{41} \}, \; \; s_2 = \{ X^{1}_{23}, X^2_{23}  \} \ . \qquad
\eea
From \eref{permKph1f0}, we see that the perfect matchings $p_i, q_i, r_i$ correspond to the external points in the toric diagram, whereas the perfect matchings $s_i$ correspond to the internal point at the origin.
In turn, we find the parameterisation of fields in terms of perfect matchings:
\bea
&& X^1_{34} = p_1 q_1 , \quad X^2_{34} = p_1 q_2 , \quad X^1_{12} = p_2 q_1, \quad X^2_{12} = p_2 q_2, \nn \\
&& X^1_{41} = r_1 s_1, \quad X^1_{23} = r_1 s_2, \quad X^2_{41} = r_2 s_1, \quad X^2_{23} = r_2 s_2~.
\eea
This is summarised in the perfect matching matrix:
\beq
P=\left(\begin{array} {c|cccccccc}
  \;& p_1 & p_2 & q_1 & q_2 & r_1 & r_2 & s_1 & s_2\\
  \hline 
    X^{1}_{34}& 1&0&1&0&0&0&0&0\\
  X^{2}_{34}& 1&0&0&1&0&0&0&0\\
  X^{1}_{12}& 0&1&1&0&0&0&0&0\\
  X^{2}_{12}& 0&1&0&1&0&0&0&0\\
  X^{1}_{41}& 0&0&0&0&1&0&1&0\\
  X^{1}_{23}& 0&0&0&0&1&0&0&1\\
  X^{2}_{41}& 0&0&0&0&0&1&1&0\\
  X^{2}_{23}& 0&0&0&0&0&1&0&1
  \end{array}
\right).
\eeq
Basis vectors of the nullspace of $P$ are given in the rows of the charge matrix:
\be
Q_F =   \left(
\begin{array}{cccccccc}
1&1&-1&-1&0&0&0&0 \\
0&0&0&0&1&1&-1&-1
\end{array}
\right)~.  \label{qfph1q111z2}
\ee
Hence, we see that the relations between the perfect matchings are given by
\bea
p_1+p_2-p_3-p_4 &=& 0~, \nn \\ 
p_5+p_6-s_1-s_2 &=& 0~. \label{relf0I}
\eea
Since the coherent component $\firr{\mathscr{S}_4}$ of the Master space is generated by the perfect matchings, subject to the relation \eref{relf0I}, it follows that 
\bea
\firr{\mathscr{S}_4} = \BC^8//Q_F~ = \CC\times \CC.  \label{firrS4}
\eea

\paragraph{The toric diagram.} We demonstrate two methods of constructing the toric diagram. 
\begin{itemize}
\item{\bf The charge matrices.}  Since the number of gauge groups is $G=4$, there are $G-2 = 2$ baryonic charges coming from the D-terms.  We collect these charges of the perfect matchings in the $Q_D$ matrix:
\bea
Q_D = \left(
\begin{array}{cccccccc}
 1 & 1 & 0 & 0 & -1 & -1 & 0 & 0 \\
 0 & 0 & 0 & 0 & -1 & -1 & 2 & 0
\end{array}
\right)~.  \label{qdph1q111z2}
\eea
From \eref{qfph1q111z2} and \eref{qdph1q111z2}, the total charge matrix is given by
\bea
Q_t = \left(
\begin{array}{cccccccc}
 1 & 1 & 0 & 0 & -1 & -1 & 0 & 0 \\
 0 & 0 & 0 & 0 & -1 & -1 & 2 & 0 \\
 1&1&-1&-1&0&0&0&0 \\
 0&0&0&0&1&1&-1&-1
\end{array}
\right)~. \label{qtph1q111z2}
\eea
We obtain the matrix $G_t$ and, after removing the first row, the columns give the coordinates of points in the toric diagram:  
\bea
G'_t =  \left(
 \begin{array}{cccccccc}
 1 & -1 & 0 & 0 & 0 & 0 & 0 & 0 \\
 0 & 0 & 1 & -1 & 0 & 0 & 0 & 0 \\
 0 & 0 & 0 & 0 & 1 & -1 & 0 & 0
\end{array}
\right)~.
\eea
The toric diagram is drawn in Figure \ref{f:torq111z2}.  
\begin{figure}[h]
\begin{center}
  \includegraphics[totalheight=6cm]{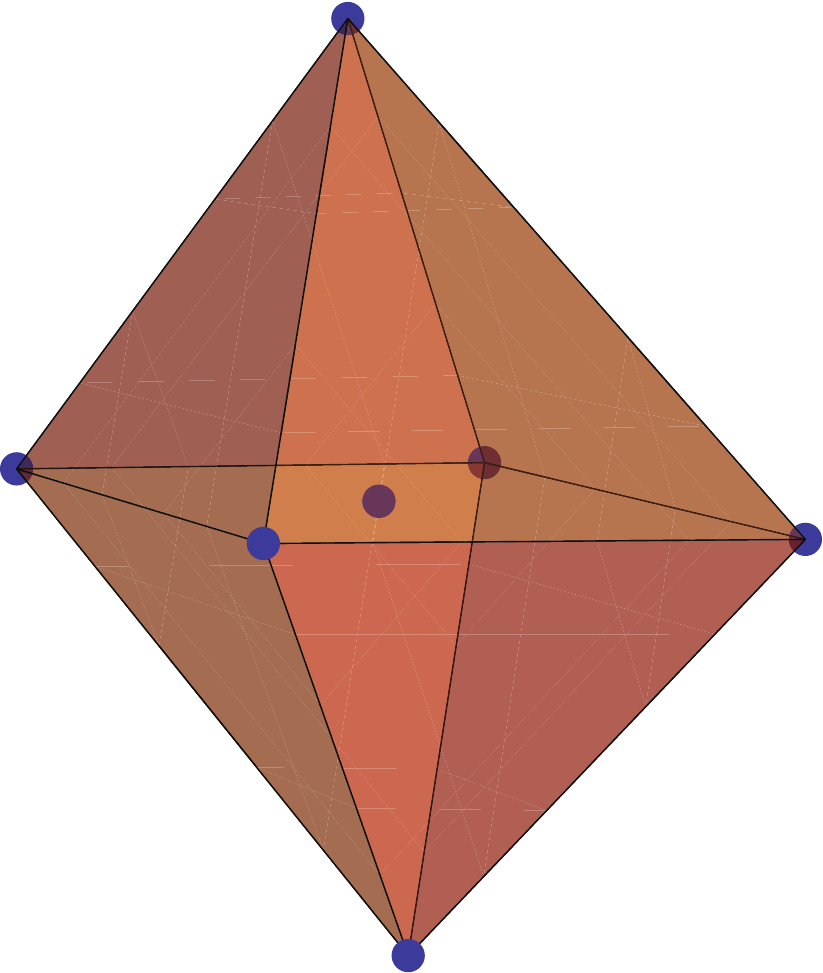}
 \caption{The toric diagram of the $Q^{1,1,1}/\BZ_2$ theory.}
  \label{f:torq111z2}
\end{center}
\end{figure}
Observe that there is an internal point (with multiplicity 2) in the toric diagram for this theory, whereas the toric diagram for the $Q^{1,1,1}$ theory (\fref{f:tdq111}) is simply 6 corners of an octahedron without an internal point.  

\item{\bf The Kasteleyn matrix.}  The powers of $x, y, z$ in each term of \eref{permKph1f0} give the coordinates of each point in the toric diagram.  We collect these points in the columns of the following $G_K$ matrix:
\bea
G_K = \left(
 \begin{array}{cccccccc}
 1 & -1 & 0 & 0 & 0 & 0 & 0 & 0 \\
 0 & 0 & 1 & -1 & 0 & 0 & 0 & 0 \\
 0 & 0 & 0 & 0 & 1 & -1 & 0 & 0
\end{array}
\right) = G'_t~.
\eea
Thus, the toric diagrams constructed from these two methods are identical.
\end{itemize}

\paragraph{The baryonic charges.}  Since the toric diagram has 6 external points, this model has precisely $6-4 = 2$ baryonic symmetries, which we shall denote by $U(1)_{B_1}$ and $U(1)_{B_2}$.  From the above discussion, we see that they arise from the D-terms.  Therefore, the baryonic charges of the perfect matchings are given by the rows of the $Q_D$ matrix. 

\paragraph{The global symmetry.}  Since the $Q_t$ matrix has 3 pairs of repeated columns, it follows that the mesonic symmetry of this model is $SU(2)^3 \times U(1)_R$.  This mesonic symmetry can also be seen from the $G_K$ (or $G'_t$) matrix by noticing that the three rows contain weights of $SU(2)$.
Since $s_1$ and $s_2$ are the perfect matchings corresponding to internal points in the toric diagram, we assign to each of them a zero R-charge.  
The remaining 6 external perfect matchings are completely symmetric and the requirement of R-charge 2 to the superpotential divides 2 equally among them, resulting in R-charge of 1/3 per each.
The global symmetry of the theory is a product of mesonic and baryonic symmetries: $SU(2)^3 \times U(1)_R \times U(1)_{B_1} \times U(1)_{B_2}$.
In Table \ref{chargeph1f0}, we present a consistent way of assigning charges to the perfect matchings under these global symmetries.
\begin{table}[h]
 \begin{center}  
  \begin{tabular}{|c||c|c|c|c|c|c|c|}
  \hline
  \;& $SU(2)_{1}$&$SU(2)_{2}$&$SU(2)_{3}$&$U(1)_R$&$U(1)_{B_1}$&$U(1)_{B_2}$& fugacity\\
  \hline \hline  
  $p_1$& $1$&$0$&$0$&$1/3$&$1$&$0$& $t b_1 x_1$ \\
  $p_2$& $-1$&$0$&$0$&$1/3$&$1$&$0$ & $t b_1/x_1$ \\
  $q_1$& $0$&$1$&$0$&$1/3$&$0$&$0$  & $t x_2$\\
  $q_2$& $0$&$-1$&$0$&$1/3$&$0$&$0$& $ t / x_2$\\
  $r_1$& $0$&$0$&$1$&$1/3$&$-1$&$-1$& $ t x_3/(b_1 b_2)$\\
  $r_2$& $0$&$0$&$-1$&$1/3$&$-1$&$-1$& $ t / (x_3  b_1 b_2)$\\
  $s_1$& $0$&$0$&$0$&$0$&$0$&$2$&  $b_2^2$ \\
  $s_2$& $0$&$0$&$0$&$0$&$0$&$0$ & $1$ \\
  $\Blue s_3$& $0$&$0$&$0$&$0$&$0$&$0$ & $1$ \\ 
  \hline
  \end{tabular}
  \end{center} \Black
  \caption{Charges under the global symmetry of the $Q^{1,1,1}/\BZ_2$ theory. Here $t$ is the fugacity of R-charge, $x_1,x_2,x_3$ are weights of $SU(2)_{1}, SU(2)_{2}, SU(2)_3$, and $b_1, b_2$ are baryonic fugacities of $U(1)_{B_1}, U(1)_{B_2}$. Note that the perfect matching $s_3$ (represented in blue) does not exist in Phase I but exists in Phase II.}
\label{chargeph1f0}
\end{table}

Below, there is a study of the Higgs mechanism of this theory.

\subsubsection{\emph{Phase II of $(\BC^2/\BZ_2) \times \BC^2$} from giving a VEV to one of $X^{i}_{12}$, $X^i_{34}$} \label{sec:ph2c2z2xc2}
\begin{figure}[ht]
\begin{center}
 \vskip 1cm
  \hskip -7cm
  \includegraphics[totalheight=5.0cm]{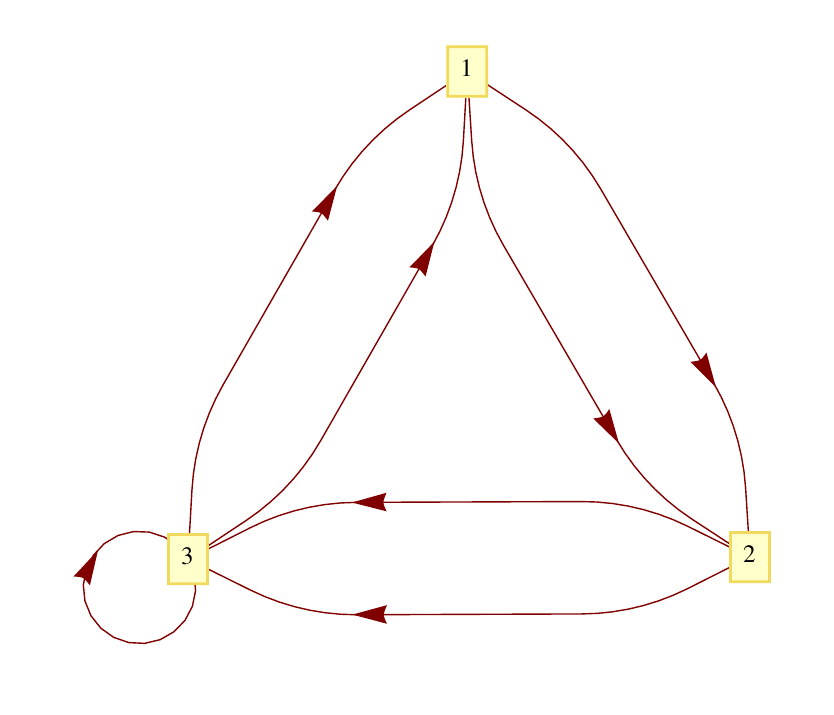}
   \vskip -5cm
  \hskip 8cm
  \includegraphics[totalheight=5.0cm]{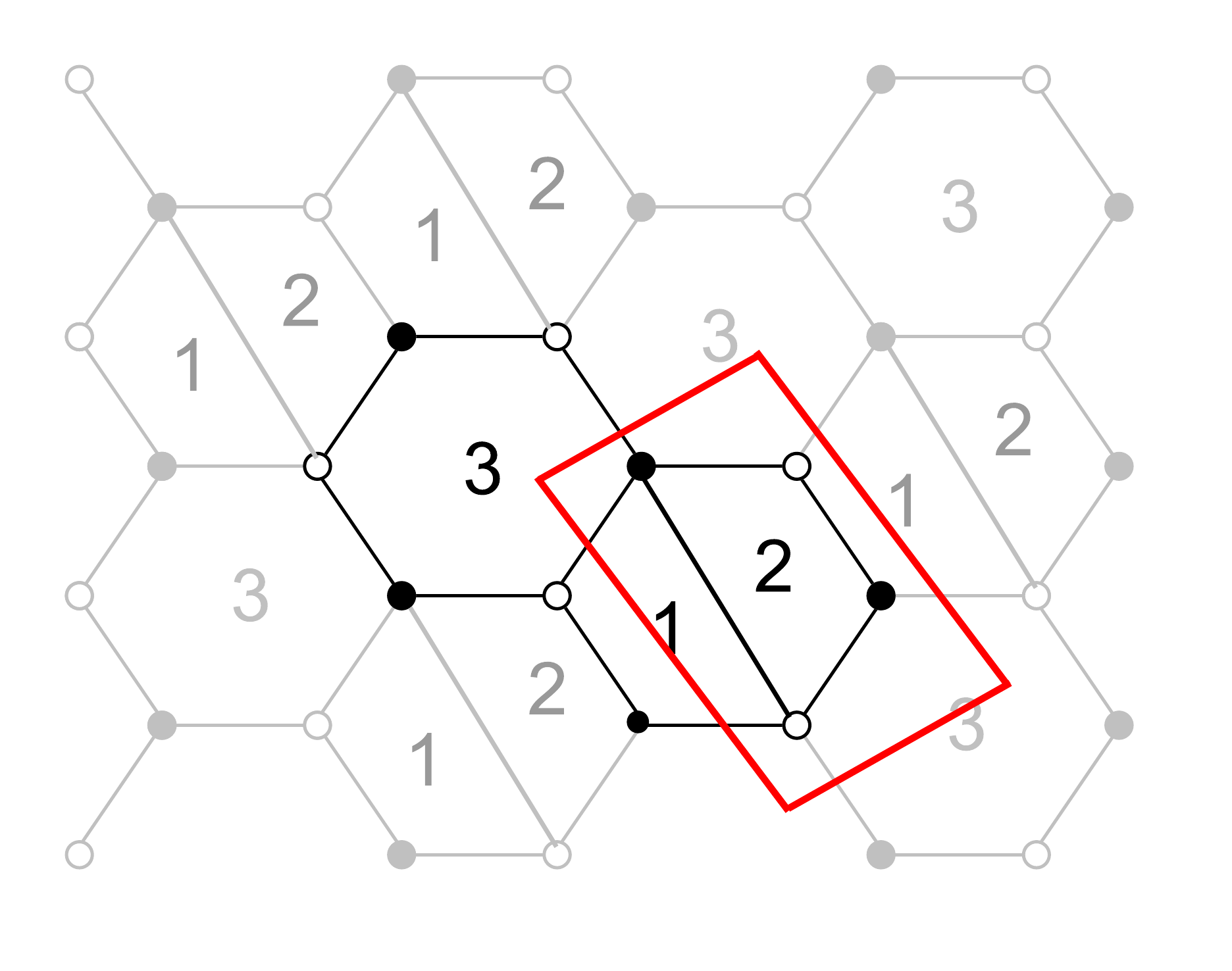}
 \caption{(i) Quiver diagram of Phase II of $\left(\BC^2/\BZ_2\right) \times \BC^2$.\ (ii) Tiling of Phase II of $\left(\BC^2/\BZ_2\right) \times \BC^2$.}
  \label{f:qtresq111z2ph3}
\end{center}
\end{figure}

By symmetry, turning on a VEV to one of the $X^{i}_{12}$, $X^i_{34}$ fields yields the same result. 
For definiteness, let us give a VEV to $X^{1}_{34}$.
This amounts to removing one of the edges that separate the squares corresponding to gauge groups 3 and 4.
As a result, these gauge groups are combined into one gauge group, identified as 3.
The quiver diagram and tiling of this model are presented in Figure \ref{f:qtresq111z2ph3}.
The superpotential is given by 
\bea
W = \epsilon_{ij} \tr(  X^1_{12} X^{i}_{23} \phi_3 X^j_{31}) - \epsilon_{kl} \tr(X^2_{12} X^k_{23} X^l_{31} )~.
\label{e:spresq111z2}
\eea
The CS levels associated with the higgsed gauge groups are added, and so the new CS levels are
\bea
k_1 = 1, \quad k_2 = -1, \quad k_3 = 0~. 
\eea
We note that this model does not give rise to a consistent tiling in 3+1 dimensions and in fact is the simplest inconsistent model in the sense of \cite{Hanany:2005ss}. It looks similar to the SPP theory but differs from it by being chiral, as opposed to the SPP quiver which is non-chiral.

\begin{figure}[ht]
\begin{center}
   \includegraphics[totalheight=7cm]{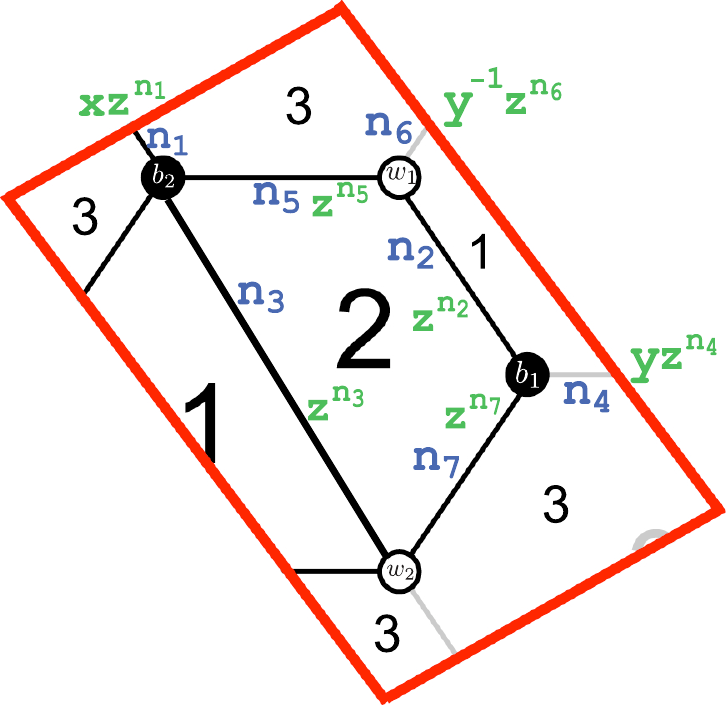}
 \caption{The fundamental domain of the tiling for Phase II of $\left(\BC^2/\BZ_2\right) \times \BC^2$: assignments of the integers $n_i$ to the edges are shown in blue and the weights for these edges are shown in green.}
  \label{f:fdresq111z2ph3}
\end{center}
\end{figure}

\paragraph{The Kasteleyn matrix.} We assign the integers $n_i$ to the edges according to Figure \ref{f:fdresq111z2ph3}.  We find that 
\bea
\text{Gauge group 1~:} \qquad k_1 &=& 1  =   n_2 + n_3 - n_4 - n_6~,  \nn \\
\text{Gauge group 2~:} \qquad k_2 &=& -1 = - n_2 - n_3 + n_5 + n_7~,  \nn \\
\text{Gauge group 3~:} \qquad k_3 &=& 0  =   n_4 - n_5 + n_6 - n_7~. 
\label{e:karesq111z2ph3}
\eea  
We choose:
\bea
n_3= 1,\qquad n_i=0 \;\;\text{otherwise}~.
\eea
We can now determine the Kasteleyn matrix. Since the fundamental domain contains 2 pairs of black and white nodes, the Kasteleyn matrix is $2 \times 2$: 
\bea
K =   \left(
\begin{array}{c|cc}
& b_1 & b_2\\
\hline
w_1 & X^2_{12} z^{n_2} & X^1_{23} z^{n_5} + X^2_{31} y^{-1} z^{n_6} \\
w_2 & X^2_{23} z^{n_7} + X^1_{31} y z^{n_4} & \phi_3 x z^{n_1} + X^1_{12} z^{n_3}
\end{array}
\right) ~.
\label{e:kastresq111z2ph3}
\eea
The permanent of the Kasteleyn matrix is given by
\bea
\mathrm{perm}~K &=&  \phi_3 X^2_{12} x z^{n_1+n_2} + X^1_{12}X^2_{12} z^{n_2+n_3} + X^{1}_{23} X^2_{23} z^{n_5+n_7}+ X^1_{31}X^2_{31}z^{n_4+ n_6}\nn \\
&& + X^1_{31}X^1_{23} y z^{n_4+n_5} + X^2_{23} X^2_{31} y^{-1} z^{n_6+n_7}\nn\\
&=& \phi_3 X^2_{12} x + X^1_{12}X^2_{12} z + X^{1}_{23} X^2_{23}+ X^1_{31}X^2_{31}+X^1_{31}X^1_{23} y +  X^2_{23} X^2_{31} y^{-1}\nn \\
&& \text{(for $n_3 = 1,~ n_i=0 \; \text{otherwise}$)}~.
\label{e:permresq111z2ph3}
\eea

\paragraph{The perfect matchings.} From (\ref{e:permresq111z2ph3}), we can take the perfect matchings to be:
\bea 
&& p_1 = \left\{\phi_3, X^2_{12}\right\}, \;\; p_2 = \left\{X^1_{12}, X^2_{12}\right\}, \;\;\; r_1 = \left\{X^1_{23}, X^1_{31}\right\}, \nn \\
&& r_2 = \left\{X^2_{23}, X^2_{31}\right\}, \; \; s_1 = \left\{X^1_{23}, X^2_{23}\right\}, \; \; s_2 = \left\{X^1_{31}, X^2_{31}\right\}~.
\eea
In turn, we find the parameterisation of fields in terms of perfect matchings:
\bea
&& \phi_3 = p_1,  \quad X^1_{12} = p_2, \quad X^2_{12} = p_1 p_2, \quad X^1_{23} = r_1 s_1 ,\nn \\
&& X^2_{23} = r_2 s_1, \quad X^1_{31} = r_1 s_2, \quad  X^2_{31} = r_2 s_2~. \label{quivfieldhigph1q111z2}
\eea
This is summarised in the perfect matching matrix:
\beq
P=\left(\begin{array} {c|cccccc}
  \;& p_1 & p_2  & r_1 & r_2 & s_1 & s_2 \\
  \hline 
  \phi_3    & 1&0&0&0&0&0\\
  X^{2}_{12}& 1&1&0&0&0&0\\
  X^{1}_{12}& 0&1&0&0&0&0\\
  X^{1}_{23}& 0&0&1&0&1&0\\
  X^{2}_{23}& 0&0&0&1&1&0\\
  X^{1}_{31}& 0&0&1&0&0&1\\
  X^{2}_{31}& 0&0&0&1&0&1\\
  \end{array}
\right).
  \label{e:presq111z2ph3}
\eeq
The basis vector of the nullspace of $P$ is given in the row of the charge matrix:
\be
Q_F =   \left(
\begin{array}{cccccc}
0, & 0, &1, & 1, &-1, &-1
\end{array}
\right)~.  \label{e:qfresq111z2ph3}
\ee
Hence, we see that the relation between the perfect matchings is given by
\bea
r_1 + r_2 - s_1 - s_2  = 0~.
\label{e:relpmresq111z2ph3}
\eea

\paragraph{The toric diagram.} We construct the toric diagram of this model using two methods.
\begin{itemize} 
\item {\bf The charge matrices.}
Since the number of gauge groups of this model is $G = 3$, there is $G-2 =1$ baryonic symmetry coming from the D-terms. The charges of the perfect matchings under this symmetry can be collected in the $Q_D$ matrix:
\be
Q_D =   \left(
\begin{array}{cccccc}
0, & 0, & 0,& 0, & -1,& 1
\end{array}
\right)~. \label{e:qdresq111z2ph3}
\ee
Note that since the CS coefficient $k_3=0$, the $Q_D$ matrix \eref{e:qdresq111z2ph3} has been chosen such that the baryonic charge of each quiver field in \eref{quivfieldhigph1q111z2} coincides with the quiver charge under gauge group 3.
We can combine (\ref{e:qfresq111z2ph3}) and (\ref{e:qdresq111z2ph3}) in a single matrix $Q_t$:
\be
Q_t = { \Blue Q_D \choose \Green Q_F \Black } =   \left( 
\begin{array}{cccccc} \Blue
0 & 0 & 0& 0 & -1& 1 \\ \Green
0 & 0& 1& 1& -1& -1  \Black
\end{array}
\right)~.
\label{e:qtresq111z2ph3}
\ee
The $G_t$ matrix is the kernel of (\ref{e:qtresq111z2ph3}) and, after removing the first row, we get a matrix whose columns represent the coordinates of the toric diagram:
\bea
G'_t =  \left(
\begin{array}{cccccc}
  1 & -1 &  0 & 0 &  0 &  0 \\
  0 & 0 &  1 & -1  &  0 & 0\\
  0 & 1 &  0 & 0 &  0 &  0
\end{array}
\right) ~. \label{e:toricdiaresq111z2ph3}
\eea
The toric diagram is presented in Figure \ref{f:tdtoricresq111z2ph3}.  
We see that this is the toric diagram of a $\BZ_2$ orbifold of $\BC^4$. The discrete symmetry $\BZ_2$ acts only on the perfect matchings $r_1, r_2$ (but not on $p_1, p_2$) and, as a result of this action, we gain a point on one of the edges (with multiplicity 2) corresponding to the perfect matchings $s_1, s_2$.  Thus, the mesonic moduli space of this model is 
\bea
\CMm = \left( \BC^2/\BZ_2 \right) \times \BC^2~, \label{meshigph1q111z2}
\eea
where the first $\BC^2$ is parametrised by the perfect matchings $r_1, r_2$, and the second $\BC^2$ is parametrised by the perfect matchings $p_1, p_2$.  We refer to this model as {\bf Phase II of $(\BC^2/\BZ_2) \times \BC^2$}.
\begin{figure}[ht]
\begin{center}
  \includegraphics[totalheight=3.0cm]{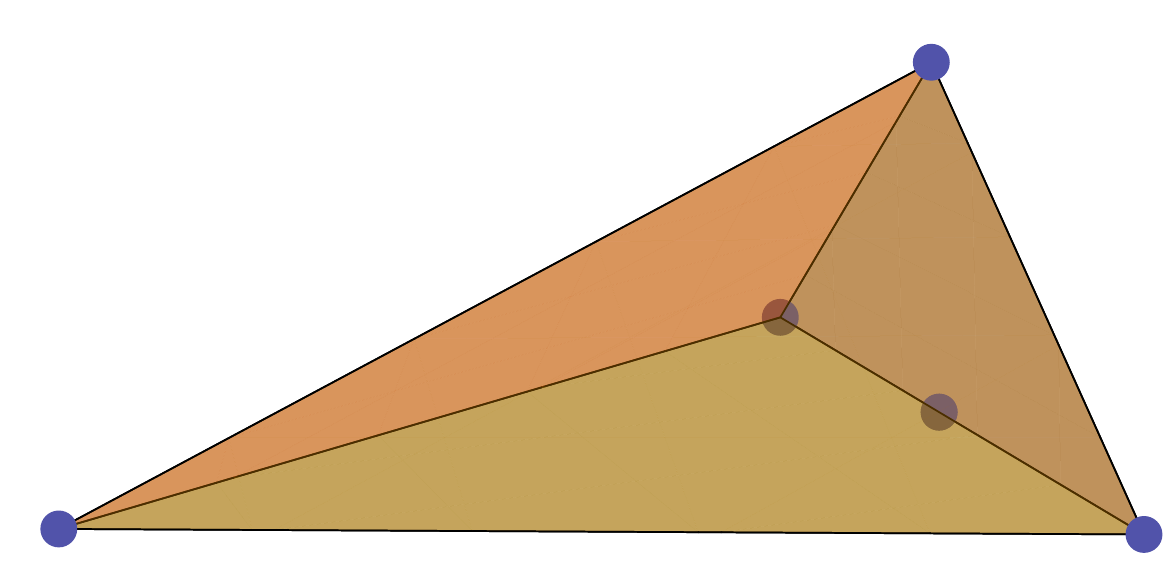}
 \caption{The toric diagram of the $\left(\BC^2/\BZ_2\right) \times \BC^2$ theory.}
  \label{f:tdtoricresq111z2ph3}
\end{center}
\end{figure}

\item {\bf The Kasteleyn matrix.}
The powers of $x, y, z$ in each term of \eref{e:kastresq111z2ph3} give the coordinates of each point in the toric diagram, which can be collected in the columns of the following $G_K$ matrix: 
\bea
G_K = \left(
\begin{array}{cccccc}
  1 & 0 &  0 & 0 &  0 &  0 \\
  0 & 0 &  1 & -1 &  0 & 0 \\
  0 & 1 &  0 & 0 &  0 &  0
\end{array}
\right)~.
\eea
Note that $G_K$ can be obtained by performing a row operation on $G'_t$; in particular, the first row of $G_K$ is derived from adding the first and the third row of $G'_t$.
\end{itemize}

\paragraph{The baryonic charges.} 
From Figure \ref{f:tdtoricresq111z2ph3}, we can appreciate that the toric diagram of this model has 4 points on the vertices and a point (with multiplicity 2) on one of the edges.  
Should the point on the edge be counted as an internal point or an external point of the toric diagram?
We know from the $Q_D$ matrix that there is 1 baryonic charge coming from the D-terms.
Therefore, if such a point were regarded as an internal point, we would have only 4 external points and, hence, $4-4=0$ baryonic charges, which is a contradiction.
Thus, the point on the edge must be regarded as an external point so that we have \emph{precisely} $5-4 = 1$ baryonic symmetry, which we shall denote as $U(1)_B$. 
The charges of the perfect matchings under this $U(1)_B$ are given by the row of the $Q_D$ matrix.

\paragraph{The global symmetry.} Since there are two pairs of repeated columns in the $Q_t$ matrix, the mesonic symmetry of this model is $SU(2)^2\times U(1)_q \times U(1)_R$. 
From the $Q_t$ matrix, we see that the perfect matchings $p_1$ and $p_2$ transform as a doublet under the first $SU(2)$, and $r_1$ and $r_2$ transform as a doublet under the second $SU(2)$.
The global symmetry of this model is the product of the mesonic and baryonic symmetries: $SU(2)^2\times U(1)_q \times U(1)_R \times U(1)_B$.
Recall from \eref{meshigph1q111z2} that the mesonic moduli space is a $\BZ_2$ orbifold of $\BC^4$.  Therefore, each of the external perfect matchings $p_1, p_2, r_1$, and $r_2$ has the same R-charge 1/2, as in the $\BC^4$ theory, and  each of the perfect matchings on the edge, $s_1, s_2$, has zero R-charge.  We assign the charges to the perfect matchings under $U(1)_q$ such that the superpotential is not charged under it and the charge vectors are linearly independent. The global charges are summarised in Table \ref{t:chresq111z2ph3}:
\begin{table}[h]
 \begin{center} 
  \begin{tabular}{|c||c|c|c|c|c|c|}
  \hline
  \;& $SU(2)_1$&$SU(2)_2$&$U(1)_R$&$U(1)_q$&$U(1)_{B}$&fugacity\\
  \hline  \hline 
  $p_1$&  1&  0 & 1/2 &  1 &   0 & $t x_1 q $\\
  \hline
  $p_2$& $-1$&  0 & 1/2 &  1 &   0 & $t q/ (x_1) $\\
  \hline
  $r_1$&  0&  1 & 1/2 & $-1$ &   0 & $t x_2 / q$ \\
  \hline
  $r_2$&  0& $-1$ & 1/2 & $-1$ &   0 & $t / (q x_2)$\\
  \hline  
  $s_1$&  0&  0 &  0  &  0 &   $-1$ & $ 1/b $\\
  \hline
  $s_2$&  0&  0 &  0  &  0 &  1 & $b$\\
  \hline
  \end{tabular}
  \end{center}
\caption{Charges of the perfect matchings under the global symmetry of the $\left( \BC^2/ \BZ_2 \right) \times \BC^2$ theory. Here $t$ is the fugacity of the R-charge (in the unit of $1/2$), $x_1,x_2$ are the fugacities of the $SU(2)$ charge, $q$ is the fugacity of the $U(1)$ symmetry and $b$ is the fugacity of the $U(1)_{B}$ symmetry.}
\label{t:chresq111z2ph3}
\end{table}

\begin{figure}[ht]
\begin{center}
  \includegraphics[totalheight=5.0cm]{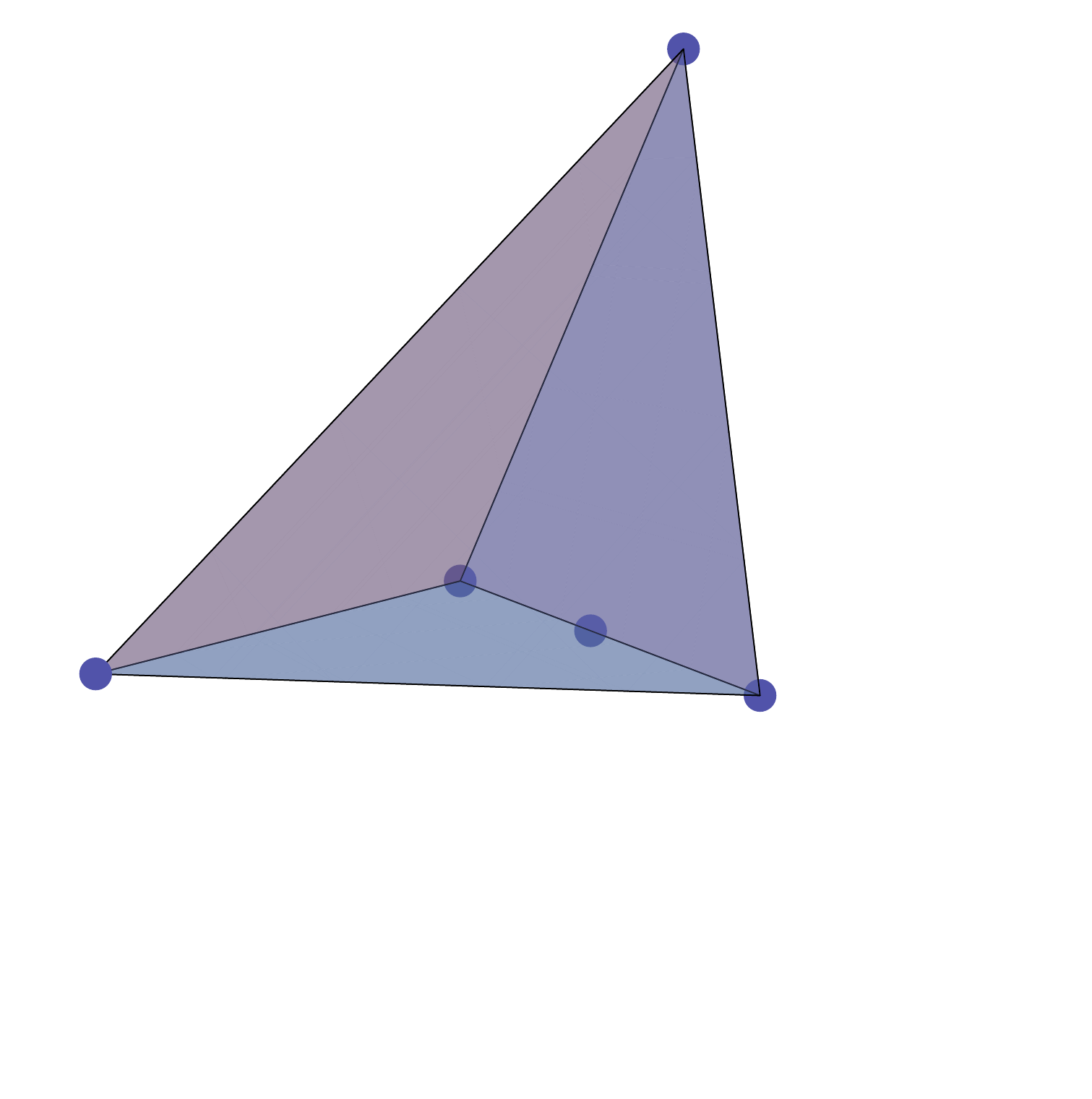}
 \caption{The lattice of generators of the $\left( \BC^2/\BZ_2 \right) \times \BC^2$ theory.}
  \label{f:latc2z2xc2}
\end{center}
\end{figure}

\paragraph{The Hilbert series.} The coherent component of the Master space is generated by the perfect matchings, which are subject to the relation (\ref{e:relpmresq111z2ph3}):
\bea
\firr{} = \BC^6//Q_F = \CC\times \BC^2~.  
\label{firrresq111z2ph3}
\eea
Thus, the Hilbert series of the coherent component of the Master space can be computed by integrating the Hilbert series of the space of perfect matchings over the fugacity $z$:
\bea
g^{\firr{}} (t, x_1, x_2, q, b) &=& \oint \limits_{|z| =1} {\frac{\ud z}{2 \pi i z }} \frac{1}{\left(1- t x_1 q\right)\left(1-\frac{t q}{x_1}\right)\left(1-\frac{t x_2 z}{q }\right)\left(1-\frac{t z}{q x_2 }\right)\left(1-\frac{ b}{z}\right)\left(1- \frac{1}{bz}\right)}\nn \\
&=& \frac{1-\frac{t^2}{q^2}}{ \left(1- t q x_1\right)\left(1-\frac{t q}{x_1}\right)\left(1-\frac{t b}{q x_2}\right)\left(1-\frac{t x_2 b}{q}\right)\left(1-\frac{t}{q x_2 b}\right)\left(1-\frac{t x_2}{q b}\right)}~.
\label{e:HSmasterresq111z2ph3}
\eea
The unrefined version of the result of the integration can be written as
\bea
g^{\firr{}} (t, 1, 1, 1, 1) &=& \frac{1+t}{\left(1-t\right)^5}~.
\eea
Integrating (\ref{e:HSmasterresq111z2ph3}) over the baryonic fugacity $b$ gives the mesonic Hilbert series
\bea
\gm (t,x_1,x_2,q) &=& \oint_{|b|=1} \frac{\ud b}{2\pi i b}\;\; g^{\firr{}} (t, x_1, x_2, q, b) \nn \\
&=&  \frac{1+\frac{t^2}{q^2}}{\left(1-\frac{t q}{x_1}\right)\left(1-t q x_1\right)\left(1-\frac{t^2}{q^2 x_2^2}\right)\left(1-\frac{t^2 x_2^2}{q^2}\right)}~.
\label{e:HSmesresq111z2ph3}
\eea
The totally unrefined Hilbert series of the mesonic moduli space can be written as
\bea
\gm (t,1,1,1) = \frac{1+t^2}{(1-t)^2 (1-t^2)^2} = \frac{1+t^2}{(1-t)^4 (1+t)^2}~. \label{gmesph2c2z2xc2}
\eea
This agrees with \eref{gmesph1c2z2xc2}.
The plethystic logarithm of (\ref{e:HSmesresq111z2ph3}) is given by
\bea
\PL[\gm (t,x_1,x_2,q )] &=& [1;0] t q  + [0;2] \frac{t^2}{q^2}- \frac{t^4}{q^4}~. \label{plc2z2xc2}
\eea
Therefore, the 5 generators of the mesonic moduli space can be written in terms of perfect matchings as
\bea
p_i, \quad r_i r_j s_1 s_2~,  \label{e:genpmresq111z2ph3}
\eea
where $i,j = 1,2$.
These generators can be represented in a lattice (\fref{f:latc2z2xc2}) by plotting the powers of each monomial in the characters of $SU(2) \times SU(2)$ and $U(1)_q$ in \eref{plc2z2xc2}.  Note that the lattice of generators is dual to the toric diagram (nodes are dual to faces and edges are dual to edges). For 
the $\left( \BC^2/\BZ_2 \right) \times \BC^2$ theory, the lattice of generators is self-dual.

\subsubsection{\emph{The $\BF_0 \times \BC$ theory} from giving a VEV to one of $X^{i}_{23}$, $X^i_{41}$}
By symmetry, turning on a VEV to either $X^i_{23}$ or $X^i_{41}$ leads to the same theory. 
For definiteness, let us give a VEV to $X^1_{23}$. 
This means that we remove the edge that separates the gauge groups 2 and 3, and merge them into one gauge group, which is identified as 3.
Let us relabel the gauge groups such that 4 becomes 2.
Then, the quiver diagram and tiling of the resulting theory are given in Figure \ref{f:qtresq111z2ph3}.
The superpotential coincides with (\ref{e:spresq111z2}).
The CS levels associated with the higgsed gauge groups are added, and so
\bea
k_1 = 1, \quad k_2 = 1, \quad k_3 = -2~. 
\eea

\paragraph{The Kasteleyn matrix.} The assignment of the integers $n_i$ is shown in Figure \ref{f:fdresq111z2ph3}. 
In order to have the desired CS, we choose the following values of $n_i$ for \eref{e:karesq111z2ph3}:
\bea
n_4=-1,\qquad n_5=1\qquad n_i=0 \;\;\text{otherwise}~.
\eea
The permanent of the Kasteleyn matrix is then
\bea
\mathrm{perm}~K &=&  \phi_3 X^2_{12} x z^{n_1+n_2} + X^1_{12}X^2_{12} z^{n_2+n_3} + X^{1}_{23} X^2_{23} z^{n_5+n_7}+ X^1_{31}X^2_{31}z^{n_4+ n_6}\nn \\
&& + X^1_{31}X^1_{23} y z^{n_4+n_5} + X^2_{23} X^2_{31} y^{-1} z^{n_6+n_7}\nn\\
&=& \phi_3 X^2_{12} x + X^1_{12}X^2_{12} + X^{1}_{23} X^2_{23} z+ X^1_{31}X^2_{31} z^{-1}+X^1_{31}X^1_{23} y +  X^2_{23} X^2_{31} y^{-1}\nn \\
&& \text{(for $n_4 = -n_5 = -1,~ n_i=0 \; \text{otherwise}$)}~.
\label{e:permresq111z2cxcz2}
\eea

\paragraph{The perfect matchings.}
From \eref{e:permresq111z2cxcz2}, the perfect matching matrix $P$ is 
\beq
P=\left(\begin{array} {c|cccccc}
  \;  & p_1 & p_2 & q_1 & q_2 & r & s \\
  \hline 
  \phi_3    &0&0&0&0& 1&0\\
  X^{2}_{12}&0&0&0&0 & 1&1\\
  X^{1}_{12}&0&0&0&0& 0&1\\
  X^{2}_{23}&1&0&1&0& 0&0\\
  X^{2}_{23}&0&1&1&0& 0&0\\
  X^{1}_{31}&1&0&0&1& 0&0\\
  X^{2}_{31}&0&1&0&1& 0&0\\
  \end{array}
\right).
  \label{e:pf0xc}
\eeq
Note that  from \eref{e:permresq111z2cxcz2}, the perfect matching $s$ corresponds to an internal point in the toric diagram, whereas the others correspond to external points.
The $Q_F$ matrix is the kernel of the $P$ matrix: 
\be
Q_F =   \left(
\begin{array}{cccccc}
1,& 1, &-1, &-1,& 0, & 0 
\end{array}
\right)~. 
\ee

\paragraph{The toric diagram.} We construct the toric diagram of this model using two methods.
\begin{itemize} 
\item {\bf The charge matrices.}
Since the number of gauge groups of this model is $G = 3$, there is $G-2 =1$ baryonic symmetry coming from the D-terms. The charges of the perfect matchings under this symmetry can be collected in the $Q_D$ matrix:
\be
Q_D =   \left(
\begin{array}{cccccc}
 1,& 1,& 0,& 0, & 0, & -2
\end{array}
\right)~. \label{e:qdresq111z2cxcz2}
\ee
We combine the $Q_F$ and $Q_D$ matrices in the total charge matrix, $Q_t$:
\be
Q_t = { \Blue Q_D \choose \Green Q_F \Black } =   \left( 
\begin{array}{cccccc} \Blue
 1& 1& 0& 0 & 0 & -2 \\ \Green
1& 1 &-1 &-1& 0 & 0   \Black
\end{array}
\right)~.
\label{e:qtresq111z2cxcz2}
\ee
We obtain the $G_t$ matrix from the kernel of (\ref{e:qtresq111z2cxcz2}) and, after removing the first row, we get a matrix with columns representing the coordinates of the toric diagram:
\bea
G'_t =  \left(
\begin{array}{cccccc}
  1 & -1 &  0 &  0 &  0 &  0 \\
  0 & 0 &  1 & -1 &  0 &  0\\
  0 & 0 &  0 &  0 &  1 & 0
\end{array}
\right) ~. \label{e:toricdiaresq111z2cxcz2}
\eea
The toric diagram is presented in Figure \ref{f:tdtoricresq111z2cxcz2}.  
We note that the perfect matching $s$ corresponds to the internal point on the base, and the others correspond to external points at the corners.
Therefore, the mesonic moduli space of this model is
\bea
\CMm = \BF_0 \times \BC ~,  
\label{meshigph1q111z2cxcz2}
\eea
where $\BF_0$, which is a $\BZ_2$ orbifold of the conifold\footnote{Note that there is another $\BZ_2$ orbifold of the conifold which is known as $L^{222}$.  The toric diagram is drawn in Figure 4a of \cite{Franco:2005sm}.  The Hilbert series of $L^{222}$ is given by
$\frac{1}{2} \left( \frac{1-t^2}{(1-t)^4} + \frac{1-t^2}{(1-t)^2(1+t)^2} \right) = \frac{1-t^4}{(1-t)^2(1-t^2)^2}$.  },  
is parametrised by $p_1, p_2, q_1, q_2, s$ (base of the pyramid in \fref{f:tdtoricresq111z2cxcz2}), and $\BC$ is parametrised by $r$ (tip of the pyramid in \fref{f:tdtoricresq111z2cxcz2}).

\begin{figure}[ht]
\begin{center}
  \includegraphics[totalheight=5.0cm]{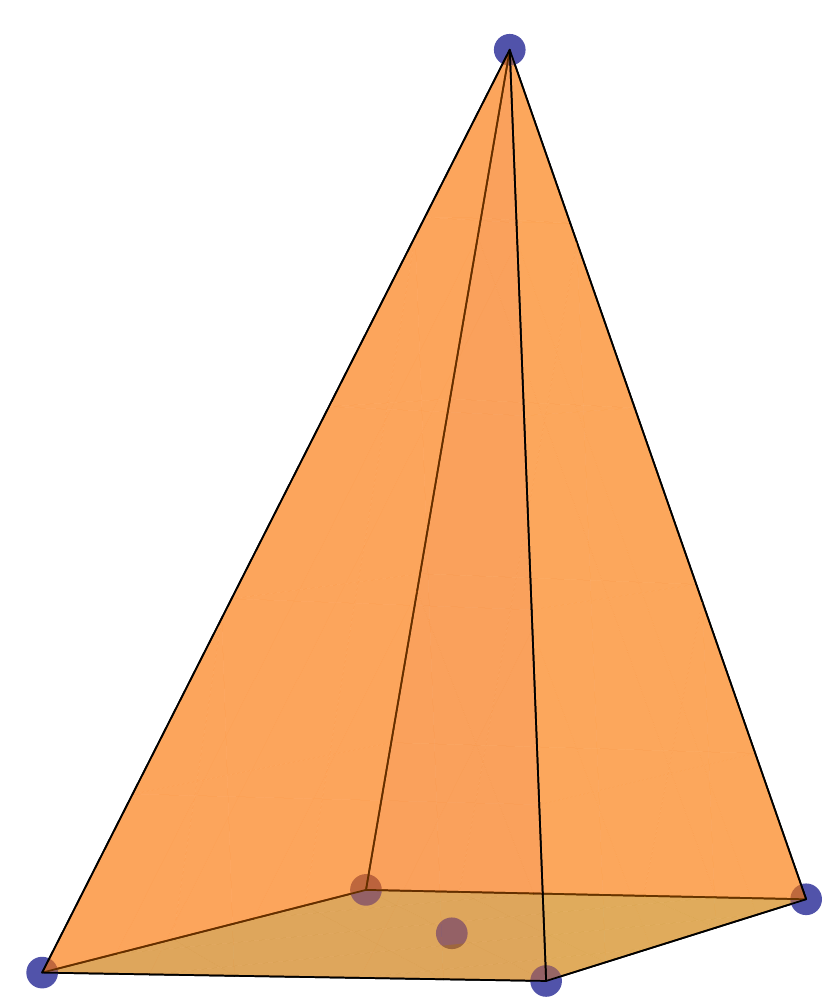}
 \caption{The toric diagram of the $\BF_0 \times \BC$ theory.}
  \label{f:tdtoricresq111z2cxcz2}
\end{center}
\end{figure}

\item {\bf The Kasteleyn matrix.}
The powers of $x, y, z$ in each term of \eref{e:permresq111z2cxcz2} give the coordinates of each point in the toric diagram.
The coordinates of the toric diagram can be collected in the columns of the following $G_K$ matrix: 
\bea
G_K = \left(
\begin{array}{cccccc}
  1 & 0 &  0 &  0 &  0 &  0 \\
  0 & 0 &  0 &  0 &  1 & -1\\
  0 & 0 &  1 & -1 &  0 &  0
\end{array}
\right)~.
\eea

\end{itemize}

\paragraph{The baryonic charges.} 
As can be seen from Figure \ref{f:tdtoricresq111z2cxcz2}, the toric diagram for this theory has 5 external and one internal points. Therefore, we have $5-4=1$ baryonic symmetry, which is referred to as $U(1)_B$. The charges of the perfect matchings under this symmetry are given by the row of the $Q_D$ matrix.

\paragraph{The global symmetry.} Since there are two pairs of repeated columns in the $Q_t$ matrix, the mesonic symmetry of this model is $SU(2)^2\times U(1)_q \times U(1)_R$. 
From the $Q_t$ matrix, we can see that the perfect matchings $p_1$ and $p_2$ transform as a doublet under the first $SU(2)$, and $r_1$ and $r_2$ transform as a doublet under the second $SU(2)$. 
The global symmetry of this model is the product of the mesonic and baryonic symmetries: $SU(2)^2\times U(1)_q \times U(1)_R \times U(1)_B$.
Recall from \eref{meshigph1q111z2cxcz2} that the mesonic moduli space is a $\BZ_2$ orbifold of the $\CC \times \BC$ theory, and so the R-charges of the external perfect matchings are the same as in the $\CC\times \BC$ theory. The internal perfect matching $s$ has 0 R-charge.
We assign the charges to the perfect matchings under $U(1)_q$ such that the superpotential is not charged under it and the charge vectors are linearly independent. The global charges are summarised in Table \ref{t:chresq111z2ph3}:
\begin{table}[h]
 \begin{center} 
  \begin{tabular}{|c||c|c|c|c|c|c|}
  \hline
  \;& $SU(2)_1$&$SU(2)_2$&$U(1)_R$&$U(1)_q$&$U(1)_{B}$&fugacity\\
  \hline  \hline 
  $p_1$&   1 &  0 & 3/8 &  1 &  0 & $t^3 x_1 q $\\
  \hline
  $p_2$& $-1$&  0 & 3/8 &  1 &  0 & $t^3 q /x_1 $\\
  \hline
  $q_1$&   0 &  1 & 3/8 &  1 &$-1$& $t^3 x_2 q/b$ \\
  \hline
  $q_2$&   0 &$-1$& 3/8 &  1 &$-1$& $t^3 q /(x_2 b)$\\
  \hline  
  $r$&   0 &  0 & 1/2 &$-4$&  0 & $t^4 / q^4 $\\
  \hline
  $s$&   0 &  0 &  0  &  0 &  2 & $b^2 $\\
  \hline
  \end{tabular}
  \end{center}
\caption{Charges of the perfect matchings under the global symmetry of the $\BF_0 \times \BC$ theory. Here $t$ is the fugacity of the R-charge, $x_1,x_2$ are the fugacities of the $SU(2)$ charge, $q$ is the fugacity of the $U(1)$ symmetry and $b$ is the fugacity of the $U(1)_{B}$ symmetry.}
\label{t:chresq111z2cxcz2}
\end{table}

\paragraph{The Hilbert series.} The coherent component of the Master space is a symplectic quotient of the space of perfect matchings by the $Q_F$ matrix:
\bea
\firr{} = \BC^6//Q_F~.  
\label{firrresq111z2cxcz2}
\eea
Thus, the Hilbert series of the coherent component of the Master space can be computed by integrating the Hilbert series of the space of perfect matchings over the baryonic fugacity $z$:
\bea
g^{\firr{}} (t_1, t_2, x_1, x_2, b; \BF_0 \times \BC) &=& \oint \limits_{|z| =1} {\frac{\ud z}{2 \pi i z }} \frac{1}{\left(1- \frac{t_1 x_1}{z}\right)\left(1-\frac{t_1}{x_1 z}\right)\left(1- \frac{t_1 x_2 z}{b}\right)\left(1- \frac{t_1 z}{x_2 b}\right)\left(1-b^2\right)\left(1- t_2\right)}\nn \\
&=& \frac{1-\frac{t_1^4}{b^2}}{\left(1- \frac{t^2_1 x_1 x_2}{ b}\right)\left(1-\frac{t^2_1 x_1}{x_2 b}\right)\left(1-\frac{t^2_1 x_2}{x_1 b}\right)\left(1-\frac{t^2_1}{x_1 x_2 b}\right)\left(1-t_2\right)\left(1-b^2\right)}~, \nn \\
\label{e:HSmasterresq111z2cxcz2}
\eea
where $t_1 = t^3 q$ and $t_2 = t^4 / q^4$.
Integrating (\ref{e:HSmasterresq111z2cxcz2}) over the baryonic fugacity $b$ gives the mesonic Hilbert series:
\bea
\gm (t_1,t_2,x_1,x_2; \BF_0 \times \BC) &=& \oint_{|b|=1} \frac{\ud b}{2\pi i b}\;\; g^{\firr{}} (t_1, t_2, x_1, x_2, b; \BF_0 \times \BC) \nn \\
& =&  \frac{\left(1-t^4_1\right)\left[1 + \left(2+\frac{1}{x^2_1} + x^2_1 + \frac{1}{x^2_2} + x^2_2\right)t^4_1 + t^8_1\right]}{\left(1-t^4_1 x^2_1 x^2_2\right)\left(1-\frac{t^4_1 x^2_1}{x^2_2}\right)\left(1-\frac{t^4_1 x^2_2}{x^2_1}\right)\left(1-\frac{t^4_1}{x^2_1 x^2_2}\right)\left(1-t_2\right)}\nn \\
&=& \sum^{\infty}_{i=0}  t^i_2 \sum^{\infty}_{n=0}  [2n;2n] t^{4n}_1 ~,
\label{e:HSmesresq111z2cxcz2}
\eea
where we note the first factor is the Hilbert series of $\BC$ and the second factor is the Hilbert series of $\BF_0$ \cite{pleth}.  
This confirms that the mesonic moduli space of this model is $\BF_0 \times \BC$.
The totally unrefined Hilbert series of the mesonic moduli space can be written as
\bea
\gm (t^3,t^4,1,1; \BF_0 \times \BC) &=& \frac{1+6t^{12}+t^{24}}{\left(1-t^{12}\right)^3} \times \frac{1}{\left(1-t^4\right)}~. \label{unrefHSf0xc}
\eea
The plethystic logarithm of (\ref{e:HSmesresq111z2cxcz2}) is given by
\bea
\PL[\gm (t_1, t_2 ,x_1,x_2 )] &=& [2;2]t^4_1 + t_2 - O(t^8_1)~. \label{plf0xc}
\eea
Therefore, the 10 generators of the mesonic moduli space can be written in terms of perfect matchings as
\bea
r, \quad p_i p_j  q_i q_j s~, 
\label{e:genpmresq111z2cxcz2}
\eea
where $i,j,l,k = 1,2$.
Note that the lattice of generators is the dual of the toric diagram (nodes are dual to faces and edges are dual to edges).
For the $\BF_0 \times \BC$ theory, the lattice of generators is self-dual.

\begin{figure}[ht]
\begin{center}
  \includegraphics[totalheight=6.0cm]{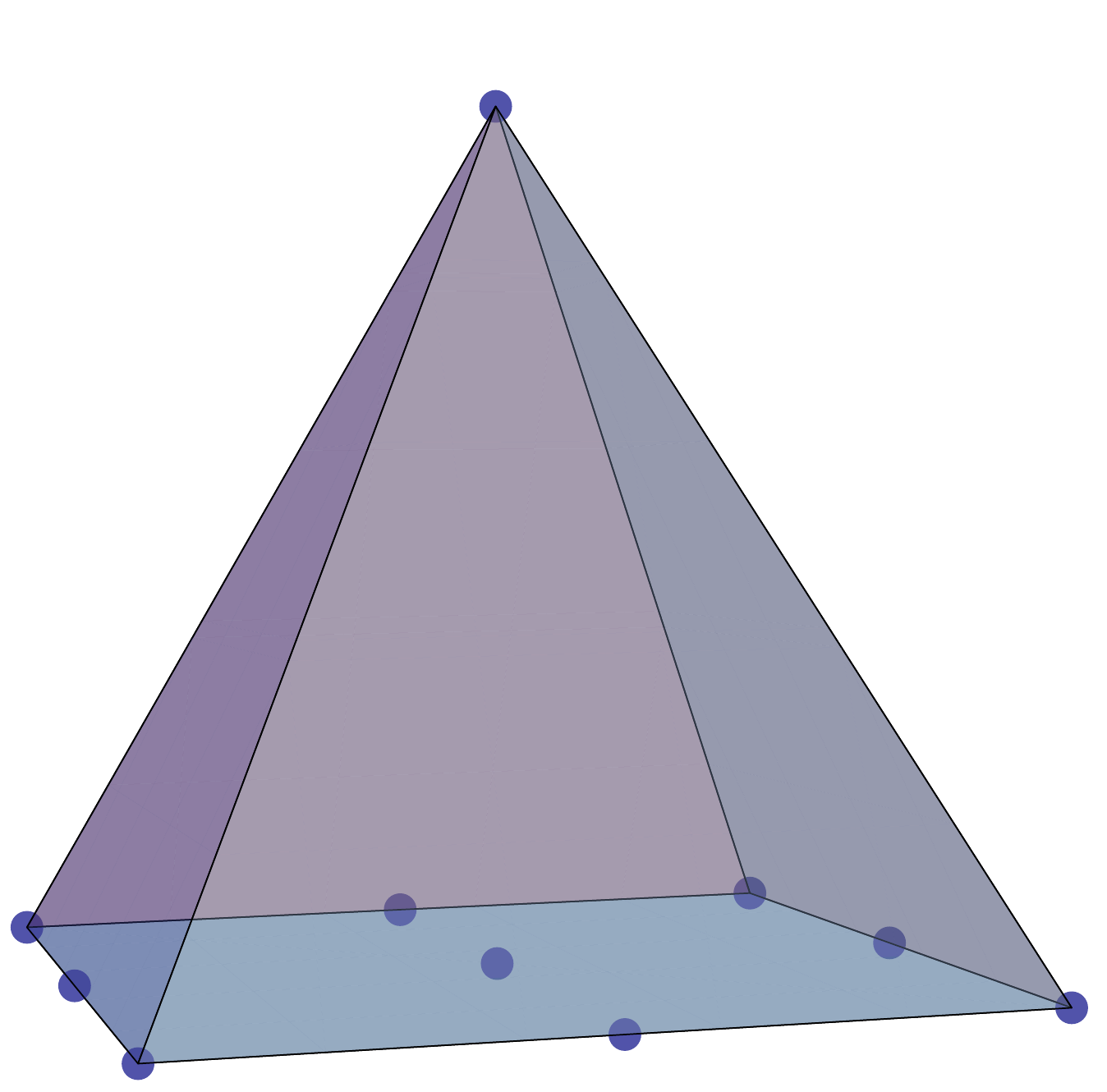}
 \caption{The lattice of generators of the $\BF_0 \times \BC$ theory.}
  \label{f:latf0xc}
\end{center}
\end{figure}

\paragraph{The $\BZ_2$ orbifold action.}
The mesonic Hilbert series of $\CC\times \BC$ is given by (4.14) of \cite{phase}:
\bea
\gm (t_1,t_2,x_1,x_2; \CC \times \BC) = \sum^{\infty}_{i=0}  t^i_2 \sum^{\infty}_{n=0}  [n;n] t^{2n}_1~. \label{meshsofconxc}
\eea
As discussed in \cite{Hanany:2008qc}, under the $\BZ_2$ orbifold action (on $\CC$), $t_1^2 \rightarrow -t_1^2$, and we need to sum over both sectors, with $t_1^2$ and with 
$-t_1^2$. Therefore, starting from \eref{meshsofconxc} and applying the $\BZ_2$ action to $t_1^2$, we are left with the terms corresponding to even $n$, and hence we obtain \eref{e:HSmesresq111z2cxcz2}. 

\subsection*{Higgsing The $\BF_0 \times \BC$ Theory}
From \fref{f:qtresq111z2ph3}, it can be seen that giving a VEV to $X^1_{12}$ of this model leads to the two-hexagon tiling with $k_1 = -k_2 = 2$.
Hence, the mesonic moduli space is a $\BZ_2$ orbifold of $\CC \times \BC$.  This CS orbifold acts on the generators as one of the gauge groups \cite{Hanany:2008cd}. Therefore, the mesonic moduli space of the resulting theory is $ \BF_0 \times \BC$, with the fully refined Hilbert series given by \eref{meshsofconxc}

Observe that in this example, the central charge does not vary as a result of the Higgs mechanism.  This indicates that one of the models, or both, does not give rise to a SCFT in $(2+1)$-dimension. It can be seen that the Higgs mechanism can be used as a consistency test, and this is the first indication of the inconsistency in $(2+1)$-dimension.

\subsection{Higgsing Phase II of $Q^{1,1,1}/\BZ_2$}
\subsection*{A summary of Phase II of $Q^{1,1,1}/\BZ_2$ (the $\mathscr{S}_2 \mathscr{O}_2$ model)}
\begin{figure}[ht]
\begin{center}
  \hskip -5cm
  \includegraphics[totalheight=3.6cm]{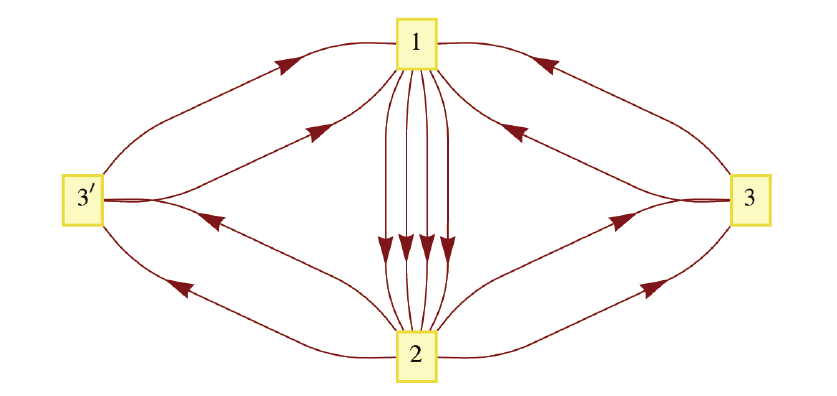}
    \vskip -3.5cm
  \hskip 8cm
  \includegraphics[totalheight=3.5cm]{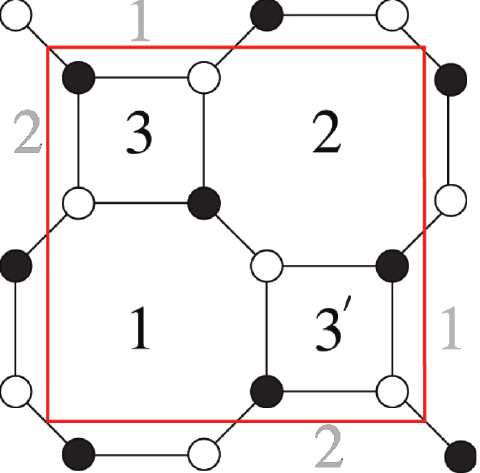}
 \caption{[Phase II of $Q^{1,1,1}/\BZ_2$]  (i) Quiver diagram for the $\mathscr{S}_2 \mathscr{O}_2$ model.\ (ii) Tiling for the $\mathscr{S}_2 \mathscr{O}_2$ model.}
  \label{f:phase2f0}
\end{center}
\end{figure}

This model, first studied in \cite{Hanany:2008fj}, and which we shall denote as $\mathscr{S}_2 \mathscr{O}_2$, has four gauge groups and bi-fundamental fields $X_{12}^{ij}$, $X_{23}^i$, $X_{23'}^i$, $X_{31}^i$ and $X^{i}_{3'1}$ (with $i,j=1,2$). From the features of this quiver gauge theory, this phase is also known as the \emph{three-block model} (see for example \cite{Benvenuti:2004dw}). The superpotential is given by 
\bea
W &=& \epsilon_{ij}\epsilon_{kl} \tr(X^{ik}_{12}X^{l}_{23} X^{j}_{31}) - \epsilon_{ij}\epsilon_{kl} \tr(X^{ki}_{12}X^{l}_{23'}X^{j}_{3'1})~.
\eea
The quiver diagram and tiling of this phase of the theory are given in Figure \ref{f:phase2f0}.  Note that in 3+1 dimensions, these quiver and tiling correspond to Phase II of the $\BF_0$ theory \cite{Feng:2000mi, Forcella:2008ng, master}.  
We choose the CS levels to be $k_1 = k_2 = -k_3 = -k_{3'}=1$.
\\

\begin{figure}[h]
\begin{center}
   \includegraphics[totalheight=6.5cm]{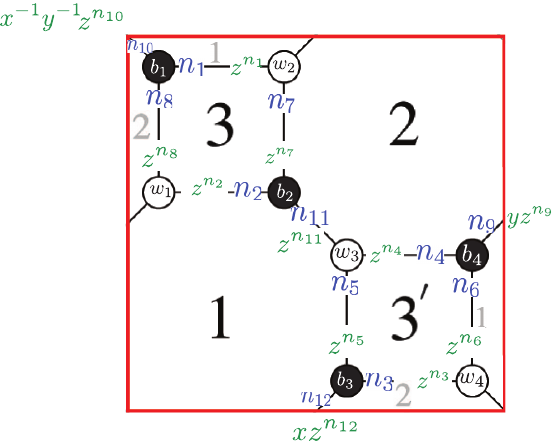}
 \caption{[Phase II of $Q^{1,1,1}/\BZ_2$]  The fundamental domain of tiling for the $\mathscr{S}_2 \mathscr{O}_2$ model: assignments of the integers $n_i$ to the edges are shown in blue and the weights for these edges are shown in green.}
  \label{f:fdph2f0}
\end{center}
\end{figure}

\paragraph{The Kasteleyn matrix.}  We assign the integers $n_i$ to the edges according to Figure \ref{f:fdph2f0}.  We find that 
\bea
\text{Gauge group 1~:} \qquad k_1 &=& 1  = - n_1 - n_2 - n_5 - n_6 +n_9 + n_{10} +n_{11} + n_{12}  ~, \nn \\
\text{Gauge group 2~:} \qquad k_2 &=& 1 =  n_3 + n_4 + n_7 +n_8 - n_9 -n_{10} - n_{11} - n_{12} ~, \nn \\
\text{Gauge group 3~:} \qquad k_3 &=& -1 =  n_1 + n_2 - n_7  - n_8 ~, \nn \\
\text{Gauge group 4~:} \qquad k_{3'} &=& -1 =  - n_3 - n_4 + n_5 + n_6 ~.
\eea  
We choose
\bea
n_2 = -1, \quad n_4 = 1,\quad n_i=0 \; ~ \text{otherwise}~.
\eea

\noindent We can now determine the Kasteleyn matrix.  Since the fundamental domain contains 4 black nodes and 4 white nodes, the Kasteleyn matrix is a $4\times 4$ matrix:
\be
K =   \left(
\begin{array}{c|cccc}
 \;& w_1&w_2&w_3&w_4 \\
  \hline 
b_1 & X^{2}_{23} z^{n_{8}} &\ X^{1}_{31} z^{n_{1}} &\ 0 &\ X^{21}_{12} x^{-1} y^{-1} {z^{n_{10}}} \\
b_2 & X^{2}_{31} z^{n_{2}} &\ X^{1}_{23} z^{n_{7}} &\ X^{12}_{12} z^{n_{11}} &\ 0  \\
b_3 & 0 &\ X^{22}_{12} x z^{n_{12}} &\ X^{1}_{3'1} z^{n_{5}} &\ X^{1}_{23'} z^{n_{3}}  \\
b_4 & X^{11}_{12} y z^{n_{9}} &\ 0 &\ X^{2}_{23'} z^{n_{4}} &\ X^{2}_{3'1} z^{n_{6}} 
\end{array}
\right) ~.
\ee
The permanent of this matrix is given by
\bea \label{permKph2f0}
\perm~K &=&  X^{1}_{31} X^{2}_{31} X^{1}_{3'1} X^{2}_{3'1} z^{n_{1} + n_{2} + n_{5} + n_{6}} + X^{1}_{23'} X^{2}_{23'} X^{2}_{23} X^{1}_{23} z^{n_3 + n_4 + n_7 + n_8}  \nn \\
&&+ X^{1}_{3'1} X^{1}_{23} X^{11}_{12} X^{21}_{12} x^{-1} z^{n_5 + n_7 + n_9 + n_{10}} + X^{2}_{3'1} X^{2}_{23} X^{12}_{12} X^{22}_{12} x {z^{n_{11} + n_{12} + n_6 + n_8}}\nn \\
&&+ X^{1}_{31} X^{1}_{23'} X^{11}_{12} X^{12}_{12} y z^{n_1 + n_3 + n_9 + n_{11}} + X^{2}_{31} X^{2}_{23'} X^{21}_{12} X^{22}_{12} y^{-1} z^{n_{2} + n_{4} + n_{10} + n_{12}} \nn \\
&&+ X^{1}_{31} X^{2}_{31} X^{1}_{23'} X^{2}_{23'} {z^{n_{1} + n_2 + n_3 + n_4}}+ X^{1}_{3'1} X^{2}_{3'1} X^{2}_{23} X^{1}_{23} z^{n_{5} + n_6 + n_7 + n_8}\nn \\
&&  + X^{11}_{12} X^{21}_{12} X^{12}_{12} X^{22}_{12} z^{(n_9 + n_{10} + n_{11} + n_{12})} \nn \\
&=& X^{1}_{31} X^{2}_{31} X^{1}_{3'1} X^{2}_{3'1} z^{-1}  +  X^{1}_{23'} X^{2}_{23'} X^{2}_{23} X^{1}_{23} z + X^{1}_{3'1} X^{1}_{23} X^{11}_{12} X^{21}_{12} x^{-1} + X^{2}_{3'1} X^{2}_{23} X^{12}_{12} X^{22}_{12} x  \nn \\
&&+ X^{1}_{31} X^{1}_{23'} X^{11}_{12} X^{12}_{12} y  + X^{2}_{31} X^{2}_{23'} X^{21}_{12} X^{22}_{12} y^{-1} + X^{1}_{31} X^{2}_{31} X^{1}_{23'} X^{2}_{23'} + X^{1}_{3'1} X^{2}_{3'1} X^{2}_{23} X^{1}_{23}  \nn \\
&&  + X^{11}_{12} X^{21}_{12} X^{12}_{12} X^{22}_{12} \qquad \text{(for $n_2 = -1, ~ n_4 = 1,~ n_i=0 \; ~ \text{otherwise}$)} ~.
\eea

\paragraph{The perfect matchings.} We summarise the correspondence between the quiver fields and the perfect matchings in the $P$ matrix as follows: 
\beq
P=\left(\begin{array} {c|ccccccccc}
  \;& p_1 & p_2 &q_1&q_2&r_1&r_2&s_1&s_2&s_3\\
  \hline
  X^{1}_{31}  & 1 & 0 & 0 & 0 & 1 & 0 & 1 & 0 & 0 \\
  X^{2}_{31}  & 1 & 0 & 0 & 0 & 0 & 1 & 1 & 0 & 0 \\
  X^{1}_{23'}  & 0 & 1 & 0 & 0 & 1 & 0 & 1 & 0 & 0 \\
  X^{2}_{23'}  & 0 & 1 & 0 & 0 & 0 & 1 & 1 & 0 & 0 \\
  X^{1}_{3'1}  & 1 & 0 & 1 & 0 & 0 & 0 & 0 & 1 & 0 \\
  X^{2}_{3'1}  & 1 & 0 & 0 & 1 & 0 & 0 & 0 & 1 & 0 \\
  X^{1}_{23}  & 0 & 1 & 1 & 0 & 0 & 0 & 0 & 1 & 0 \\
  X^{2}_{23}  & 0 & 1 & 0 & 1 & 0 & 0 & 0 & 1 & 0 \\
  X^{11}_{12} & 0 & 0 & 1 & 0 & 1 & 0 & 0 & 0 & 1 \\
  X^{21}_{12} & 0 & 0 & 1 & 0 & 0 & 1 & 0 & 0 & 1 \\
  X^{12}_{12} & 0 & 0 & 0 & 1 & 1 & 0 & 0 & 0 & 1\\
  X^{22}_{12} & 0 & 0 & 0 & 1 & 0 & 1 & 0 & 0 & 1 
  \end{array}
\right)~.
\eeq
From \eref{permKph2f0}, we see that the perfect matchings $p_i, q_i, r_i$ correspond to external points in the toric diagram, whereas the perfect matchings $s_i$ correspond to the internal point at the origin.
The basis vectors of the null space of $P$ are given in the rows of the charge matrix:
\be
Q_F =   \left(
\begin{array}{ccccccccc}
 1 & 1 & 0 & 0 & 0 & 0 & -1 & -1 & 0\\
 0 & 0 & 1 & 1 & 0 & 0 & 0 & -1 & -1 \\
 0 & 0 & 0 & 0 & 1 & 1 & -1 & 0 & -1
 \end{array}
\right)~. \label{qfph2f0}
\ee
Hence, we see that the relations between the perfect matchings are given by
\bea
p_1 + p_2 - s_1 - s_2 &=& 0~, \nn\\
q_1 + q_2 - s_2 - s_3 &=& 0~, \nn \\ \
r_1 + r_2 - s_1 - s_3 &=& 0~. \label{relf0ii}
\eea
Since the coherent component of the Master space is generated by the perfect matchings, subject to the relations \eref{relf0ii}, it follows that 
\bea
\firr{\mathscr{S}_2 \mathscr{O}_2} = \BC^9//Q_F~.  \label{firrph2f0}
\eea

\paragraph{The toric diagram.} We demonstrate two methods of constructing the toric diagram. 
\begin{itemize}
\item{\bf The charge matrices.}    Since the number of gauge groups is $G=4$, there are $G-2 = 2$ baryonic symmetries coming from the D-terms.  We collect the baryonic charges of the perfect matchings in the $Q_D$ matrix:
\bea
Q_D = \left(
\begin{array}{ccccccccc}
 1 & 1 & 0 & 0 & -1 & -1 & 0 & 0 & 0 \\
 0 & 0 & 0 & 0 & -1 & -1 & 2 & 0 & 0
\end{array}
\right)~. \label{qdph2f0}
\eea
From \eref{qfph2f0} and \eref{qdph2f0}, the total charge matrix is given by
\bea
Q_t = \left(
\begin{array}{ccccccccc}
 1 & 1 & 0 & 0 & -1 & -1 & 0 & 0 & 0 \\
 0 & 0 & 0 & 0 & -1 & -1 & 2 & 0 & 0\\
  1 & 1 & 0 & 0 & 0 & 0 & -1 & -1 & 0\\
 0 & 0 & 1 & 1 & 0 & 0 & 0 & -1 & -1 \\
 0 & 0 & 0 & 0 & 1 & 1 & -1 & 0 & -1
\end{array}
\right)~.
\eea
The matrix $G_t$ is obtained by finding the kernel of $Q_t$ and, after removing the first row, the columns give the coordinates of points in the toric diagram:  
\bea
G'_t = \left(
\begin{array}{ccccccccc}
 1 & -1 & 0 & 0 & 0 & 0 & 0 & 0 & 0\\
 0 & 0 & 1 & -1 & 0 & 0 & 0 & 0 & 0 \\
 0 & 0 & 0 & 0 & 1 & -1 & 0 & 0 & 0 
\end{array}
\right)~.
\eea
We see that the toric diagram is given by Figure \ref{f:torq111z2}, with an internal point (with multiplicity 3) at the centre. 

\item {\bf The Kasteleyn matrix.} The powers of $x, y, z$ in each term of the permanent of the Kasteleyn matrix give the coordinates of each point in the toric diagram.  We collect these points in the columns of the following $G_K$ matrix:
\bea
G_K = \left(
\begin{array}{ccccccccc}
 1 & -1 & 0 & 0 & 0 & 0 & 0 & 0 & 0\\
 0 & 0 & 1 & -1 & 0 & 0 & 0 & 0 & 0 \\
 0 & 0 & 0 & 0 & 1 & -1 & 0 & 0 & 0 
\end{array}
\right) = G'_t~.
\eea
Thus, the toric diagrams constructed from these two methods are indeed identical.
\end{itemize}

\paragraph{The baryonic charges.}
Since the toric diagram has 6 external points, this model has precisely $6-4 = 2$ baryonic symmetries, which we shall denote by $U(1)_{B_1}$ and $U(1)_{B_2}$.  From the above discussion, we see that they arise from the D-terms.  Therefore, the baryonic charges of the perfect matchings are given by the rows of the $Q_D$ matrix. 

\paragraph{The global symmetry.} From the $Q_t$ matrix, we notice that the charge assignment breaks the symmetry of the space of perfect matchings to $SU(2)^3\times U(1)_R$. This mesonic symmetry can also be seen from the $G_K$ (or $G'_t$ ) matrix by noticing that the three rows contain weights of $SU(2)$.
Since $s_1, s_2, s_3$ are the perfect matchings corresponding to the internal point in the toric diagram, we assign to each of them zero R-charge.  
The remaining 6 external perfect matchings are completely symmetric and the requirement of R-charge 2 to the superpotential divides 2 equally among them, resulting in R-charge 1/3 each. 
The global symmetry of the theory is a product of mesonic and baryonic symmetries: $SU(2)^3 \times U(1)_R \times U(1)_{B_1} \times U(1)_{B_2}$.
In Table \ref{chargeph1f0}, we give a consistent charge assignment for the perfect matchings under the global symmetries.

Below, there is a study of the Higgs mechanism of this theory.

\subsubsection{\emph{Phase III of $(\BC^2/\BZ_2) \times \BC^2$} from giving a VEV to one of $X^{ij}_{12}$}  \label{sec:ph3c2z2xc2}
\begin{figure}[ht]
\begin{center}
 \vskip 1cm
  \hskip -7cm
  \includegraphics[totalheight=2.5cm]{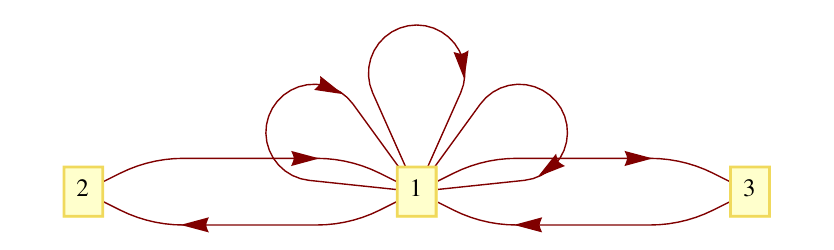}
   \vskip -3.5cm
  \hskip 8cm
  \includegraphics[totalheight=6.0cm]{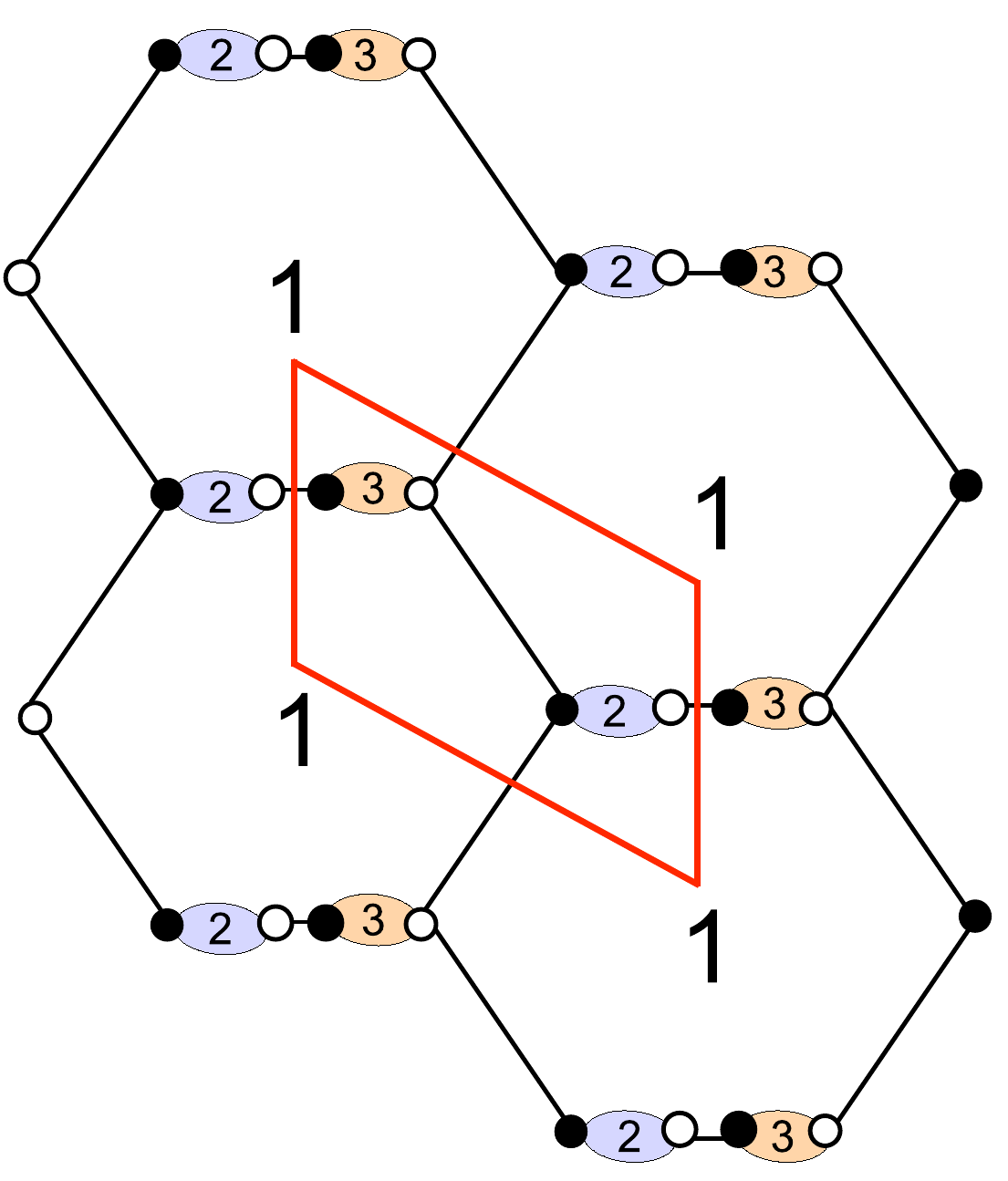}
 \caption{(i) Quiver diagram of Phase III of $\left(\BC^2/\BZ_2\right) \times \BC^2$.\ (ii) Tiling of Phase III of $\left(\BC^2/\BZ_2\right) \times \BC^2$.}
  \label{f:qtresq111z2ph4}
\end{center}
\end{figure}

By symmetry, we see that turning on a VEV to any of the  $X^{ij}_{12}$ fields yields the same result.  
For definiteness, let us consider the case of  $X^{12}_{12}$.  
This amounts to removing one of the edges that separate the faces corresponding to gauge groups 1 and 2, and collapsing the two vertices adjacent to a bivalent vertex into a single vertex of higher valence.
As a result, the gauge groups 1 and 2 are combined into one gauge group, identified as 1, 
and the edges corresponding to $X^1_{23}, X^2_{31}$ and $X^2_{23'}, X^1_{3'1}$ are removed.
For convenience, let us relabel gauge group $3'$ as $2$.
The quiver diagram and tiling of this model are presented in Figure \ref{f:qtresq111z2ph4}.
Let us denote the adjoint fields by $\phi^i$ (with $i=1,2,3$).
The superpotential can be written as 
\bea
W = \tr\left[( \phi^2 - \phi^3 \phi^1 ) X_{12} X_{21} + (\phi^1 \phi^3 - \phi^2)X_{13}X_{31} \right]~.
\label{e:spresq111z2ph4}
\eea
The CS levels associated with the higgsed gauge groups are added, and so
\bea
k_1= 2, \quad k_2 = -1, \quad k_3 = -1 ~.
\eea

\begin{figure}[ht]
\begin{center}
   \includegraphics[totalheight=7cm]{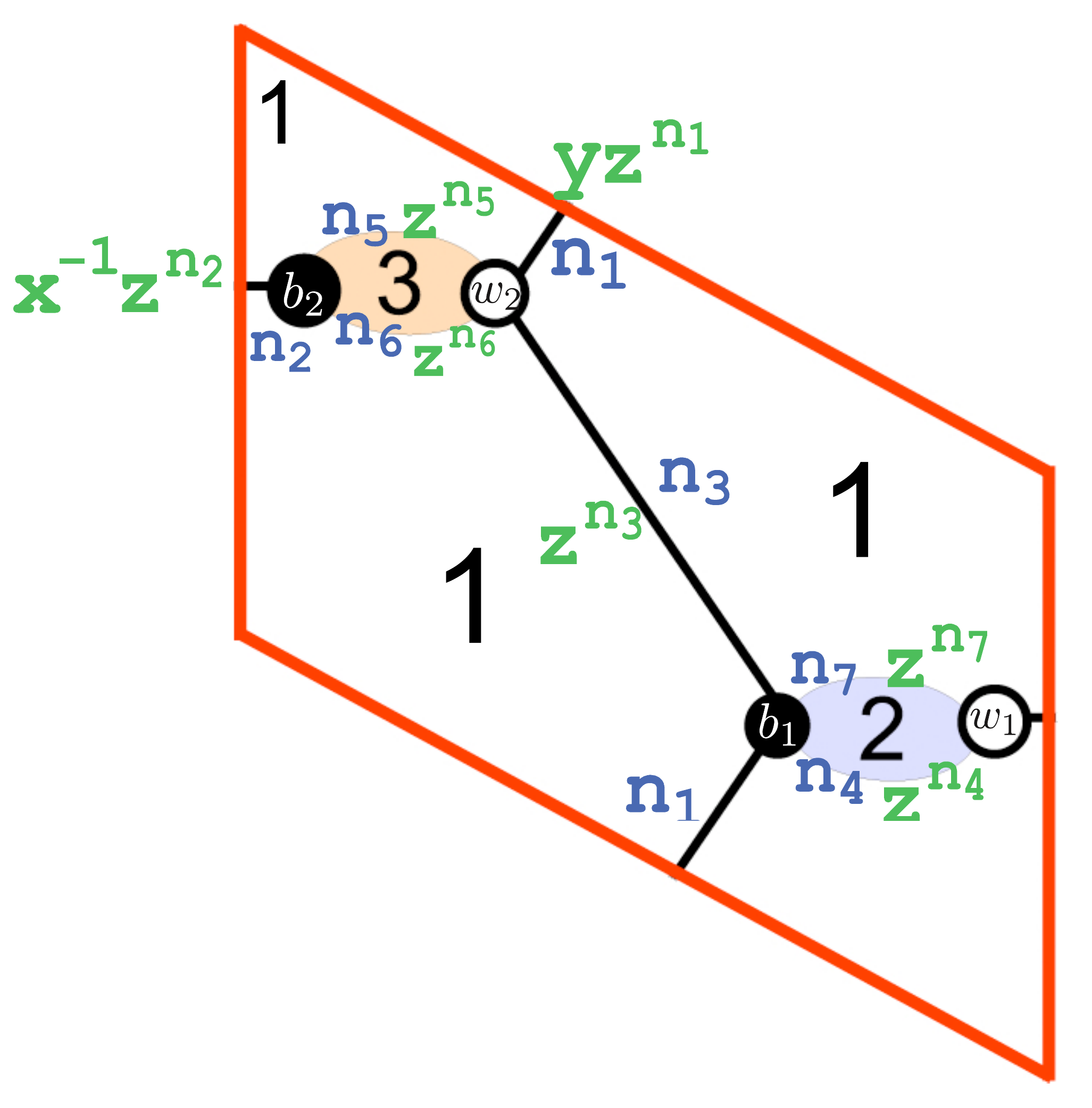}
 \caption{The fundamental domain of the tiling for Phase III of $\left(\BC^2/\BZ_2\right) \times \BC^2$: assignments of the integers $n_i$ to the edges are shown in blue and the weights for these edges are shown in green.}
  \label{f:fdresq111z2ph4}
\end{center}
\end{figure}

\paragraph{The Kasteleyn matrix.} 
We assign the integers $n_i$ to the edges of the tiling as shown in Figure \ref{f:fdresq111z2ph4}. We have
\bea
\text{Gauge group 1~:} \qquad k_1 &=& 2  =  n_4 - n_5 +n_6 -n_7~,  \nn \\
\text{Gauge group 2~:} \qquad k_2 &=&  -1 =    -n_4 + n_7~,  \nn \\
\text{Gauge group 3~:} \qquad k_3 &=&  -1  = n_5 - n_6~. 
\label{e:karesq111z2ph4}
\eea  
We choose:
\bea
n_4= -n_5= 1,\qquad n_i=0 \;\;\text{otherwise}~.
\eea
We can now determine the Kasteleyn matrix. Since the fundamental domain contains 2 pairs of black and white nodes, the Kasteleyn matrix is $2 \times 2$:
\bea
K =   \left(
\begin{array}{c|cc}
& b_1 & b_2\\
\hline
w_1 & X_{12} z^{n_4} + X_{21}z^{n_7} & \phi^1 y z^{n_1} + \phi^3 z^{n_3} \\
w_2 & \phi^2 x^{-1} z^{n_2} & X_{13} z^{n_6} + X_{31} z^{n_5}
\end{array}
\right) ~.
\label{e:kastresq111z2ph4}
\eea
The permanent is given by
\bea
\mathrm{perm}~K &=&  \phi^1 \phi^2 y x^{-1} z^{n_1+n_2} + \phi^2 \phi^3 x^{-1} z^{n_2+ n_3} + X_{12} X_{31} z^{n_4+n_5} + X_{21}X_{13} z^{n_6+n_7} \nn \\
&& +  X_{12} X_{13} z^{n_4+n_6}+ X_{21} X_{31} z^{n_5+n_7}\nn\\
&=& \phi^1 \phi^2 y x^{-1} + \phi^2 \phi^3 x^{-1} +  X_{12} X_{31} + X_{21}X_{13} + X_{12} X_{13} z +  X_{21} X_{31} z^{-1}\nn \\
&& \text{(for $n_4 = -n_5 = 1,~ n_i=0 \; \text{otherwise}$)}~.
\label{e:permresq111z2ph4}
\eea

\paragraph{The perfect matchings.} 
From (\ref{e:permresq111z2ph4}), we can take the perfect matchings to be
\bea 
p_1 &=& \left\{\phi^1, \phi^2 \right\}, \;\; p_2 = \left\{\phi^2, \phi^3 \right\}, \;\; s_1 = \left\{X_{12}, X_{31}\right\}, \;\; s_2 = \left\{X_{21}, X_{13}\right\}, \nn \\  
 r_1 &=& \left\{X_{12}, X_{13}\right\}, \;\;  r_2 = \left\{X_{21}, X_{31}\right\}\ . \qquad
\eea
In turn, we can parametrise the chiral fields in terms of perfect matchings:
\bea
\phi^1 &=&  p_1, \quad \phi^2 = p_1 p_2, \quad  \phi^3 = p_2, \quad  X_{12} = s_1 r_1,  \nn \\
\quad X_{31} &=& s_1 r_2, \quad X_{13} = s_2 r_1, \quad X_{21} = s_2 r_2~.
\eea
We can collect all these pieces of information in the perfect matching matrix:
\beq
P=\left(\begin{array} {c|cccccc}
  \;& p_1 & p_2 & s_1 & s_2 & r_1 & r_2\\
  \hline 
  \phi^1& 1&0&0&0&0&0\\
  \phi^2& 1&1&0&0&0&0\\
  \phi^3& 0&1&0&0&0&0\\
  X_{12}  & 0&0&1&0&1&0\\  
  X_{31}  & 0&0&1&0&0&1\\
  X_{13}  & 0&0&0&1&1&0\\
  X_{21}  & 0&0&0&1&0&1\\
  \end{array}
\right).
  \label{e:presq111z2ph4}
\eeq
Note that the perfect matching matrices (\ref{e:presq111z2ph4}) and (\ref{e:presq111z2ph3}) coincide.
Thus, the $Q_F$ matrix is given by \eref{e:qfresq111z2ph3}:
\be
Q_F =   \left(
\begin{array}{cccccc}
0, &0, &1, &1, &-1, &-1
\end{array}
\right)~.  \label{e:qfresq111z2ph4}
\ee

\paragraph{The toric diagram.} We construct the toric diagram of this model using two methods.
\begin{itemize}

\item {\bf The charge matrices.}
Since the number of gauge groups of this model is $G = 3$, there is $G-2 =1$ baryonic symmetry coming from the D-terms. The charges of the perfect matchings under this baryonic symmetry can be collected in the $Q_D$ matrix:
\be
Q_D =   \left(
\begin{array}{cccccc}
0, & 0, & 1,& -1,& 0,& 0
\end{array}
\right)~. \label{e:qdresq111z2ph4}
\ee
We can combine (\ref{e:qfresq111z2ph4}) and (\ref{e:qdresq111z2ph4}) in a single matrix, $Q_t$, that contains all the baryonic charges of the perfect matchings:
\be
Q_t = { \Blue Q_D \choose \Green Q_F \Black } =   \left( 
\begin{array}{cccccc} \Blue
0 & 0&  1& -1&  0&  0 \\ \Green
0 & 0&  1&  1& -1& -1  \Black
\end{array}
\right)~.
\label{e:qtresq111z2ph4}
\ee
Note that the $Q_t$ matrix \eref{e:qtresq111z2ph4} coincides with (\ref{e:qtresq111z2ph3}). Hence, the $G'_t$ matrix is given by \eref{e:toricdiaresq111z2ph3}. The toric diagram is given in Figure \ref{f:tdtoricresq111z2ph3}.  Thus, the mesonic moduli space of this model is 
\bea
\CMm= \left( \BC^2/\BZ_2 \right) \times \BC^2~.
\eea
We refer to this model as {\bf Phase III of the $ \left( \BC^2/\BZ_2 \right) \times \BC^2$ theory}.

\item {\bf The Kasteleyn matrix.} 
The coordinates of each point in the toric diagram are given by the powers of $x,y$ and $z$ in each term of \eref{e:permresq111z2ph4}.
We collect these points in the columns of the following $G_K$ matrix:
\bea
G_K = \left(
\begin{array}{cccccc}
  -1 & -1 &  0 & 0 & 0 & 0 \\
   0 &  1 &  0 & 0 & 0 & 0\\
   0 &  0  & 1 &-1 &  0 & 0
\end{array}
\right)~.
\eea
The toric diagrams constructed from the $G_K$ and $G'_t$ matrices are the same up to a transformation {\footnotesize $\CT = \left( \begin{array}{ccc} -1&0&-2\\0&0&1\\0&1&0 \end{array} \right) \in GL(3, \BZ)$}, where we have $G_K = \CT \cdot G'_t$.
\end{itemize}

\paragraph{The baryonic charges.} Since the toric diagram of this model has 5 external points, there is exactly $5-4 = 1$ baryonic symmetry, which we shall denote as $U(1)_{B}$. The charges of the perfect matchings under this symmetry come from the row of the $Q_D$ matrix \eref{e:qdresq111z2ph4}.

\paragraph{The global symmetry.}  Since there are two pairs of repeated columns in the $Q_t$ matrix, the mesonic symmetry of this model is $SU(2)^2\times U(1)_q \times U(1)_R$. 
The global symmetry of this model is the product of the mesonic and baryonic symmetries: $SU(2)^2\times U(1)_q \times U(1)_R \times U(1)_B$.
From the $Q_t$ matrix, we see that the perfect matchings $p_1$ and $p_2$ transform as a doublet under the first $SU(2)$, and $r_1$ and $r_2$ transform as a doublet under the second $SU(2)$. 
A consistent charge assignments for the perfect matchings is given in Table \ref{t:chresq111z2ph3}.

\paragraph{The Hilbert series.} The coherent component of the Master space is the symplectic quotient of the space of perfect matchings by the $Q_F$ matrix:
\bea
\firr{} = \BC^6//Q_F = \BC^6// (0, ~ 0, ~ 1, ~ 1, ~ -1, ~ -1)~.  
\label{e:firrresq111z2ph4}
\eea
Thus, the Master space of this model is the same as that of Phase II of $\left(\BC^2/\BZ_2\right) \times \BC^2$, and its Hilbert series is given by \eref{e:HSmasterresq111z2ph3}.
The mesonic moduli space is the quotient of the Master space by the $Q_D$ matrix.  Since the $Q_D$ matrices of this model and that of Phase II of $\left(\BC^2/\BZ_2\right) \times \BC^2$ coincide, the mesonic moduli spaces of both models are identical.  The mesonic Hilbert series is given by (\ref{e:HSmesresq111z2ph3}).  The 5 generators of the mesonic moduli space written in terms of perfect matchings are presented in (\ref{e:genpmresq111z2ph3}).

\subsubsection{\emph{Phase II of $(\BC^2/\BZ_2) \times \BC^2$} from giving a VEV to one of $X^i_{31}$, $X^i_{23}$, $X^i_{3'1}$, $X^i_{23'}$} \label{sec:anoph2c2z2xc2}
By symmetry, giving a VEV to any of the $X^i_{13}, X^i_{23}, X^i_{13'}$ and $X^i_{23'}$ fields leads to the same theory.  
For definiteness, let us examine the case in which $X^2_{23'}$ acquires a VEV. 
This amounts to removing one of the edges separating the gauge groups 2 and $3'$ in Figure \ref{f:phase2f0}. 
Thus, the octagon corresponding to gauge group 2 combines with the square corresponding to gauge group $3'$ to form a decagon, which we shall label as $2$.
As a result of integrating out massive fields, the two vertices adjacent to a bivalent vertex collapse into a single vertex of higher valence.
Therefore, decagons become hexagons and the remaining octagons become squares. 
It follows that the resulting tiling is simply two hexagons with one diagonal.
The CS levels associated with the higgsed gauge groups are added, and so the new CS level is
\bea
k_1 = 1, \quad k_2 = 0, \quad k_3 = -1~. 
\eea
Therefore, the resulting theory is Phase II of the $\left( \BC^2/\BZ_2 \right) \times \BC^2$ theory, with the quiver diagram and tiling presented in Figure \ref{f:qtresq111z2ph3}. The toric diagram is drawn in Figure \ref{f:tdtoricresq111z2ph3}.

\section{Higgsing The $\left( \BC^2/\BZ_2 \right) \times \BC^2$ Theory}
\paragraph{A summary.}
Aspects of this theory have been studied in \cite{Imamura:2008nn, Imamura:2008ji}.  Below, we summarise important information of the three phases discussed in earlier sections.

\begin{table}[htdp]
\begin{center}
\begin{tabular}{|c||c|c|c|}
\hline
\; & Phase I & Phase II &Phase III \\
\hline
Section & \sref{sec:ph1c2z2xc2} & \sref{sec:ph2c2z2xc2} and  \sref{sec:anoph2c2z2xc2} & \sref{sec:ph3c2z2xc2} \\
\hline
Tiling & $\sD_1 \sH_1$ (\fref{f:phase2c4})  & $\sH_2 \partial_1$ (\fref{f:qtresq111z2ph3}) & $\sD_2 \sH_1$ (\fref{f:qtresq111z2ph4}) \\
\hline
CS levels &  $k_1 = 2, ~ k_2 = -2$ & $k_1= 1, ~ k_2 = -1, ~ k_3 =0$ & $k_1=2, ~ k_2 = -1, ~ k_3 = -1$ \\
\hline
Global charges & \multicolumn{3}{c|}{\tref{t:chresq111z2ph3}} \\
\hline
Toric diagram & \multicolumn{3}{c|}{\fref{f:tdtoricresq111z2ph3}} \\
\hline
Generators & \multicolumn{3}{c|}{\eref{e:genpmresq111z2ph3} and \fref{f:latc2z2xc2}} \\
\hline
\end{tabular}
\caption{A summary of three phases of the $\left( \BC^2/\BZ_2 \right) \times \BC^2$ theory.}
\label{t:summary}
\end{center}
\end{table}

\paragraph{Higgsing Phase I of $\left( \BC^2/\BZ_2 \right) \times \BC^2$.}
Giving a VEV to $X_{12}$ or $X_{21}$ leads to the one-hexagon tiling with $k=0$.
The tiling suggests that there is a branch of the moduli space which is $\BC^3$.  
In the presence of a gauge kinetic term, there is an additional complex degree of freedom.
Therefore, the mesonic moduli space is $\BC^4$. 

\paragraph{Higgsing Phase II of $\left( \BC^2/\BZ_2 \right) \times \BC^2$.}
Giving a VEV to $X^1_{12}$ leads to the two hexagon tiling with $k_1 = k_2 = 0$.
Looking at the tiling, it seems like a branch of the moduli space is $\left(\BC^2/\BZ_2 \right) \times \BC$.
In the presence of a gauge kinetic term, there is an additional complex degree of freedom, and the mesonic moduli space is $\left(\BC^2/\BZ_2 \right) \times \BC^2$.

\paragraph{Higgsing Phase III of $\left( \BC^2/\BZ_2 \right) \times \BC^2$.}
Giving a VEV to either $X_{12}$, $X_{21}$, $X_{13}$ or $X_{31}$ leads to a one double-bonded one-hexagon tiling with $k_1= -k_2 = 1$.
Thus, the mesonic moduli space of the resulting theory is Phase II of $\BC^4$.
  
\appendix
\section{Connections between Theories via Massive Deformation} \label{massivedeform}
In this appendix, we consider another type of connection between M2-brane theories via massive deformation.  This has previously been studied in \cite{Franco:2008um, Kim:2007ic}.   The starting point is an M2-brane theory with two or more adjoint fields.  Let us denote the two adjoint fields of interest by $\phi_1$ and $\phi_2$.  The renormalisation group (RG) flow is triggered by the introduction of a mass term $\Delta W = \phi_2^2 -  \phi_1^2$ in the superpotential.  Integrating out $\phi_1$ and $\phi_2$ yields the superpotential of the resulting theory in the IR. 

This process can be realised on the brane tiling as follows.  The mass term gives an R-charge 1 to each of the adjoint fields.  Consequently, the corresponding edges shrink to zero size \cite{Hanany:2005ss}. The nodes at the ends of each edge combine into a single node of higher valence.  One of the combined nodes becomes white and the other becomes black.

\paragraph{Phase II of $\CC \times \BC$ to Phase I of $\BC^4$.} 
We consider the adjoint fields $\phi^1$ and $\phi^2$ from the tiling \fref{f:phase2conxc}.  As a result of the massive deformation, the end point of the RG flow is Phase I of $\BC^4$, whose tiling is given by \fref{f:con}.


\section{M-Theory on $\mathrm{CY}_4$ from Type IIA Theory on $\mathrm{CY}_3$ with RR fluxes} \label{app:IIA}
In this section, we discuss string theory origins of M2-brane theories.  A theory on D2-branes probing a $\mathrm{CY}_3$ singularity, with non-zero RR fluxes, can be lifted to a theory on M2-branes probing a $\mathrm{CY}_4$ singularity \cite{Aganagic:2009zk, ucsbtalk}.  Below, we demonstrate this in various examples.

\subsection{The conifold theory with $k$ units of RR 2-form flux on $\BP^1$}
The conifold has one vanishing 2-cycle $\BP^1$.  There are 2 nodes in the quiver diagram (\fref{f:con}): one node corresponds to a gauge field living on a bound state of a D2-brane and a D4-brane wrapped on $\BP^1$, and the other node corresponds to a gauge field living on a bound state of a D2-brane and a D4-brane wrapped on $\BP^1$ with opposite orientation.
By symmetry, the D2-brane contributes $+1$ unit of charge to each node, whereas the D4-brane on $\BP^1$ contributes $+1$ unit of charge to node 1 and $-1$ to node 2.
If the flux on the D4-brane wrapped on $\BP^1$ is $k$, then the fluxes on nodes 1 and 2 are $k$ and $-k$, respectively.  
These fluxes induce CS interactions.

To see the effect of the flux, let us consider the WZ interaction term on the D4-brane:
\bea
\int_{D4} F_{RR} \wedge A \wedge dA~,
\eea
where $F_{RR}$ is the RR 2-form field strength of Type IIA and $A$ is the gauge field on the D4-brane.
The RR flux on the D4-brane wrapped on $\BP^1$  is given by
\bea
k = \int_{\BP^1} F_{RR}~.
\eea
This gives rise to the CS levels $\pm k$, as described above.

T-dualising the D2-brane on $\mathrm{CY}_3$ with RR fluxes on compact cycles yields the Type IIB description given in \cite{Bergman:1999na}. T-dualising once more results in the Type IIA brane tilings \cite{phase, Imamura:2008qs}.  The CS levels become the intersection numbers between ($p$,$q$)-branes in Type IIB under the first T-duality, and become the 1-form field strength flux in the tiling under the second T-duality. 

Thus, the Type IIA conifold theory with $k$ units of RR 2-form flux on $\BP^1$ can be lifted to \emph{Phase I of $\BC^4$} with the CS levels $(k, -k)$.  The Hilbert series is derived in \cite{Hanany:2008cd}:
\bea
\gm (t, \BC^4/\BZ_k) = \frac{1+t^2+2kt^k - 2kt^{k+2} - t^{2k} - t^{2k+2}}{(1-t^2)^3(1-t^k)^2}~, \label{c4modzk}
 \eea
 where $t$ is the fugacity for R-charges (in the unit of 1/2).  In the large $k$ limit, we recover the Hilbert series of the conifold:
\bea
\frac{1+t^2}{(1-t^2)^3}~.
\eea
The plethystic logarithm of the Hilbert series \eref{c4modzk} can be written in terms of representations of $SU(2) \times SU(2) \times U(1)$ as
\bea
\PL [\gm (t, x,y; \BC^4/\BZ_k) ] =   \left \{
\begin{array}{ll}
\left( [1;0] + [0;1] \right) t   & \text{for  } k = 1 \\
\left( [k;0] + [0;k] \right) t^k +  [1;1] t^2 - t^4 + \ldots & \text{for } k > 1 \\
\left[1;1\right] t^2-t^4 & \text{for } k \rightarrow \infty ~.
\end{array} \right .
\eea
For any $k$, there always exist operators at order $t^k$ transforming in the $[k;0] + [0;k]$ representation; for $k>1$ there always exist operators at order $t^2$ in the $[1;1]$ representation. 

\subsection{The $(\BC^2/\BZ_2) \times \BC$ theory with $k$ units of RR 2-form flux on $\BP^1$}
The Calabi-Yau $(\BC^2/\BZ_2) \times \BC$ has one vanishing 2-cycle $\BP^1$.  The quiver diagram is presented in \fref{f:phase1conxc}. The situation is very similar to the conifold theory described above.   This theory  can be lifted to \emph{Phase II of $\CC \times \BC$} with the CS levels $(k, -k)$. The Hilbert series is derived in \cite{Hanany:2008cd}:
\bea
\gm (t, (\CC/\BZ_k) \times \BC) = \frac{1+t^{12}+2 k t^{6 k}-t^{12 k}-2 k t^{6 (2+ k)}-t^{12 (1+ k)}}{\left(1-t^4\right) \left(1-t^{12}\right)^2 \left(1-t^{6 k}\right)^2}~, \label{HSConxck}
 \eea
 where $t$ is the fugacity for R-charges (in units of 1/8)\footnote{Since we grade the chiral ring by the R-charges, the powers of $t$ appearing in the Hilbert series \eref{HSConxck} are different from those in (4.11) of  \cite{Hanany:2008cd}}.  Note that for $k=2$, we recover the Hilbert series \eref{unrefHSf0xc} for $\BF_0 \times \BC$, namely
\bea
\gm (t, \BF_0 \times \BC) = \frac{1+6 t^{12}+t^{24}}{\left(1-t^4\right) \left(1-t^{12} \right)^3}~. \nn
\eea
In the large $k$ limit, we obtain the Hilbert series of the $(\BC^2/\BZ_2) \times \BC$ theory:
\bea
\frac{1+t^{12}}{\left(1-t^4\right) \left(1-t^{12}\right)^2}~.
\eea

The fully refined plethystic logarithms are given by
\bea
\PL [\gm (t_1,t_2, x,y;  (\CC/\BZ_k) \times \BC) ]=   \left \{
\begin{array}{ll}
t_2 +[1;1] t_1^2-t_1^{4} & \text{for  } k = 1 \\
t_2 + [2;2] t_1^{4} - t_1^{8} + \ldots & \text{for } k =2~. \\
\end{array} \right .
\eea
where $t_1 =  t^3/q,~t_ 2 = t^4/q^4$ (with $q$ being the $U(1)$ fugacity and $t$ being the $U(1)_R$ fugacity). 
For $k=2$, we recover the plethystic logarithm \eref{plf0xc} for $\BF_0 \times \BC$.
Note that for $k=1, 2$ the orbifolds preserve the mesonic global symmetry $SU(2) \times SU(2) \times U(1) \times U(1)$.
For $k>2$, the plethystic logarithm can be written as
\bea
\PL [\gm (t_1, t_2, x,y;  (\CC/\BZ_k) \times \BC) ] = t_2 + [k] \left(t_1^2 x \right)^k + [k] \left(t_1^2/x \right)^k  + [2] t_1^{4}  - t_1^{8} + \ldots ~ (\text{for}~k > 2)~.\qquad
\eea
It can be seen that the mesonic global symmetry is broken down to $SU(2) \times U(1) \times U(1) \times U(1)$, where $x$ is now regarded as a $U(1)$ fugacity.
As $k \rightarrow \infty$, the plethystic logarithm becomes
\bea
\PL [\gm (t_1, t_2, y; (\BC^2/\BZ_2) \times \BC) ] = t_2 + [2] t_1^{4}  - t_1^{8} ~,\qquad
\eea
and the mesonic symmetry is $SU(2) \times U(1) \times U(1)$.

\subsection{The $SPP$ theory with $k$ units of RR 2-form fluxes on both $\BP^1$'s}
The SPP theory has two vanishing 2-cycles, $\BP^1_1$ and $\BP^1_2$.  One vanishing $\BP^1$ gives locally a conifold singularity $\BP^1_1$, while the second gives locally a $\BZ_2$ singularity, $\BP^1_2$. Each of the 3 nodes in the quiver diagram (\fref{f:phase2D3}) corresponds to a collection of fractional branes.  The fractional brane charges can be determined as follows.  
The toric diagram and $(p,q)$-web of the $SPP$ (or $L^{1,2,1}$) theory are given in Figure 4 of \cite{Franco:2005sm}.  The external legs of the $(p,q)$-web are given by
\bea
(0, -1),~(-1,0),~(1,0),~(1,0),~(-1,1)~.
\eea
We may take linear combinations of these charges to form the $(p,q)$-charges corresponding to the 3 nodes:
\bea
(p_1, q_1) = (1, -1), \quad (p_2, q_2)=(-1,1), \quad (p_3,q_3) = (0,0)~.
\eea
Then, the fractional brane charges can be obtained using the method described in \cite{Hanany:2001py}:
\bea
\mathrm{ch} (F_1) = (-1, 0 ,0), \quad \mathrm{ch}(F_2) = (0, 1, 0), \quad \mathrm{ch} (F_3) = (1, -1, 1)~,
\eea
where the 3 entries are the D4-brane charge on $\BP^1_1$, the D4-brane charge on $\BP^1_2$, and the D2-brane charge respectively. 
A flux $k_1$ on $\BP^1_1$, and a flux $k_2$ on $\BP^1_2$ give a contribution to the CS levels
\bea
\vec{k} = (-k_1 , k_2, k_1 - k_2)~.
\eea
Below, we see that setting $k_1 = 1, k_2 =0$ (\emph{i.e.} turning on only a flux on $\BP^1_1$) gives the $\BC^2/\BZ_2 \times \BC^2$ theory, whereas setting $k_1 = 0, k_2 =1$ (\emph{i.e.} turning on only a flux on $\BP^1_2$) gives the $D_3$ theory.

\paragraph{The $D_3$ theory or the $\BC^2/\BZ_2 \times \BC^2$ theory?}
Let us consider a Type IIB setup. The relevant brane configuration is given in \cite{Uranga:1998vf, Erlich:1999rb}, \cite{Amariti:2009rb,Imamura:2008nn}. There are 3 NS5-branes: two parallel branes, which are referred to as $\text{NS}_1$ and $\text{NS}_2$, along the directions $012389$, and the other, which is referred to as $\text{NS}'$, along the directions $012345$. These NS5-branes are separated along the compact direction 6. Also, there are $N$ D3-branes along the directions 0126.  On the interval of D3-branes between two NS5-branes, there is a $U(N)$ adjoint chiral multiplet.  If the two 5-branes are not parallel, the adjoint chiral field on the interval between these 5-branes becomes massive.

The distance between two NS5-branes gives the gauge coupling of each group, which is the $B$-field on the $\BP^1$.  Since there are three $\BP^1$'s with a condition that the sum is homologically trivial,  the parallel NS-branes have a $\BP^1$ which is locally $\BZ_2$ singularity (\emph{i.e.} $\BP^1_2$), and the $\text{NS}-\text{NS}'$ branes have a $\BP^1$ which is locally the conifold singularity (\emph{i.e.} $\BP^1_1$).  

Now let us insert $k$ D5-branes in the directions $012789$ such that they intersect one of the NS-branes (let's say $\text{NS}_2$) along $012389$.  The intermediate $(1,k)$-brane is along $012[37]_\theta89$. Since the $(1,k)$-branes and $\text{NS}_1$-brane are parallel in the $89$ direction, there is a massless adjoint chiral field on the interval between these two 5-branes.  The difference between the RR-charges on $(1,k)$ and $\text{NS}_1$-branes gives the CS level $k$ (see \emph{e.g.} (2.8) of \cite{Imamura:2008nn}) for the gauge group with adjoint field. Note that this is also the flux on $\BP^1_2$.  Similarly, for the gauge groups without adjoint field, the CS levels are $-k$ and $0$. Thus, the gauge theory is $D_3/\BZ_k$.


Instead, if we add $k$ D5-branes in the directions $012457$ such that they intersect the $\text{NS}'$-brane along $012345$, the $(1,k)$-brane along $ 012 [37]_\theta 45$ is not parallel to the NS-branes. Hence there is no massless adjoint chiral field on the intervals between these 5-branes.  The $(1,k)$-brane and each of the NS-branes together induce the CS levels $-k$ and $k$ for the gauge groups without adjoint field.  Note that one of these is also the flux on $\BP^1_1$.  Similarly, the CS level for the gauge group with adjoint field is $0$.  Therefore, the corresponding gauge theory is $\left( \BC^2/\BZ_2 \times \BC^2 \right)/\BZ_k$.


\paragraph{The Hilbert series for $D_3/\BZ_k$.} Note that the $Q_D$ matrix \eref{qdph2d3} contains the charges under $U(1)_3$.  Therefore, the $\BZ_k$ orbifold acts on the generators like, let's say, $U(1)_1$ and the Hilbert series can be obtained by inserting $\omega^j$ into appropriate places in the mesonic Hilbert series (5.15) of \cite{phase}:
\bea
\gm (t, D_3/\BZ_k) = \frac{1}{k} \sum_{j=0}^{k-1} \frac{1-t^6}{(1-\omega^j t^3)(1-\omega^{-j} t^3)(1-\omega^j t^2)(1-\omega^{-j} t^2)(1-t^2)}~,
\eea
where $\omega^k =1$ and $t$ is the fugacity of R-charge in the unit of $1/3$.
This can be written in a closed form as
\bea
\gm (t, D_3/\BZ_k) &=& \frac{1}{(1-t) \left(1-t^4\right) \left(1-t^5\right) \left(1+t^k\right)\left(1-t^{3 k}\right)} \times (1-t+t^2-t^3+ \nn \\
&& + t^4+t^k+2 t^{2 k}+t^{3 k}+t^{4 k}-t^{1+k}+t^{2+k}-t^{3+k}+t^{4+k}+2 t^{2+2 k}+2 t^{4+2 k} \nn \\
&& -t^{1+3 k}+t^{2+3 k}-t^{3+3 k}+t^{4+3 k}-t^{1+4 k}+t^{2+4 k}-t^{3+4 k}+t^{4+4 k})~. \label{d3hsk}
\eea
Setting $k=1$, we obtain (5.38) of \cite{phase}.  As $k \rightarrow \infty$, we recover the Hilbert series of $SPP$:
\bea
\frac{1+t^5}{\left(1-t^2\right)\left(1-t^5\right)  \left(1-t^4\right) }~.
\eea
The plethystic logarithm of the Hilbert series \eref{d3hsk} can be written as
\bea
\PL [\gm (t;  D_3/\BZ_k )]=  \left \{
\begin{array}{ll}
3 t^2+2 t^3-t^6 & \text{for  } k = 1 \\
t^2+t^4 + 2t^5+ 2 t^{2k} + 2t^{2k+1} -t^{10} +\ldots  & \text{for } k > 1\\
t^2+t^4 + 2t^5 -t^{10} & \text{for } k \rightarrow \infty~.
\end{array} \right .
\eea

\paragraph{The Hilbert series for $\left( \BC^2/\BZ_2 \times \BC^2 \right)/\BZ_k$.} Note that the $Q_D$ matrix \eref{e:qdresq111z2ph3} contains the charges under $U(1)_3$.  Therefore, the $\BZ_k$ orbifold acts on the generators like, let's say $U(1)_1$, namely $(0,1,-1)$ on $(p_1,p_2,r_ir_js_1s_2)$ and the Hilbert series is
\bea \label{hsc2z2c2k}
\gm \left(t; \left( \BC^2/\BZ_2 \times \BC^2 \right)/\BZ_k \right) &=& \frac{1}{k} \sum_{j=0}^{k-1} \frac{1- \omega^{-2j} t^4}{\left(1-t\right)\left(1- \omega^j t \right)\left(1- \omega^{-j} t^2 \right)^3} \nn \\
&=& \frac{1+t^3+t^k+2 k t^{2 k}-t^{3 k}-t^{4 k}+t^{3+k}-2 k t^{3+2 k}-t^{3+3 k}-t^{3+4 k}}{(1-t) \left(1-t^3\right)^2 \left(1-t^{2 k}\right)^2}~, \nn\\
\eea
where $\omega^k =1$ and $t$ is the fugacity of R-charge in the unit of $1/2$.
Setting $k=1$, we recover \eref{gmesph2c2z2xc2}.  As $k \rightarrow \infty$, we obtain
\bea
\frac{1-t^6}{(1-t)(1-t^3)^3}~. 
\eea
The plethystic logarithm of the Hilbert series \eref{hsc2z2c2k} can be written as
\bea
\PL [\gm (t; \left( \BC^2/\BZ_2 \times \BC^2 \right)/\BZ_k ]=  \left \{
\begin{array}{ll}
2t+3t^2-t^4 & \text{for  } k = 1 \\
t^k + (2k+1)t^{2k} + t+3 t^3-t^6 + \ldots  & \text{for } k > 1\\
t+3 t^3-t^6 & \text{for } k \rightarrow \infty~.
\end{array} \right .
\eea

\subsection{The $dP_0$ (or $\BC^3/\BZ_3$) theory with 
$k$ units of 4-form flux on $\BP^2$}
The Calabi-Yau $\BC^3/\BZ_3$ has one vanishing 4-cycle, $\BP^2$, and one vanishing 2-cycle, $\BP^1$ inside $\BP^2$.  Each of the 3 nodes in the quiver diagram (\fref{f:m111}) corresponds to a gauge field living on a bound state of D6-branes wrapped on $\BP^2$, D4-branes wrapped on $\BP^1$, and D2-branes.  The fractional brane charges associated with these 3 nodes are given in (8.2) of \cite{Franco:2002ae}:
\bea \label{fracchargedp0}
\mathrm{ch} (F_1) = (2, -1, -1/2), \quad \mathrm{ch} (F_2) = (-1, 1, -1/2), \quad \mathrm{ch} (F_3) =  (-1,0,0)~.
\eea
Let us denote the RR 4-form flux on the $\BP^2$ by $k_4$ and the RR 2-form flux on the $\BP^1$ by $k_2$. Thus, their contributions to the CS levels are determined by \eref{fracchargedp0} as follows:
\bea
\vec{k} = (2k_4 - k_2, -k_4+k_2, -k_4)~.
\eea
Setting $k_2 = 0$ and $k_4 = k$, we obtain
\bea
\vec{k} = (2k, -k, -k)~.
\eea
Therefore, the $\BC^3/\BZ_3$ background with $0$ units of 2-form flux on $\BP^1$ and $k$ units of 4-form flux on $\BP^2$ theory can be lifted to $M^{1,1,1}/\BZ_k$.

\paragraph{The Hilbert series.}  From the $Q_D$ matrix \eref{e:qdfano24}, it can be seen that the D-term charges come from $U(1)_3 - U(1)_1$.  Hence, the D-term corresponding to $U(1)_2$ is non-zero.  Therefore, the $\BZ_k$ orbifold acts on the perfect matchings like $U(1)_2$, namely $(0,0,0, -1,1,0)$ on $(p_1,p_2,p_3,r_1,r_2,s_1)$.
The Hilbert series is given by
\bea
\gm (t; M^{1,1,1}/\BZ_k)=  \frac{1}{k} \sum_{j=0}^{k-1} \oint \limits_{|z| =1} {\frac{\ud z}{2 \pi i z }}\oint \limits_{|b| =1} {\frac{\ud b}{2 \pi i b }} \frac{1}{\left(1- t^4 z\right)^3 \left(1-\frac{t^3}{b z } \omega^{-j} \right)\left(1-\frac{t^3}{b z} \omega^{j} \right)\left(1- \frac{b^2}{z}\right)}~,  \nn \\
\label{e:highkHStrfano24}
\eea
where $\omega^k =1$ and $t$ is the fugacity of R-charge in units of $1/9$.  This expression can be written in a closed form as
\bea \label{m111hsk}
\gm (t; M^{1,1,1}/\BZ_k; \text{odd $k$})&=& \frac{1}{\left(1-t^{18}\right)^3 \left(1-t^{18 k}\right)^3} \times (1+7 t^{18}+t^{36}-t^{18 k}+9 k t^{18 k}+9 k^2 t^{18 k}  \nn\\
&& -t^{36 k}-9 k t^{36 k} +9 k^2 t^{36 k}+t^{54 k}-7 t^{18+18 k}-18 k^2 t^{18+18 k}-t^{36+18 k} \nn \\
&& -9 k t^{36+18 k}+9 k^2 t^{36+18 k} -7 t^{18+36 k}-18 k^2 t^{18+36 k}-t^{36+36 k}\nn \\
&& +9 k t^{36+36 k} +9 k^2 t^{36+36 k}+7 t^{18+54 k}+t^{36+54 k}) := f(k) ~, \nn \\
\gm (t; M^{1,1,1}/\BZ_k; \text{even $k$})&=& f(k/2)~. 
\eea
Setting $k=1$, we recover \eref{unrefHSM111}, as expected.  In the limit $k \rightarrow \infty$, we find the Hilbert series of the $dP_0$ theory \cite{pleth}:
\bea
\frac{1+7 t^{18}+t^{36}}{\left(1-t^{18}\right)^3}~.
\eea

The fully refined plethystic logarithm of the $\BZ_k$ CS orbifold (for $k>2$)\footnote{The plethystic logarithm for $k=1$ is given by \eref{plm111} and the expression for $k=2$ is identical.} of $M^{1,1,1}$ can be written as
{\small
\bea 
\PL [\gm (t,x, y_1,y_2;  M^{1,1,1}/\BZ_k) ]=  \left \{
\begin{array}{ll}
 \left[3k,0\right] (t^9 x)^{2k} + \left[3k,0\right] (t^9/x)^{2k}+  [3,0] t^{18}- [2,2] t^{36} + \ldots & \text{for odd $k$}\\
  \left[\frac{3}{2}k,0\right] (t^9 x)^{k} + \left[\frac{3}{2}k,0\right] (t^9/x)^{k}+  [3,0] t^{18}- [2,2] t^{36} + \ldots & \text{for even $k$}~. \qquad
\end{array} \right .
\eea}
For $k=1,2$ the global mesonic symmetry is $SU(3) \times SU(2) \times U(1)$, but when $k>2$ it can be seen that this symmetry is broken down to $SU(3) \times U(1) \times U(1)$, where $x$ is now regarded as a $U(1)$ fugacity.
As $k \rightarrow \infty$, the plethystic logarithm becomes
\bea
\PL [\gm (t,y_1,y_2;  dP_0) ]= [3,0] t^{18}- [2,2] t^{36} + ([1,4]+[4,1]+[1,1]+[2,2])t^{54} + \ldots~, \qquad \quad
\eea
and the global mesonic symmetry is $SU(3) \times U(1)$.

\subsection{The $\BF_0$ theory with $k$ units of RR 4-form flux on $\BP^1 \times \BP^1$}
The $\BF_0$ theory has 1 vanishing 4-cycle, $\BP^1 \times \BP^1$, and 2 vanishing 2-cycles, $\BP^1_1$ and $\BP^1_2$.  Each of the 4 nodes in the quiver diagram (\fref{f:phase1f0}) corresponds to a collection of fractional branes, whose charges are given respectively as (see (4.15) of \cite{Hanany:2001py}):
\bea
\mathrm{ch}(F_1) = (-1, 1,0, 0), \quad \mathrm{ch}(F_2) = (1, -1, 1, -1), \quad \mathrm{ch}(F_3) = (1, 0, -1, 0), \quad \mathrm{ch}(F_4) = (-1,0, 0,0)~, \nn \\
\eea
Let us denote by $k_1$ and $k_2$ the RR 2-form fluxes on $\BP^1_1$ and $\BP^1_2$, and by $k_4$ the RR 4-form flux on $\BP^1 \times \BP^1$. Then, the CS levels are
\bea
\vec{k} = (-k_4 + k_1, k_4+k_2-k_1, k_4-k_2, -k_4)~.
\eea
Setting $k_1 = 0, k_2 = 0, k_4 =-k$, we obtain
\bea
\vec{k} = (k,-k,-k,k)~.
\eea
Thus, the $\BF_0$ theory with $-k$ units of 4-form flux on $\BP^1 \times \BP^1$ can be lifted to Phase I of the $Q^{1,1,1}/\BZ_2$ with the CS levels $(k,-k,-k,k)$.

\paragraph{The Hilbert series.} 
From the $Q_D$ matrix \eref{qdph1q111z2}, it can be seen that the D-term charges comes from $U(1)_1+U(1)_3$ and $U(1)_3+U(1)_4$.  Hence, the D-term corresponding to $U(1)_2$ is non-zero.  Therefore, the $\BZ_k$ orbifold acts on the perfect matchings like $U(1)_2$, namely $(0,-1,0, 0,1,1,-1,0)$ on $(p_1,p_2,q_1,q_2,r_1,r_2,s_1,s_2)$. Then, the Hilbert series is given by
{\small
\bea
\gm (t, Q^{1,1,1}/(\BZ_k \times \BZ_2)) &=&  \frac{1}{k} \sum_{j=0}^{k-1} \left( \prod_{i=1}^2 \oint \limits_{|z_i| =1} {\frac{\ud z_i}{2 \pi i z_i }}\oint \limits_{|b_i| =1} \frac{\ud b_i}{2 \pi i b_i } \right)  \frac{1}{(1-t b_1 z_1)\left(1-t b_1 z_1 \omega ^{-j}\right)(1- \frac{t}{z_1})^2} \nn\\
&& \times \frac{1}{ \left(1-t \omega ^j \frac{z_2}{b_1 b_2}\right)^2\left(1- \omega ^{-j} \frac{b_2^2}{ z_2}\right)\left(1-\frac{1}{z_2}\right)}~,
\eea}
where $\omega^k =1$ and $t$ is the fugacity of R-charge in the unit of $1/3$.  This expression can be written in a closed form as
\bea
\gm (t, Q^{1,1,1}/(\BZ_k \times \BZ_2)) &=& \frac{1}{(1 - t^6)^3 (1 - t^{6 k})^3} \times \big(1+6 t^6+t^{12}-t^{6 k}+8 k t^{6 k}+8 k^2 t^{6 k}-\nn \\
&& t^{12 k}-8 k t^{12 k}+ 8 k^2 t^{12 k}+t^{18 k}-6 t^{6+6 k}-16 k^2 t^{6+6 k}-t^{12+6 k}- \nn \\
&& 8 k t^{12+6 k}+8 k^2 t^{12+6 k}-6 t^{6+12 k}-16 k^2 t^{6+12 k}-t^{12+12 k}+  \nn \\
&& 8 k t^{12+12 k}+8 k^2 t^{12+12 k}+6 t^{6+18 k}+t^{12+18 k} \big)~.  \label{q111z2hsk}
\eea
Setting $k=1$, we recover the formula (6.23) of \cite{phase}. In the limit $k \rightarrow \infty$, we obtain
\bea
\frac{1+6 t^6+t^{12}}{\left(1-t^6\right)^3}~,
\eea
which is the mesonic Hilbert series of $\BF_0$ (see (3.26) of \cite{pleth}).

The fully refined plethystic logarithm of the $\BZ_k$ CS orbifold (for $k >1$)\footnote{The formula for $k=1$ is given in (6.24) of \cite{phase}.} of $Q^{1,1,1}/\BZ_2$ can be written as
\bea 
\PL \left[\gm \left(t,x,y,z;  \frac{Q^{1,1,1}}{\BZ_k \times \BZ_2}\right)\right] &=& [2k;2k] (t^3 x)^{2k}+[2k;2k] (t^3/x)^{2k}+ \left[2;2 \right] t^{6} -([0;0] \nn \\
&& +[2;2]+[4;0]+[0;4]) t^{12}+ \ldots ~. \qquad
\eea
For $k=1$ the global mesonic symmetry is $SU(2) \times SU(2) \times SU(2) \times U(1)$, but when $k>1$ it can be seen that this symmetry is broken down to $SU(2) \times SU(2) \times U(1) \times U(1)$, where $x$ is now regarded as a $U(1)$ fugacity.
As $k \rightarrow \infty$, the plethystic logarithm becomes
\bea
\PL \left[\gm \left(t,y,z; \BF_0 \right)\right] &=& \left[2;2 \right] t^{6} -([0;0]+[2;2]+[4;0]+[0;4]) t^{12} + ([2;4]\nn \\
&& + [4;2]+2[2;2] +[4;0]+[0;4]+[2;0]+[0;2])t^{18} + \ldots~, \qquad \quad
\eea
and the mesonic symmetry is $SU(2) \times SU(2) \times U(1)$.

\acknowledgments
We are indebted to Yang-Hui He and Alberto Zaffaroni for a closely related collaboration.  
J.~D.~ would like to thank the STFC for his studentship.
A.~H.~ would like to thank the kind hospitality of the KITP in Santa Barbara, the Galileo Galilei Institute for Theoretical Physics, the INFN, the Benasque Center for Theoretical Physics, the Institute for Advanced Study in Princeton, and the Simons Center for Geometry and Physics
during the various stages of this work.
This research was supported in part by the National Science Foundation under Grant No. PHY05-51164. 
N.~M.~ is grateful to the 49th Cracow School of Theoretical Physics, the Galileo Galilei Institute for Theoretical Physics, Anannit Sumawong and Edward O'Reilly for their kind hospitality during the completion of this work. He also thanks Alexander Shannon for useful discussions as well as his family for the warm encouragement and support. 
This research is supported by the DPST project, the Royal Thai Government, and the Imperial College node of the Marie Curie Research and Training Network: MRTN-CT-2004-005616 (ENRAGE).
G.~T.~ wants to express his deep gratitude to his family for the great support during the preparation of this work. He also wants to thank Elisa Rebessi for her endless tenderness and wonderful intelligence, which are incredible sources of joy and encouragement for his life.

%
%


\begin{thebibliography}{99}

\bibitem{BL}
  J.~Bagger and N.~Lambert,
  ``Modeling multiple M2's,''
  Phys.\ Rev.\  D {\bf 75}, 045020 (2007)
  [arXiv:hep-th/0611108].
  ``Gauge Symmetry and Supersymmetry of Multiple M2-Branes,''
  Phys.\ Rev.\  D {\bf 77}, 065008 (2008)
  [arXiv:0711.0955 [hep-th]].
  ``Comments On Multiple M2-branes,''
  JHEP {\bf 0802}, 105 (2008)
  [arXiv:0712.3738 [hep-th]].

\bibitem{gus}
A.~Gustavsson,
  ``Algebraic structures on parallel M2-branes,''
  arXiv:0709.1260 [hep-th]. 
  ``Selfdual strings and loop space Nahm equations,''
  JHEP {\bf 0804}, 083 (2008)
  [arXiv:0802.3456 [hep-th]].

\bibitem{VanRaamsdonk:2008ft}
  M.~Van Raamsdonk,
  ``Comments on the Bagger-Lambert theory and multiple M2-branes,''
  JHEP {\bf 0805}, 105 (2008)
  [arXiv:0803.3803 [hep-th]].

\bibitem{Aharony:2008ug}
  O.~Aharony, O.~Bergman, D.~L.~Jafferis and J.~Maldacena,
  ``N=6 superconformal Chern-Simons-matter theories, M2-branes and their
  gravity duals,''
  arXiv:0806.1218 [hep-th].
  

 \bibitem{Martelli:2008si}
  D.~Martelli and J.~Sparks,
  ``Moduli spaces of Chern-Simons quiver gauge theories and AdS(4)/CFT(3),''
  arXiv:0808.0912 [hep-th].
  
 \bibitem{Ueda:2008hx}
  K.~Ueda and M.~Yamazaki,
  ``Toric Calabi-Yau four-folds dual to Chern-Simons-matter theories,''
  arXiv:0808.3768 [hep-th].
  
\bibitem{Hanany:2008cd}
  A.~Hanany and A.~Zaffaroni,
  ``Tilings, Chern-Simons Theories and M2 Branes,''
  arXiv:0808.1244 [hep-th].
  
 \bibitem{Hanany:2008fj}
  A.~Hanany, D.~Vegh, A.~Zaffaroni,
  ``Brane Tilings and M2 Branes,''
  arXiv:0809.1440.
  
  \bibitem{phase}
  J.~Davey, A.~Hanany, N.~Mekareeya and G.~Torri,
  ``Phases of M2-brane Theories,''
  arXiv:0903.3234 [hep-th].
  
\bibitem{Hanany:2009vx}
  A.~Hanany and Y.~H.~He,
  ``Chern-Simons: Fano and Calabi-Yau,''
  arXiv:0904.1847 [hep-th].

\bibitem{Petrini:2009ur}
  M.~Petrini and A.~Zaffaroni,
  ``N=2 solutions of massive type IIA and their Chern-Simons duals,''
  arXiv:0904.4915 [hep-th].
  
\bibitem{Fabbri:1999hw}
  D.~Fabbri, P.~Fre', L.~Gualtieri, C.~Reina, A.~Tomasiello, A.~Zaffaroni and A.~Zampa,
  ``3D superconformal theories from Sasakian seven-manifolds: New  nontrivial
  evidences for AdS(4)/CFT(3),''
  Nucl.\ Phys.\  B {\bf 577}, 547 (2000)
  [arXiv:hep-th/9907219].
  

\bibitem{Hanany:2005ve}
  A.~Hanany and K.~D.~Kennaway,
  ``Dimer models and toric diagrams,'' hep-th/0503149.

\bibitem{Franco:2005rj}
  S.~Franco, A.~Hanany, K.~D.~Kennaway, D.~Vegh and B.~Wecht,
  ``Brane dimers and quiver gauge theories,''
  JHEP {\bf 0601}, 096 (2006)
  [arXiv:hep-th/0504110].
  
\bibitem{Hanany:2005ss}
  A.~Hanany and D.~Vegh,
  ``Quivers, tilings, branes and rhombi,''
  JHEP {\bf 0710}, 029 (2007)
  [arXiv:hep-th/0511063].

\bibitem{Franco:2005sm}
  S.~Franco, A.~Hanany, D.~Martelli, J.~Sparks, D.~Vegh and B.~Wecht,
  ``Gauge theories from toric geometry and brane tilings,''
  JHEP {\bf 0601}, 128 (2006)
  [arXiv:hep-th/0505211].

\bibitem{Feng:2005gw}
  B.~Feng, Y.~H.~He, K.~D.~Kennaway and C.~Vafa,
  ``Dimer models from mirror symmetry and quivering amoebae,''
  Adv.\ Theor.\ Math.\ Phys.\  {\bf 12}, 3 (2008)
  [arXiv:hep-th/0511287].
  
\bibitem{Broomhead:2008an}
  N.~Broomhead,
  ``Dimer models and Calabi-Yau algebras,''
  arXiv:0901.4662 [math.AG].
  
 
\bibitem{Kennaway:2007tq}
  K.~D.~Kennaway,
  ``Brane Tilings,''
  Int.\ J.\ Mod.\ Phys.\  A {\bf 22}, 2977 (2007)
  [arXiv:0706.1660 [hep-th]].
  
\bibitem{Yamazaki:2008bt}
  M.~Yamazaki,
  ``Brane Tilings and Their Applications,''
  Fortsch.\ Phys.\  {\bf 56}, 555 (2008)
  [arXiv:0803.4474 [hep-th]].
   


\bibitem{Feng:2000mi}
  B.~Feng, A.~Hanany and Y.~H.~He,
  ``D-brane gauge theories from toric singularities and toric duality,''
  Nucl.\ Phys.\  B {\bf 595}, 165 (2001)
  [arXiv:hep-th/0003085].
  
\bibitem{Feng:2001xr}
  B.~Feng, A.~Hanany and Y.~H.~He,
  ``Phase structure of D-brane gauge theories and toric duality,''
  JHEP {\bf 0108}, 040 (2001)
  [arXiv:hep-th/0104259].

\bibitem{Feng:2002zw}
  B.~Feng, S.~Franco, A.~Hanany and Y.~H.~He,
  ``Symmetries of toric duality,''
  JHEP {\bf 0212}, 076 (2002)
  [arXiv:hep-th/0205144].

\bibitem{Feng:2002fv}
  B.~Feng, S.~Franco, A.~Hanany and Y.~H.~He,
  ``Unhiggsing the del Pezzo,''
  JHEP {\bf 0308}, 058 (2003)
  [arXiv:hep-th/0209228].
  
  \bibitem{Feng:2001bn}
  B.~Feng, A.~Hanany, Y.~H.~He and A.~M.~Uranga,
  ``Toric duality as Seiberg duality and brane diamonds,''
  JHEP {\bf 0112}, 035 (2001)
  [arXiv:hep-th/0109063].\\
   C.~E.~Beasley and M.~R.~Plesser,
  ``Toric duality is Seiberg duality,''
  JHEP {\bf 0112}, 001 (2001)
  [arXiv:hep-th/0109053].
  
\bibitem{Franco:2003ea}
  S.~Franco, A.~Hanany and Y.~H.~He,
  ``A trio of dualities: Walls, trees and cascades,''
  Fortsch.\ Phys.\  {\bf 52}, 540 (2004)
  [arXiv:hep-th/0312222].
    
\bibitem{Franco:2003ja}
  S.~Franco, A.~Hanany, Y.~H.~He and P.~Kazakopoulos,
  ``Duality walls, duality trees and fractional branes,''
  arXiv:hep-th/0306092.

\bibitem{Franco:2002mu}
  S.~Franco and A.~Hanany,
  ``Toric duality, Seiberg duality and Picard-Lefschetz transformations,''
  Fortsch.\ Phys.\  {\bf 51}, 738 (2003)
  [arXiv:hep-th/0212299].

\bibitem{Feng:2002kk}
  B.~Feng, A.~Hanany, Y.~H.~He and A.~Iqbal,
  ``Quiver theories, soliton spectra and Picard-Lefschetz transformations,''
  JHEP {\bf 0302}, 056 (2003)
  [arXiv:hep-th/0206152].

\bibitem{Forcella:2008ng}
  D.~Forcella, A.~Hanany and A.~Zaffaroni,
  ``Master Space, Hilbert Series and Seiberg Duality,''
  arXiv:0810.4519 [hep-th].
  
  
\bibitem{Franco:2008um}
  S.~Franco, A.~Hanany, J.~Park and D.~Rodriguez-Gomez,
  ``Towards M2-brane Theories for Generic Toric Singularities,''
  JHEP {\bf 0812}, 110 (2008)
  [arXiv:0809.3237 [hep-th]].

\bibitem{taxonomy}
  A.~Hanany and Y.~H.~He,
  ``M2-Branes and Quiver Chern-Simons: A Taxonomic Study,''
  arXiv:0811.4044 [hep-th].
  
\bibitem{Franco:2009sp}
  S.~Franco, I.~R.~Klebanov and D.~Rodriguez-Gomez,
  ``M2-branes on Orbifolds of the Cone over $Q^{1,1,1}$,''
  arXiv:0903.3231 [hep-th].
  
\bibitem{Amariti:2009rb}
  A.~Amariti, D.~Forcella, L.~Girardello and A.~Mariotti,
  ``3D Seiberg-like Dualities and M2 Branes,''
  arXiv:0903.3222 [hep-th].
  

  
  
\bibitem{Lee:2006hw}
  S.~Lee,
  ``Superconformal field theories from crystal lattices,''
  Phys.\ Rev.\  D {\bf 75}, 101901 (2007)
  [arXiv:hep-th/0610204].

\bibitem{Lee:2007kv}
  S.~Lee, S.~Lee and J.~Park,
  ``Toric AdS(4)/CFT(3) duals and M-theory crystals,''
  JHEP {\bf 0705}, 004 (2007)
  [arXiv:hep-th/0702120].
  
\bibitem{Kim:2007ic}
  S.~Kim, S.~Lee, S.~Lee and J.~Park,
  ``Abelian Gauge Theory on M2-brane and Toric Duality,''
  Nucl.\ Phys.\  B {\bf 797}, 340 (2008)
  [arXiv:0705.3540 [hep-th]].
  
 \bibitem{Imamura:2008qs}
  Y.~Imamura and K.~Kimura,
  ``Quiver Chern-Simons theories and crystals,''
  arXiv:0808.4155 [hep-th].


\bibitem{Beasley:1999uz}
  C.~Beasley, B.~R.~Greene, C.~I.~Lazaroiu and M.~R.~Plesser,
  ``D3-branes on partial resolutions of abelian quotient singularities of Calabi-Yau threefolds,''
  Nucl.\ Phys.\  B {\bf 566}, 599 (2000)
  [arXiv:hep-th/9907186].
  
\bibitem{Park:1999ep}
  J.~Park, R.~Rabadan and A.~M.~Uranga,
  ``Orientifolding the conifold,''
  Nucl.\ Phys.\  B {\bf 570}, 38 (2000)
  [arXiv:hep-th/9907086].
  
\bibitem{Gubser:1998vd}
  S.~S.~Gubser,
  ``Einstein manifolds and conformal field theories,''
  Phys.\ Rev.\  D {\bf 59}, 025006 (1999)
  [arXiv:hep-th/9807164].
  
\bibitem{Butti:2005vn}
  A.~Butti and A.~Zaffaroni,
  ``R-charges from toric diagrams and the equivalence of a-maximization and
  Z-minimization,''
  JHEP {\bf 0511}, 019 (2005)
  [arXiv:hep-th/0506232].

  

\bibitem{pleth}
  S.~Benvenuti, B.~Feng, A.~Hanany and Y.~H.~He,
  ``Counting BPS operators in gauge theories: Quivers, syzygies and plethystics,''
  JHEP {\bf 0711}, 050 (2007)
  [arXiv:hep-th/0608050].
  
  A.~Hanany and C.~Romelsberger,
  ``Counting BPS operators in the chiral ring of N = 2 supersymmetric gauge
  theories or N = 2 braine surgery,''
  Adv.\ Theor.\ Math.\ Phys.\  {\bf 11}, 1091 (2007)
  [arXiv:hep-th/0611346].
  
  B.~Feng, A.~Hanany and Y.~H.~He,
  ``Counting gauge invariants: The plethystic program,''
  JHEP {\bf 0703}, 090 (2007)
  [arXiv:hep-th/0701063].
  
  D.~Forcella, A.~Hanany and A.~Zaffaroni,
  ``Baryonic generating functions,''
  JHEP {\bf 0712}, 022 (2007)
  [arXiv:hep-th/0701236].
  
  J.~Gray, A.~Hanany, Y.~H.~He, V.~Jejjala and N.~Mekareeya,
  ``SQCD: A Geometric Apercu,''
  JHEP {\bf 0805}, 099 (2008)
  [arXiv:0803.4257 [hep-th]].
  
   A.~Hanany and N.~Mekareeya,
  ``Counting Gauge Invariant Operators in SQCD with Classical Gauge Groups,''
  JHEP {\bf 0810}, 012 (2008)
  [arXiv:0805.3728 [hep-th]].
  
  A.~Hanany, N.~Mekareeya and G.~Torri,
  ``The Hilbert Series of Adjoint SQCD,''
  arXiv:0812.2315 [hep-th].

\bibitem{Hanany:2008qc}
  A.~Hanany, N.~Mekareeya and A.~Zaffaroni,
  ``Partition Functions for Membrane Theories,''
  JHEP {\bf 0809}, 090 (2008)
  [arXiv:0806.4212 [hep-th]].
  
\bibitem{master}
 D.~Forcella, A.~Hanany, Y.~H.~He and A.~Zaffaroni,
  ``The Master Space of N=1 Gauge Theories,''
  JHEP {\bf 0808}, 012 (2008)
  [arXiv:0801.1585 [hep-th]]; 
  `Mastering the Master Space,''
  Lett.\ Math.\ Phys.\  {\bf 85}, 163 (2008)
  [arXiv:0801.3477 [hep-th]].
    D.~Forcella,
  ``Master Space and Hilbert Series for N=1 Field Theories,''
  arXiv:0902.2109 [hep-th].
  
\bibitem{Kim:2009wb}
  S.~Kim,
  ``The complete superconformal index for N=6 Chern-Simons theory,''
  arXiv:0903.4172 [hep-th].

  \bibitem{Butti:2007jv}
  A.~Butti, D.~Forcella, A.~Hanany, D.~Vegh and A.~Zaffaroni,
  ``Counting Chiral Operators in Quiver Gauge Theories,''
  JHEP {\bf 0711}, 092 (2007)
  [arXiv:0705.2771 [hep-th]].
  
\bibitem{Benvenuti:2004dw}
  S.~Benvenuti and A.~Hanany,
  ``New results on superconformal quivers,''
  JHEP {\bf 0604}, 032 (2006)
  [arXiv:hep-th/0411262].

\bibitem{Franco:2006gc}
  S.~Franco and D.~Vegh,
  ``Moduli spaces of gauge theories from dimer models: Proof of the
  correspondence,''
  JHEP {\bf 0611} (2006) 054
  [arXiv:hep-th/0601063].
  
\bibitem{Davide}
  A.~Amariti, D.~Forcella, L.~Girardello and A.~Mariotti,
  ``3D Seiberg-like Dualities and M2 Branes,''
  arXiv:0903.3222 [hep-th].

\bibitem{Seba}
S.~Franco, I.~Klebanov and D.~Rodriguez-Gomez,
``M2-branes on Orbifolds of the Cone over $Q^{1,1,1}$,''
arXiv:0903.3231 [hep-th].

\bibitem{Bergman:1999na}
  O.~Bergman, A.~Hanany, A.~Karch and B.~Kol,
  ``Branes and supersymmetry breaking in 3D gauge theories,''
  JHEP {\bf 9910}, 036 (1999)
  [arXiv:hep-th/9908075].
  
\bibitem{Aganagic:2001ug}
  M.~Aganagic and C.~Vafa,
  ``G(2) manifolds, mirror symmetry and geometric engineering,''
  arXiv:hep-th/0110171.
  
\bibitem{Imamura:2009ur}
  Y.~Imamura,
  ``Monopole operators in N=4 Chern-Simons theories and wrapped M2-branes,''
  arXiv:0902.4173 [hep-th].
  
\bibitem{Imamura:2008nn}
  Y.~Imamura and K.~Kimura,
  ``On the moduli space of elliptic Maxwell-Chern-Simons theories,''
  Prog.\ Theor.\ Phys.\  {\bf 120}, 509 (2008)
  [arXiv:0806.3727 [hep-th]].
  
\bibitem{Imamura:2008ji}
  Y.~Imamura and S.~Yokoyama,
  ``N=4 Chern-Simons theories and wrapped M5-branes in their gravity duals,''
  arXiv:0812.1331 [hep-th].
  

\bibitem{Aganagic:2009zk}
  M.~Aganagic,
  ``A Stringy Origin of M2 Brane Chern-Simons Theories,''
  arXiv:0905.3415 [hep-th].

\bibitem{ucsbtalk}
A.~Hanany,
``Finding M Theory Duals to Type IIA Backgrounds with RR Fluxes''
(talk at \emph{Fundamental Aspects of Superstring Theory}),
May 29, 2009, Kavli Institute for Theoretical Physics (KITP): 
{\sf http://online.itp.ucsb.edu/online/strings09/hanany2/}
  
\bibitem{Franco:2002ae}
  S.~Franco and A.~Hanany,
  ``Geometric dualities in 4d field theories and their 5d interpretation,''
  JHEP {\bf 0304}, 043 (2003)
  [arXiv:hep-th/0207006].
  
\bibitem{Hanany:2001py}
  A.~Hanany and A.~Iqbal,
  ``Quiver theories from D6-branes via mirror symmetry,''
  JHEP {\bf 0204}, 009 (2002)
  [arXiv:hep-th/0108137].
  
\bibitem{Uranga:1998vf}
  A.~M.~Uranga,
  ``Brane Configurations for Branes at Conifolds,''
  JHEP {\bf 9901}, 022 (1999)
  [arXiv:hep-th/9811004].
  
\bibitem{Erlich:1999rb}
  J.~Erlich, A.~Hanany and A.~Naqvi,
  ``Marginal deformations from branes,''
  JHEP {\bf 9903}, 008 (1999)
  [arXiv:hep-th/9902118].
  


\end{thebibliography}
\end{document}